\documentclass[12pt]{article}
\usepackage{amsfonts}
\usepackage{amsmath}
\usepackage{amssymb}
\usepackage{amsthm}
\usepackage{accents}
\usepackage{youngtab}
\usepackage{braket}
\usepackage{graphicx}
\usepackage[svgnames]{xcolor}
\usepackage{ytableau,tikz}
\usepackage{here}
\usepackage{comment}
\usepackage{simplewick}
\usepackage{tikz}
\usepackage{feynmf}

\newcommand{\ba}{\begin{eqnarray}}
\newcommand{\ea}{\end{eqnarray}}
\newcommand{\bal}{\begin{align}}
\newcommand{\eal}{\end{align}}
\newcommand{\nn}{\nonumber}

\newcommand{\cN}{\mathcal{N}}
\newcommand{\cW}{\mathcal{W}}

\newcommand{\cX}{\mathcal{X}}

\newcommand{\cY}{\mathcal{Y}}
\newcommand{\lt}{\left}
\newcommand{\rt}{\right}

\newcommand{\cob}{\color{blue}}
\newcommand{\gammah}{\hat{\gamma}}

 \DeclareMathOperator*{\Res}{Res}

 \newcommand{\uwidehat}[1]{%
  \mathpalette\douwidehat{#1}%
}

\makeatletter
\newcommand{\douwidehat}[2]{%
  \sbox0{$\m@th#1\widehat{\hphantom{#2}}$}%
  \sbox2{$\m@th#1x$}
  \sbox4{$\m@th#1#2$}
  \dimen0=\ht0
  \advance\dimen0 -.8\ht2
  \dimen2=\dp4
  \rlap{%
    \raisebox{\dimexpr\dimen0-\dimen2}{%
      \scalebox{1}[-1]{\box0}%
    }%
  }%
  {#2}%
}
\makeatother

 \makeatletter
\@addtoreset{equation}{section}

\makeatother

\usepackage[colorlinks=true,linkcolor=black,citecolor=black,urlcolor=black]{hyperref}

\begin{document}

\begin{titlepage}
\vspace*{-2cm}
	\begin{flushright}
		DIAS-STP-19-06
	\end{flushright}
	
	\vskip 1cm
	
	\begin{center}
		\textbf{\LARGE Web Construction of\\[.5em] ABCDEFG and Affine Quiver Gauge Theories}
	\end{center}
	\vskip 2.5cm
		\begin{center}
			{\Large Taro Kimura$^\clubsuit$ and Rui-Dong Zhu$^\spadesuit$}
			\\[.6cm]
			{\em  $^\clubsuit$Department of Physics, Keio University, Kanagawa, Japan}\\
		        {\em  $^\spadesuit$School of Theoretical Physics, Dublin Institute for Advanced Studies}\\
			{\em  10 Burlington Road, Dublin, Ireland}
			\\[.4cm]
		\end{center}
	\vfill
	\begin{abstract}
	The topological vertex formalism for 5d $\cN=1$ gauge theories is not only a convenient tool to compute the instanton partition function of these theories, but it is also accompanied by a nice algebraic structure that reveals various kinds of nice properties such as dualities and integrability of the underlying theories. The usual refined topological vertex formalism is derived for gauge theories with $A$-type quiver structure (and $A$-type gauge groups). In this article, we propose a construction with a web of vertex operators for all $ABCDEFG$-type and affine quivers by introducing several new vertices into the formalism, based on the reproducing of known instanton partition functions and qq-characters for these theories. 
	\end{abstract}
	\vfill
\end{titlepage}

\noindent\hrulefill
\tableofcontents
\noindent\hrulefill

\section{Introduction}

The exact computation of partition functions on various curved spaces via localization for supersymmetric Yang-Mills theories starting from \cite{Pestun} (see \cite{loc-review} for a nice review) provides extremely powerful ways to examine different types of dualities derived from the superstring theory. Among them, the Alday-Gaiotto-Tachikawa (AGT) relation \cite{AGT,Wyllard09}, which states that the Nekrasov partition function of a 4d $\cN=2$ gauge theory with gauge group SU($N$) can be encoded in terms of some conformal block in 2d CFT with $\cW_N$ symmetry, is in particular intriguing for us. It can be even uplift to gauge theories with eight supercharges in 5d \cite{5dAGT} and 6d \cite{ellip-Virasoro,IKY,Kimura:2016dys}, where the corresponding 2d CFT is respectively $q$-deformed and elliptically deformed. Thanks to the string duality \cite{Leung-Vafa} between topological string theory and 5d $\cN=1$ gauge theory, we can reformulate the 5d version of the AGT relation in the language of topological vertices \cite{AKMV,Awata:2005fa,IKV} on the toric Calabi-Yau geometry when we restrict our attention to 5d gauge theories constructed from $(p,q)$ 5-brane webs introduced in~\cite{AHK}. In this context, one finds \cite{AFS} a beautiful algebraic structure of the (refined) topological vertex, which is often called the Ding-Iohara-Miki (DIM) algebra \cite{DI,Miki} or the quantum toroidal algebra of $\mathfrak{gl}_1$, and the dual $q$-deformed $\cW$-algebra can be obtained as a truncation of this algebra due to the finite number of D5 branes in the system. One can also perform a fiber-base duality~\cite{fiber-base} to the brane web to find a dual $q$-deformed $\cW_M$-algebra associated to the gauge theory, where $M$ is the number of NS5 branes in the construction. This $\cW_M$-algebra is nothing but the quiver $\cW$-algebra discovered in \cite{Kimura-Pestun}. 

The ADHM construction \cite{ADHM}, which lies behind the localization calculation \cite{LMNS,Moore:1997dj,NekrasovInstanton} on $\Omega$-background, can be extended to SO and Sp type gauge groups \cite{Marino:2004cn,ABCD-instanton}. We expect the AGT relation holds for any gauge group, and indeed it has been examined for $ABCDEFG$-type (i.e. all Lie algebraic) gauge groups at one instanton level in \cite{ABCDEFG-instanton} in 4d that this is true. The examination in 5d is even more difficult to perform and more non-trivial. One thus may want to rely on the topological vertex formalism and the associated algebraic structure to check and understand the underlying principle of this duality. The topological vertex formalism for SO and Sp type gauge groups (with orientifold inserted in the brane web) is proposed in \cite{Kim-Yagi}, and it has been applied to realize the $G$-type gauge group via Higgsing an SO(7) gauge group \cite{G-type} and various kinds of matter contents beyond the fundamental representation \cite{Sp-antisym,rank-3-antisym}. On the other hand, one can consider instead the fiber-base dual setup, where we can still use $A$-type gauge groups but with more general quiver structures. The advantage to consider this setup is that we have a closed-form expression for the Nekrasov partition function at each node and the quiver structure tells us how to sew these factors together with bifundamental contributions. Simply-laced quivers, i.e. $ADE$-type quivers (including $\widehat{ADE}$-type ones), make perfect sense in this context and the topological vertex formalism for $D$-type quivers has been considered in \cite{D-type}. Quiver structure beyond $ADE$-classification has not gathered much attention so far (see the discussion for $B$-type quivers in 3d in \cite{3d-B-type} and general quivers in the context of little strings in \cite{ABCDEFG-littlestring}), but the partition function for a general quiver gauge theory (with $A$-type gauge group) can be written down based on the correspondence with quiver $\cW$-algebras \cite{Fractional-KP}.

In this article, we initiate a program to complete the brane web construction for all Lie-algebraic and affine quivers. The guiding principle is to use the DIM algebra, which is expected to be there from the AGT relation for $A$-type gauge groups, to glue appropriate vertex operators together in the preferred direction (or equivalently the D5 brane direction) of the brane web, so that the partition function and the qq-characters, introduced by Nekrasov~\cite{BPS/CFT}, derived from a Ward identity approach (proposed in \cite{BMZ} and \cite{5dBMZ}) are correctly reproduced. The answer given in this article is more or less ``phenomenological'', and schematically let us denote the vertex operator that corresponds to the refined topological vertex as 
\ba
\exp\lt(\sum_{n=1}^\infty\frac{1}{n}b_n a_{-n}z^n\rt)\exp\lt(\sum_{n=1}^\infty\frac{1}{n}c_n a_{n}z^{-n}\rt),\nn
\ea
where $ a_n$ satisfying $\lt[ a_n, a_m\rt]=n\delta_{n+m,0}$ is the free boson oscillator, then new vertices we need to introduce in this article for non-simply-laced quivers take the form 
\ba
\exp\lt(\sum_{n=1}^\infty\frac{1}{n}b_nd_n a_{-n}z^n\rt)\exp\lt(\sum_{n=1}^\infty\frac{1}{n}c_nd_n^{-1} a_{n}z^{-n}\rt),\nn
\ea
for some appropriate factor $d_n$, which obeys the same contraction rule with itself as the usual refined topological vertex. We also found a similar idea to introduce such a new ``twisted'' vertex applies to the construction of $D$-type quivers, and by establishing the Awata-Feigin-Shiraishi (AFS) property for this new vertex, we see that this new vertex can be interpreted as a merged object of a topological vertex and the effect of the orientifold ${\bf ON}^-$. The realization of $E$-type quivers in the language of brane construction has long been a challenge (c.f. see a recent progress \cite{Magnetic-Quiver} for how to obtain the affine $E^{(1)}_8$ quiver for 3d theories from a D6-D8 system), and it remains to be difficult with all the new vertices introduced in the way described above. We propose a novel construction in this article, based on the observation that topological vertices (including the newly introduced vertices) in the web diagram correspond to the simple roots of the quiver Lie algebra. We adopt the realization of the $E_8$ root system with 8 unit vectors, in which there is a simple root constructed with coefficients $-\frac{1}{2}$, 
\ba
\alpha_7=-\frac{1}{2}e_1-\frac{1}{2}e_2-\frac{1}{2}e_3-\frac{1}{2}e_4-\frac{1}{2}e_5-\frac{1}{2}e_6-\frac{1}{2}e_7-\frac{1}{2}e_8,
\ea
and correspondingly we introduce the ``square-root'' vertex built from the vertex operator, 
\ba
\exp\lt(\frac{1}{2}\sum_{n=1}^\infty\frac{1}{n}b_n a_{-n}z^n\rt)\exp\lt(\frac{1}{2}\sum_{n=1}^\infty\frac{1}{n}c_n a_{n}z^{-n}\rt),\nn
\ea
for the web construction of $E$-type quivers (including affine $E$-type quivers). 

This article is organized as follows: we first review the topological vertex formalism for $A$-type quiver gauge theories in section \ref{s:rev-A}, and the generalized topological vertex formalism to include orientifolds for $D$-type quivers in section \ref{s:rev-D}. We review on the (instanton) partition function obtained in \cite{Fractional-KP} in section \ref{s:Frac-Part} and rewrite it in a form that is more convenient for construction with a web of vertex operators. The web construction for non-simply-laced quivers will then be given in section \ref{s:BCFG}. In sections \ref{s:square-root}, \ref{s:trivalent} and \ref{s:affine}, we make use of the observed correspondence between web construction and simple-root system in the quiver Lie algebra to generalize our work to $E$-type and affine quivers. At the end, we conclude this article by writing down the AFS properties for the new vertices introduced in this article as potential mathematical definitions of them in the DIM algebra. 

\paragraph{Notations} Let us list some repeatedly used notations in this article. For a partition $\lambda=\{\lambda_k \in \mathbb{Z}_{\ge 0} \mid \lambda_{k}\geq \lambda_{k+1}\}_{k=1}^\infty$, which is also graphically represented as a Young diagram, we often want to assign a complex parameter $v_i$, which will be referred to as the exponentiated Coulomb moduli (or Coulomb moduli for short) in this article, to it, and then an important characteristic set of complex numbers defined by 
\ba
\cX_i:=\{v_i q_1^{d_i(k-1)}q_2^{\lambda_k}\}^\infty_{k=1},
\ea
where $q_{1,2}$ are $\Omega$-background parameters and $d_i$ is a positive integer weight number associated to the node $\lambda$ belongs to in the quiver diagram\footnote{Note that the label $i$ here is for the node in the quiver diagram, and the setup here corresponds to having a U(1) gauge group in the $i$-th node. More generally, we can have several Young diagrams, $\lambda^{(1)}$, $\lambda^{(2)}$, \dots, in one node, and consider the characteristic sets $\cX^{(1)}_i$, $\cX^{(2)}_i$, \dots, with respect to Coulomb moduli, $v_i^{(1)}$, $v_i^{(2)}$, \dots }, will frequently appear in the context of instanton counting. Adding and removing boxes from the Young diagram play a key role in the computation in this article, and for a box $s$ added to the $j$-th column of $\lambda$, we denote $\lambda+s=\{\lambda_k+\delta_{k,j}\}_{k=1}^\infty$ and use a short-hand notation for the characteristic number set of $\lambda+s$, 
\ba
\cX_{i+s}:=\{v_i q_1^{d_i(k-1)}q_2^{(\lambda+s)_k}\}^\infty_{k=1}.
\ea
A shift of the Coulomb moduli sometimes appears in the calculation, and we denote the characteristic number set after a shift of $v_i$ by $q$ as
\ba
q\cX_i:=\{qv_i q_1^{d_i(k-1)}q_2^{\lambda_k}\}^\infty_{k=1}. 
\ea
When $\lambda=\emptyset$, we simply use 
\ba
\cX_i^\emptyset :=\{v_i q_1^{d_i(k-1)}\}^\infty_{k=1}.
\ea
A similar notation not to be confused with $\cX_i$ is the coordinate system for boxes in the Young diagram: $\chi_s=v_iq_1^{j-1}q_2^{k-1}$ is the coordinate for the box $s=(j,k)$ in $\lambda$, which is the $q$-deformation of the content $\log v + \epsilon_1(j-1) + \epsilon_2 (k-1)$ for a box $(j,k) \in \lambda$.

An important rational function, 
\ba
S(z):=\frac{(1-q_1z)(1-q_2z)}{(1-z)(1-q_1q_2z)}.
\ea
with the property $S(z^{-1})=S(q_1^{-1}q_2^{-1}z)$, appears a lot in the context of DIM algebra, as it controls the commutation relations of the algebra. With the $S$-function, one can define the following $\cY$-function, 
\ba
\cY_\lambda(z):=(1-v/z)\prod_{x\in\lambda}S(\chi_x/z)=\frac{\prod_{x\in A(\lambda)}1-\chi_x/z}{\prod_{y\in R(\lambda)}1-\chi_y q_3^{-1}/z},\label{def-Y}
\ea
where $A(\lambda)$ is the set of boxes that are allowed to be added to the Young diagram $\lambda$ to obtain another Young diagram, and $R(\lambda)$ is the set of boxes that can be removed from $\lambda$ to form a Young diagram, as a building block of the Nekrasov factor.
Remark that these $A(\lambda)$ and $R(\lambda)$ are also denoted as the outer and inner boundaries $\partial_\pm \lambda$, e.g., in~\cite{NPS}.

For later convenience, we also define $q_3:=q_1^{-1}q_2^{-1}=:\gamma^2$. 

For a vertex operator of the form, 
\ba
V=\exp\lt(\sum_{n=1}^\infty\frac{1}{n}b_n a_{-n}z^n\rt)\exp\lt(\sum_{n=1}^\infty\frac{1}{n}c_n a_{n}z^{-n}\rt),
\ea
we use its $m$-th power $V^m$ to denote 
\ba
V^m:=\exp\lt(m\sum_{n=1}^\infty\frac{1}{n}b_n a_{-n}z^n\rt)\exp\lt(m\sum_{n=1}^\infty\frac{1}{n}c_n a_{n}z^{-n}\rt),
\ea
which reduces to 
\ba
:\underbrace{VV\cdots V}_{\hbox{\footnotesize m}}:,
\ea
when $m$ is a positive integer. We often consider (vertex) operators acting on the tensor product of several vector spaces, and denote the operator $V$ acting only on the $i$-th vector space interchangeably as $V_i$ or $V^{(i)}$.

\section{Review on $A$-type Quiver Gauge Theories}\label{s:rev-A}

In this article, we focus on 5d $\cN=1$ supersymmetric gauge theories with SU-type gauge groups and $\mathfrak{g}$-type quiver structure. The construction of $A$-type quiver gauge theories is well studied in the literature, and the $(p,q)$ 5-brane web construction \cite{AHK} from string theory is one of the most convenient realization of this class of theories. It is dual to M-theory compactified on toric Calabi-Yau, and the instanton partition function can be calculated as the partition function of topological string on the same toric Calabi-Yau. The partition function of topological string on toric Calabi-Yau can be computed in a Feynman-diagram-like way with the topological vertex \cite{AKMV, Awata:2005fa, IKV}. In this section, we give a brief review on the AFS rewriting \cite{AFS} of this topological vertex formalism, and generalize it to a web of vertex operators encoded in the so-called Ding-Iohara-Miki (DIM) algebra in later sections. 

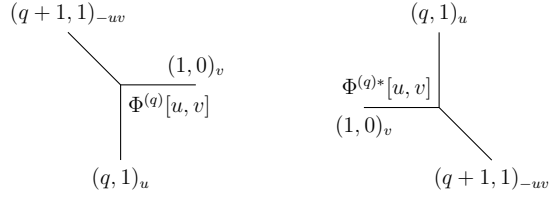
\begin{figure}
	\begin{center}
		\begin{tikzpicture}
		\begin{scope}[xscale = -1]
		\node[below,scale=0.7] at (1,-1) {$(q,1)_u$};
		\node[above,scale=0.7] at (0,0) {$(1,0)_{ v}$};
		\node[above,scale=0.7] at (1.7,0.7) {$(q+1,1)_{-uv}$};
		\node[below right,scale=0.7] at (1,0) {$\Phi^{(q)}[u, v]$};
		\draw (0,0) -- (1,0) -- (1.7,0.7);
		\draw (1,0) -- (1,-1);
		\end{scope}
		\end{tikzpicture}
		\hspace{1cm}
		\begin{tikzpicture}
		\begin{scope}[xscale = -1]
		\node[below,scale=0.7] at (-0.7,-1.7) {$(q+1,1)_{-uv}$};
		\node[above,scale=0.7] at (0,0) {$(q,1)_u$};
		\node[below,scale=0.7] at (1,-1) {$(1,0)_{v}$};
		\node[above left,scale=0.7] at (0,-1) {$\Phi^{(q)\ast}[u,v]$};
		\draw (0,0) -- (0,-1) -- (1,-1);
		\draw (0,-1) -- (-0.7,-1.7);
		\end{scope}
		\end{tikzpicture}
	\end{center}
	\caption{$\Phi^{(q)}[u, v]$ and $\Phi^{(q)\ast}[u, v]$}
	\label{fig_vertex}
\end{figure}

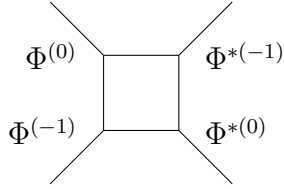
\begin{figure}
\begin{center}
\begin{tikzpicture}
\draw (-0.71,0.71)--(0,0);
\draw (0,0)--(1,0);
\draw (0,0)--(0,-1);
\draw (1,0)--(1,-1);
\draw (0,-1)--(1,-1);
\draw (1,0)--(1.71,0.71);
\draw (1,-1)--(1.71,-1.71);
\draw (0,-1)--(-0.71,-1.71);
\node at (-0.2,0) [left] {$\Phi^{(0)}$};
\node at (-0.2,-1) [left] {$\Phi^{(-1)}$};
\node at (1.2,0) [right] {$\Phi^{\ast(-1)}$};
\node at (1.2,-1) [right] {$\Phi^{\ast(0)}$};
\end{tikzpicture}
\end{center}
\caption{The web diagram of 5d $\cN=1$ pure SU(2) gauge theory and the assignment of topological vertices.}
\label{f:SU(2)}
\end{figure}

A gauge theory of $A$-type quiver is specified by a web diagram of 5-branes (see Table \ref{t:brane-web} for the brane configuration), where a $(p,q)$ 5-brane is drawn as a line, whose angle $\theta$ with the 5-axis is $\tan \theta=q/p$ (the convention here is that D5-brane is denoted as a $(1,0)$ 5-brane and NS5-brane corresponds to a $(0,1)$ 5-brane). When a D5-brane intersects with a $(q,1)$ 5-brane and a $(q+1,1)$ 5-brane at some vertex, we assign to the vertex a refined topological vertex $\Phi^{(q)}$ or $\Phi^{\ast(q)}$, according to the rule that if the D5-brane is stretching to the left (negative direcition of 5-axis), we assign $\Phi^{\ast (q)}$ to the vertex, and if the D5-brane is stretching to the right (positive direcition of 5-axis), we assign $\Phi^{(q)}$ to it (see Figure \ref{fig_vertex}). Each vertex contains two parameters, $v$ and $u$, which respectively describes the position of the D5 and $(q,1)$ 5-brane attached to the vertex\footnote{We will in particular refer $v$'s as the Coulomb moduli in this article.}. The full partition function is given by the vacuum expectation value of the product of all these vertices. For example, the partition function of 5d $\cN=1$ pure SU(2) gauge theory at Chern--Simons level $\kappa = 0$, whose web diagram is shown in Figure \ref{f:SU(2)}, can be written down as 
\ba
Z_{{\rm SU(2)}}&=&\begin{array}{c}
\widehat{\underline{\ 0\ }}\\
\Phi^{(0)}[u,v_2]\\
\Phi^{(-1)}[-u/v_1,v_1]\\
\uwidehat{\overline{\ 0\ }}\\
\end{array}
\begin{array}{c}
\\
\cdot\\
\cdot\\
\\
\end{array}
\begin{array}{c}
\widehat{\underline{\ 0\ }}\\
\Phi^{\ast(-1)}[-u^\ast/v_2,v_2]\\
\Phi^{\ast(0)}[u^\ast,v_1]\\
\uwidehat{\overline{\ 0\ }}\\
\end{array},\label{ex-SU2}
\ea
where we used a ``two-dimensional'' notation to represent in the horizontal direction the multiplication of topological vertices in the preferred direction and other products in the vertical direction, with the convention that $\widehat{\underline{\ 0\ }}$ denotes the bra vacuum state and $\uwidehat{\overline{\ 0\ }}$ corresponds to the ket vacuum state.

\begin{table}
\begin{center}
\begin{tabular}{|c|c|c|c|c|c|c|c|c|c|c|}
\hline
& 0 & 1 & 2 & 3 & 4& 5 & 6 & 7 & 8 & 9 \\
\hline
D5 & $-$ & $-$ & $-$ & $-$ & $-$ & $-$ & $\bullet$ & $\bullet$ & $\bullet$ & $\bullet$ \\
\hline
NS5 & $-$ & $-$ & $-$ & $-$ & $-$ & $\bullet$ & $-$ & $\bullet$ & $\bullet$ & $\bullet$ \\
\hline
7-brane & $-$ & $-$ & $-$ & $-$ & $-$ & $\bullet$ & $\bullet$ & $-$ & $-$ & $-$ \\
\hline
\end{tabular}
\caption{Configuration of branes in the brane web construction. Bar represents the direction branes stretch along, and dot means the point-like direction for branes. General 5-branes lie as a line in the 5-6 plane, and the web diagram encodes all these information on the 5-6 plane.}
\label{t:brane-web}
\end{center}
\end{table}

To be more precise about the meaning of the multiplication of topological vertices and the vacuum state, let us recall that there attaches a Fock-space degree of freedom to each leg of the topological vertex (refer to \cite{melting-crystal} and the formulation of the refined topological vertex itself given in~\cite{IKV}), one can insert a complete basis of the Fock space in the preferred direction of each horizontal internal line and project a topological vertex to a vertex operator that connects the remaining two Fock spaces associated to the topological vertex. A convenient basis $\ket{v,\lambda}$ of the Fock space in the preferred direction, where $v$ labels the position of the corresponding D5 brane and $\lambda$ is a Young diagram label of the states in the Fock space, was found in \cite{AFS} to be the Macdonald basis and can be identified with the basis of the $(1,0)_v$ representation of the so-called Ding-Iohara-Miki (DIM) algebra \cite{DI, Miki}. More concretely, the vertices $\Phi$ and $\Phi^\ast$ are in fact respectively found in \cite{AFS} to intertwine among three representations labeled by $(p,q)$ of the Ding-Iohara-Miki algebra\footnote{Note that the convention of labels $(p,q)$ for representations of DIM is different from those in mathematical contexts such as \cite{AFS}. We chose it to match with the convention in physical contexts and the direction of the corresponding brane in our web diagram.}, which are all isomorphic to Fock spaces,  
\ba
\Phi^{(p)}[u,v]:\ (p,1)_u\otimes (1,0)_v\rightarrow (p+1,1)_{-uv},\\
\Phi^{(p)\ast}[u,v]:\ (p+1,1)_{-uv}\rightarrow (p,1)_u\otimes (1,0)_v.
\ea
The label of the corresponding representation to each leg matches the axio-dilaton charge in the brane web, and in terms of the DIM algebra, it parameterizes two central charges in the algebra as 
\ba
(p,q):\ (\psi^+_0/\psi^-_0,\hat{\gamma})\mapsto (\gamma^{-2p},\gamma^q),
\ea

 The vertex operators acting on the Fock space in the non-preferred direction are given by
\ba
\Phi^{(n)}[u,v]\ket{v,\lambda}=:\Phi^{(n)}_{\lambda}[u,v]=t_n(\lambda,u,v):\Phi_\emptyset (v)\prod_{x\in \lambda} \eta(\chi_x):,
\ea
with $t_{n}(\lambda,u,v)=\lt(-uv\rt)^{|\lambda|}\prod_{x\in\lambda}(\gamma/\chi_x)^{n+1}$, and 
\ba
\bra{v,\lambda}\Phi^{\ast(n)}[u,v]=:\Phi^{\ast(n)}_{\lambda}[u,v]=t^\ast_{n}(\lambda,u,v):\Phi^\ast_\emptyset (v)\prod_{x\in\lambda}\xi(\chi_x):,
\ea
with $t^\ast_{n}(\lambda,u,v)=(\gamma u)^{-|\lambda|}\prod_{x\in\lambda}(\chi_x/\gamma)^n$, where $\chi_x=vq_1^{i-1}q_2^{j-1}$ is the coordinate for the box $x=(i,j)$ in $\lambda$, and the vertex operators are written in terms of free boson oscillators $ a_n$, satisfying 
\ba
\lt[ a_n, a_m\rt]=n\delta_{n+m,0},
\ea
as
\ba
\Phi_\emptyset[v]:=\exp\lt(-\sum_{n=1}^\infty \frac{v^n}{n}\frac{1}{1-q_1^{-n}} a_{-n}\rt)\exp\lt(\sum_{n=1}^\infty \frac{1}{n}\frac{v^{-n}}{1-q_2^{-n}} a_n\rt),\label{shift-vert-ele-1}\\
\Phi^\ast_\emptyset[v]:=\exp\lt(\sum_{n=1}^\infty \frac{1}{n}\frac{\gamma^nv^n}{1-q_1^{-n}} a_{-n}\rt)\exp\lt(-\sum_{n=1}^\infty \frac{1}{n}\frac{\gamma^nv^{-n}}{1-q_2^{-n}} a_n\rt),\label{shift-vert-ele-2}
\ea
\ba
\eta(z)=\exp\lt(-\sum_{n=1}^\infty \frac{1}{n}q_1^n(1-q_2^{n})z^n a_{-n}\rt)\exp\lt(-\sum_{n=1}^\infty \frac{1}{n}(1-q_1^{-n})z^{-n} a_{n}\rt),\label{shift-vert-ele-3}\\
\xi(z)=\exp\lt(\sum_{n=1}^\infty \frac{1}{n}q_1^n(1-q_2^{n})z^n\gamma^n a_{-n}\rt)\exp\lt(\sum_{n=1}^\infty \frac{1}{n}(1-q_1^{-n})z^{-n}\gamma^n a_{n}\rt).\label{shift-vert-ele-4}
\ea
With the above explicit expression of vertices, we can insert the identity operator ${\bf 1}=\sum_\lambda a_\lambda \ket{v,\lambda}\bra{v,\lambda}$ in internal lines in the preferred direction and project the vertices at two ends of the internal lines to vertex operators to reduce the calculation to the evaluation of VEVs of vertex operators. $a_\lambda$ here denotes the inversion of the normalization factor of the basis in the $(1,0)_v$-representation of DIM, 
\ba
a_\lambda=\frac{1}{\bra{v,\lambda}v,\lambda\rangle}=Z^{\rm vec}(\lambda)(-\gamma v)^{-|\lambda|}\prod_{x\in\lambda}\chi_x,
\ea
where $Z^{\rm vec}(\lambda)$ is given by the inverse of the Nekrasov factor, 
\ba
Z^{\rm vec}(\lambda)=N^{-1}_{\lambda\lambda}(1,q_1,q_2),
\ea
where 
\ba
N_{\lambda\nu}(Q=v_1/v_2;q_1,q_2)=\prod_{(x,x')\in \cX_1\times\cX_2}\frac{\lt(q_2x'/x;q_2\rt)_\infty}{\lt(q_1q_2x'/x;q_2\rt)_\infty}\times \prod_{(x,x')\in \cX_1^\emptyset\times\cX_2^\emptyset}\frac{\lt(q_1q_2x'/x;q_2\rt)_\infty}{\lt(q_2x'/x;q_2\rt)_\infty}\nn\\
=\prod_{x\in\lambda}(1-\chi_xq_1q_2/v_2)\prod_{y\in\nu}(1-v_1/\chi_y)\prod_{x\in\lambda,y\in \nu}S(\chi_x/\chi_y),\label{Nekra-S}
\ea
with $\cX_1$ and $\cX_2$ the characteristic set attached to the Young diagrams $\lambda$ and $\nu$. Note that the Nekrasov factor is normalized to $N_{\emptyset\emptyset}(Q;q_1,q_2)=1$. 
We then take the vacuum expectation value of the vertex operators obtained from projection using the contraction rules, 
\ba
&&\contraction{}{\Phi_\mu[u_2,v_2]}{}{\Phi_\lambda[u_1,v_1]}\Phi_\mu[u_2,v_2]\Phi_\lambda[u_1,v_1]={\cal G}^{-1}(v_1,v_2)Z_{bfd}(v_1,v_2;\lambda,\mu|1)^{-1}:\Phi_\mu[u_2,v_2]\Phi_\lambda[u_1,v_1]:,\label{contra-1}\\
&&\contraction{}{\Phi^\ast_\mu[u_2,v_2]}{}{\Phi^\ast_\lambda[u_1,v_1]}\Phi^\ast_\mu[u_2,v_2]\Phi^\ast_\lambda[u_1,v_1]={\cal G}^{-1}(v_1,v_2q_3^{-1})Z_{bfd}(v_1,v_2q_3^{-1};\lambda,\mu|1)^{-1}:\Phi^\ast_\mu[u_2,v_2]\Phi^\ast_\lambda[u_1,v_1]:,\label{contra-2}
\ea
\ba
\contraction{}{\Phi_\mu[u_2,v_2]}{}{\Phi^\ast_\lambda[u_1,v_1]}\Phi_\mu[u_2,v_2]\Phi^\ast_\lambda[u_1,v_1]={\cal G}(v_1,v_2\gamma^{-1})Z_{bfd}(v_1,v_2\gamma^{-1};\lambda,\mu|1):\Phi_\mu[u_2,v_2]\Phi^\ast_\lambda[u_1,v_1]:,\label{contra-3}\\
\contraction{}{\Phi^\ast_\mu[u_2,v_2]}{}{\Phi_\lambda[u_1,v_1]}\Phi^\ast_\mu[u_2,v_2]\Phi_\lambda[u_1,v_1]={\cal G}(v_1,v_2\gamma^{-1})Z_{bfd}(v_1,v_2\gamma^{-1};\lambda,\mu|1):\Phi^\ast_\mu[u_2,v_2]\Phi_\lambda[u_1,v_1]:\label{contra-4}.
\ea
where the prefactor ${\cal G}$ is given by the $q$-double gamma function as 
\ba
{\cal G}(v_1,v_2)
= \exp\lt(\sum_{n>0}\frac{1}{n}\frac{(v_1/v_2)^n}{(1-q_1^{-n})(1-q_2^{-n})}\rt)
 = \prod_{n,m \ge 0} \left( 1 - \frac{v_1}{v_2} q_1^{-n} q_2^{-m} \right)^{-1},
\ea
and the U(1)$\times$U(1) bifundamental contribution is also given by the Nekrasov factor as 
\ba
Z^{\rm bf}(v_1,v_2;\lambda,\mu)=N_{\lambda \mu}(v_1/v_2,q_1,q_2).
\ea
Let us complete our evaluation of the partition function of the pure SU(2) gauge theory, (\ref{ex-SU2}), 
\ba
Z_{{\rm SU(2)}}&=&\sum_{\lambda_1,\lambda_2}a_{\lambda_1}a_{\lambda_2}\begin{array}{c}
\widehat{\underline{\ 0\ }}\\
\Phi^{(0)}_{\lambda_2}[u,v_2]\\
\Phi^{(-1)}_{\lambda_1}[-u/v_1,v_1]\\
\uwidehat{\overline{\ 0\ }}\\
\end{array}
\begin{array}{c}
\widehat{\underline{\ 0\ }}\\
\Phi^{\ast(-1)}_{\lambda_2}[-u^\ast/v_2,v_2]\\
\Phi^{\ast(0)}_{\lambda_1}[u^\ast,v_1]\\
\uwidehat{\overline{\ 0\ }}\\
\end{array}\nn\\
&=&{\cal G}^{-1}(v_1,v_2){\cal G}^{-1}(v_1,v_2q_3^{-1})\sum_{\lambda_1,\lambda_2}\frac{(-\gamma v_1)^{-|\lambda_1|}(-\gamma v_2)^{-|\lambda_2|}\prod_{x\in\lambda_1}\chi_x\prod_{y\in\lambda_2}\chi_y}{N_{\lambda_1\lambda_1}(1,q_1,q_2)N_{\lambda_2\lambda_2}(1,q_1,q_2)}\nn\\
&&\times\frac{(uv^2_2/(u^\ast \gamma))^{|\lambda_2|}(u/(u^\ast \gamma))^{|\lambda_1|}\prod_{y\in\lambda_2}(\gamma/\chi_y)^2}{N_{\lambda_1\lambda_2}(v_1/v_2,q_1,q_2)N_{\lambda_1\lambda_2}(v_1q_3/v_2,q_1,q_2)}\nn\\
&=&{\cal G}^{-1}(v_1,v_2){\cal G}^{-1}(v_1,v_2q_3^{-1})\sum_{\lambda_1,\lambda_2}\lt(\frac{uv_2}{u^\ast v_1}\rt)^{|\lambda_1|+|\lambda_2|}\frac{1}{N_{\lambda_1\lambda_1}(1,q_1,q_2)N_{\lambda_2\lambda_2}(1,q_1,q_2)}\nn\\
&&\times \frac{1}{N_{\lambda_1\lambda_2}(v_1/v_2,q_1,q_2)N_{\lambda_1\lambda_2}(v_1q_3/v_2,q_1,q_2)}.
\ea
In the above computation, ${\cal G}$-factors correspond to the 1-loop part of the full Nekrasov partition function of the gauge theory, and the remaining gives the instanton corrections. 

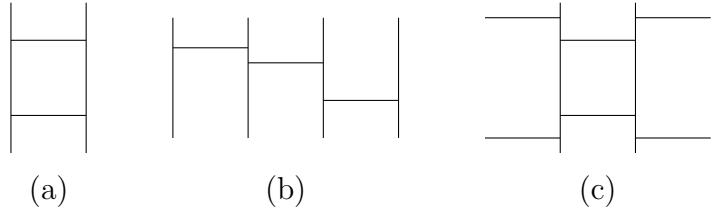
\begin{figure}
\begin{center}
\begin{tikzpicture}
\draw(0,1)--(0,-1);
\draw(1,1)--(1,-1);
\draw(0,0.5)--(1,0.5);
\draw(0,-0.5)--(1,-0.5);
\node at (0.5,-1.5) {(a)};
\end{tikzpicture}
\hskip 1cm
\begin{tikzpicture}
\draw(0,0.8)--(0,-0.8);
\draw(0,-0.3)--(1,-0.3);
\draw(1,0.8)--(1,-0.8);
\draw(-1,0.8)--(-1,-0.8);
\draw(-1,0.2)--(0,0.2);
\draw(-2,0.8)--(-2,-0.8);
\draw (-2,0.4)--(-1,0.4);
\node at (-0.5,-1.5) {(b)};
\end{tikzpicture}
\hskip 1cm
\begin{tikzpicture}
\draw(0,1)--(0,-1);
\draw(1,1)--(1,-1);
\draw(0,0.5)--(1,0.5);
\draw(0,-0.5)--(1,-0.5);
\node at (0.5,-1.5) {(c)};
\draw (1,0.8)--(2,0.8);
\draw (1,-0.8)--(2,-0.8);
\draw (0,0.8)--(-1,0.8);
\draw (0,-0.8)--(-1,-0.8);
\end{tikzpicture}
\end{center}
 \caption{Examples of simplified webs: (a) SU(2) gauge theory, (b) $A_3$ quiver U(1) gauge theory, (c) SU(2) gauge theory with four flavors.}
 \label{f:simp-web}
\end{figure}

In the purpose of this article, we only focus on the information of the quiver structure of the gauge theory, and it is thus not necessary to distinguish the Fock spaces used to construct representations of the DIM algebra with different values of central charges. It is convenient to introduce a simplified web diagram which only distinguishes the Fock spaces from the representation space of the DIM algebras in the preferred direction. We represent each representation space in the preferred direction with a horizontal line again, and all Fock spaces in non-preferred directions with vertical lines. Several simple examples are presented in Figure \ref{f:simp-web}. It is easy to see that adding vertical lines (Fock spaces) in the simplified diagram lifts up the rank of linear quiver in the current construction, and adding horizontal lines inside the chamber between two vertical lines raises the rank of gauge group. Since we focus on the quiver structure of gauge theories in this article, we set all gauge groups to be U(1) for simplicity in the remaining part of this article, but we note that it is always straightforward to raise the rank of gauge groups by adding more D5 branes (horizontal lines) or using the generalized vertex introduced in \cite{BFHMZ} instead of the topological vertex used here. 

\begin{figure}
\begin{center}
\begin{tikzpicture}
\draw(0,1)--(0,-1);
\draw(1,1)--(1,-1);
\draw(0,0.5)--(1,0.5);
\draw(0,-0.5)--(1,-0.5);
\node at (0.5,-1.5) {(a)};
\draw [fill=yellow] (0.5,0) circle (0.6); node {\textbf{2}};
\node at (0.5,0) {{\bf 2}};
\end{tikzpicture}
\hskip 1cm
\begin{tikzpicture}
\draw(0,0.8)--(0,-0.8);
\draw(0,-0.3)--(1,-0.3);
\draw(1,0.8)--(1,-0.8);
\draw(-1,0.8)--(-1,-0.8);
\draw(-1,0.2)--(0,0.2);
\draw(-2,0.8)--(-2,-0.8);
\draw (-2,0.4)--(-1,0.4);
\node at (-0.5,-1.5) {(b)};
\draw [ultra thick] (-1.5,0.4)--(-0.5,0.2);
\draw [ultra thick] (0.5,-0.3)--(-0.5,0.2);
\begin{scope}
\end{scope} 
\draw [fill=yellow] (-0.5,0.2) circle (0.4);
\draw [fill=yellow] (-1.5,0.4) circle (0.4);
\draw [fill=yellow] (0.5,-0.3) circle (0.4);
\end{tikzpicture}
\hskip 1cm
\begin{tikzpicture}
\draw(0,1)--(0,-1);
\draw(1,1)--(1,-1);
\draw(0,0.5)--(1,0.5);
\draw(0,-0.5)--(1,-0.5);
\node at (0.5,-1.5) {(c)};
\draw (1,0.8)--(2,0.8);
\draw (1,-0.8)--(2,-0.8);
\draw (0,0.8)--(-1,0.8);
\draw (0,-0.8)--(-1,-0.8);
\draw [ultra thick] (-1,0)--(0.5,0);
\draw [ultra thick] (0.5,0)--(2,0);
\draw [fill=yellow] (0.5,0) circle (0.6);
\draw [fill=yellow] (-1.6,0.6) rectangle (-0.4,-0.6);
\draw [fill=yellow] (2.6,0.6) rectangle (1.4,-0.6);
\node at (0.5,0) {{\bf 2}};
\node at (-1,0) {{\bf 2}};
\node at (2,0) {{\bf 2}};
\end{tikzpicture}
\end{center}
\label{f:simp-web-quiver}
\caption{Quiver structure in simplified webs: (a) SU(2) gauge theory, (b) $A_3$ quiver U(1) gauge theory, (c) SU(2) gauge theory with four flavors. When the gauge group is U(1), we omit the number in the correspondong node for the gauge group information. }
\end{figure}
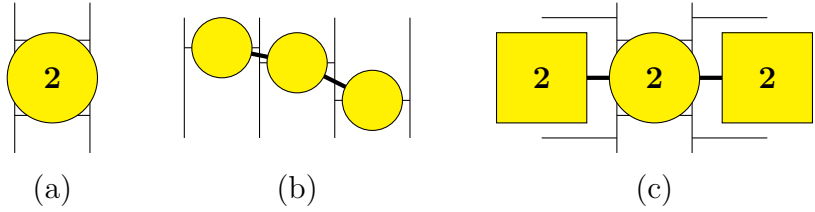

The qq-characters form an infinite family of physical observables (for example they are argued in \cite{Kim-qq,Agarwal:2018tso} to be realized as Wilson lines/surfaces in 5d/6d theories), and they are deeply related to the integrability of the underlying gauge theory. See also~\cite{Kimura:2017auj} for a realization in topological string theory. They span the quiver $\cW$-algebra \cite{Kimura-Pestun}, where the qq-character corresponding to the fundamental representation reduces to the generating current of the algebra, and in the Nekrasov-Shatashvili limit gives rise to the T-operator of the corresponding quantum integrable system. We review how to derive the qq-character, which is an important quantity to check, in the web diagram (or simplified diagram) by using the Ward identity in DIM algebra following \cite{5dBMZ,BFHMZ} in Appendix \ref{a:qq-character}. We will argue later that this prescription gives the correct expressions of the qq-characters in the web diagram proposed in this article for ABCDEFG-type quiver. 

\section{$D$-type Quiver Construction with Orientifold}\label{s:rev-D}

\begin{figure}
\begin{center}
	\begin{tikzpicture}
	\draw (0,0)--(-0.71,0.71);
	\draw (0,0)--(0,-1);
	\draw (0,-1)--(-0.71,-1.71);
	\draw (0,-1)--(2.2,-1);
	\draw (2.2,-0.8)--(1.8,-0.8);
	\draw (0,0)--(1,0);
	\draw (1,0)--(1.8,-0.8);
	\draw (1.8,-0.8)--(1.8,-0.95);
	\draw (1.8,-1.05)--(1.8,-1.8);
	\draw (1,0)--(1,1);
	\draw[dashed] (2.2,1)--(2.2,-1.8);
	\node at (3,-1) { ${\bf ON}^-$};
	\end{tikzpicture}
\end{center}
\caption{The web diagram for the $D_2$ quiver theory.}
\label{fig_D2construction}
\end{figure}
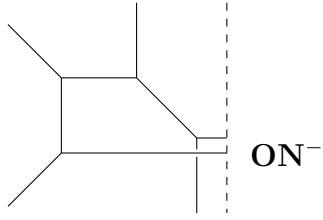

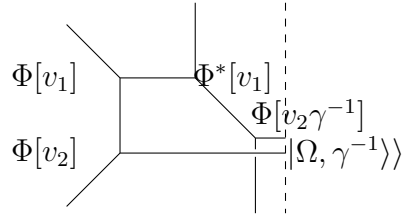
\begin{figure}
\begin{center}
	\begin{tikzpicture}
	\draw (0,0)--(-0.71,0.71);
	\draw (0,0)--(0,-1);
	\draw (0,-1)--(-0.71,-1.71);
	\draw (0,-1)--(2.2,-1);
	\draw (2.2,-0.8)--(1.8,-0.8);
	\draw (0,0)--(1,0);
	\draw (1,0)--(1.8,-0.8);
	\draw (1.8,-0.8)--(1.8,-0.95);
	\draw (1.8,-1.05)--(1.8,-1.8);
	\draw (1,0)--(1,1);
	\draw[dashed] (2.2,1)--(2.2,-1.8);
	\node at (-1,0) {$\Phi[v_1]$};
	\node at (-1,-1) {$\Phi[v_2]$};
	\node at (1.5,0) {$\Phi^\ast[v_1]$};
	\node at (2.5,-0.5) {$\Phi[v_2\gamma^{-1}]$};
	\node at (3,-1) { $\ket{\Omega, \gamma^{-1}}\rangle$};
	\end{tikzpicture}
\end{center}
\caption{The vertex assignment for the $D_2$ quiver theory.}
\label{fig_D2vert}
\end{figure}

Now we go to the next simplest example, $D$-type quiver gauge theories. It was reported in \cite{D-type} that the topological vertex formalism can also be applied to the brane webs proposed in \cite{HKLTY} constructed with an ${\bf ON}^-$ orientifold plane for $D$-type quiver theories~\cite{Kapustin:1998fa,Hanany:1999sj}. The web diagram of the simplest case, $D_2\simeq A_1\times A_1$ quiver, for example is given in Figure \ref{fig_D2construction}. The orientifold ${\bf ON}^-$ in the diagram is assigned with the reflection state defined by 
\ba
\ket{\Omega, \alpha}\rangle:=\sum_{\lambda}a_\lambda\ket{v,\lambda}\otimes \ket{v\alpha, \lambda},
\ea
with the rule 
\begin{align}
\begin{tikzpicture}
	\draw (0,0.3)--(-1,0.3);
	\draw (0,0)--(-1,0);
	\draw[dashed] (0,1)--(0,-1);
	\node at (3,0) {$\ket{\Omega,\gamma^{-1}}\rangle$.};
	\draw [<->] (0.5,0)--(2,0);
\end{tikzpicture}
\end{align}
The explicit assignment of vertices is presented in Figure \ref{fig_D2vert} for $D_2$ quiver theory and can be easily generalized to higher rank $D$-type quiver algebras.

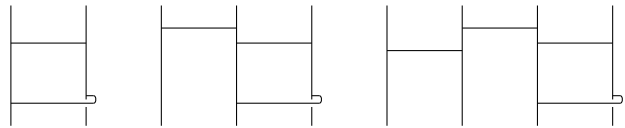
\begin{figure}
\begin{center}
\begin{tikzpicture}
\draw(0,0.8)--(0,-0.8);
\draw(0,0.3)--(1,0.3);
\draw(0,-0.5)--(1.1,-0.5);
\draw(1,0.8)--(1,-0.45);
\draw(1,-0.55)--(1,-0.8);
\draw(1.1,-0.5) to [out=0,in=0] (1.1,-0.4);
\draw(1.1,-0.4)--(1,-0.4);
\draw(3,0.8)--(3,-0.8);
\draw(3,0.3)--(4,0.3);
\draw(3,-0.5)--(4.1,-0.5);
\draw(4,0.8)--(4,-0.45);
\draw(4,-0.55)--(4,-0.8);
\draw(4.1,-0.5) to [out=0,in=0] (4.1,-0.4);
\draw(4.1,-0.4)--(4,-0.4);
\draw(2,0.8)--(2,-0.8);
\draw(2,0.5)--(3,0.5);
\draw(7,0.8)--(7,-0.8);
\draw(7,0.3)--(8,0.3);
\draw(7,-0.5)--(8.1,-0.5);
\draw(8,0.8)--(8,-0.45);
\draw(8,-0.55)--(8,-0.8);
\draw(8.1,-0.5) to [out=0,in=0] (8.1,-0.4);
\draw(8.1,-0.4)--(8,-0.4);
\draw(6,0.8)--(6,-0.8);
\draw(6,0.5)--(7,0.5);
\draw(5,0.8)--(5,-0.8);
\draw(5,0.2)--(6,0.2);
\end{tikzpicture}
\end{center}
\caption{Examples for simpler brane diagrams for $D_2$, $D_3$ and $D_4$ quiver theories.}
\label{fig_simpD}
\end{figure}

We also introduce simplified webs for these constructions with orientifold as shown in Figure \ref{fig_simpD}, where the ``reflection'' of D5-brane is represented by a bended brane in the preferred direction. 

\begin{figure}
\begin{center}
\begin{tikzpicture}
\draw(0,0.8)--(0,-0.8);
\draw(0,0.3)--(1,0.3);
\draw(0,-0.5)--(1.1,-0.5);
\draw(1,0.8)--(1,-0.45);
\draw(1,-0.55)--(1,-0.8);
\draw(1.1,-0.5) to [out=0,in=0] (1.1,-0.4);
\draw(1.1,-0.4)--(1,-0.4);
\draw(3,0.8)--(3,-0.8);
\draw(3,0.3)--(4,0.3);
\draw(3,-0.5)--(4.1,-0.5);
\draw(4,0.8)--(4,-0.45);
\draw(4,-0.55)--(4,-0.8);
\draw(4.1,-0.5) to [out=0,in=0] (4.1,-0.4);
\draw(4.1,-0.4)--(4,-0.4);
\draw(2,0.8)--(2,-0.8);
\draw(2,0.5)--(3,0.5);
\draw [fill=yellow] (0.5,0.4) circle (0.4);
\draw [fill=yellow] (0.5,-0.6) circle (0.4);
\draw [ultra thick] (2.5,0.5)--(3.5,0.4);
\draw [ultra thick] (2.5,0.5)--(3.5,-0.6);
\draw [fill=yellow] (3.5,0.4) circle (0.4);
\draw [fill=yellow] (3.5,-0.6) circle (0.4);
\draw [fill=yellow] (2.5,0.5) circle (0.4);
\draw(7,0.8)--(7,-0.8);
\draw(7,0.3)--(8,0.3);
\draw(7,-0.5)--(8.1,-0.5);
\draw(8,0.8)--(8,-0.45);
\draw(8,-0.55)--(8,-0.8);
\draw(8.1,-0.5) to [out=0,in=0] (8.1,-0.4);
\draw(8.1,-0.4)--(8,-0.4);
\draw(6,0.8)--(6,-0.8);
\draw(6,0.5)--(7,0.5);
\draw(5,0.8)--(5,-0.8);
\draw(5,0.2)--(6,0.2);
\draw [ultra thick] (6.5,0.5)--(7.5,0.4);
\draw [ultra thick] (6.5,0.5)--(7.5,-0.6);
\draw [ultra thick] (5.5,0.2)--(6.5,0.5);
\draw [fill=yellow] (7.5,0.4) circle (0.4);
\draw [fill=yellow] (7.5,-0.6) circle (0.4);
\draw [fill=yellow] (6.5,0.5) circle (0.4);
\draw [fill=yellow] (5.5,0.2) circle (0.4);
\end{tikzpicture}
\end{center}
\caption{Quiver structures in simpler brane diagrams for $D_2$, $D_3$ and $D_4$ quiver theories.}
\label{fig_quiverD}
\end{figure}
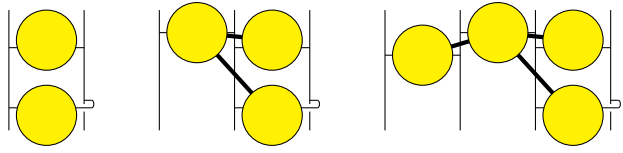

From the explicit computation of the $D_2$ quiver partition function, 
\ba
Z_{D_2}=\sum_{\lambda_2} a_{\lambda_2}\begin{array}{c}
\widehat{\underline{\ 0\ }}\\
\Phi^{(n_1)}[u,v_2]\\
\cdot\\
\Phi^{(n_1-1)}[-u/v_1,v_1]\\
\uwidehat{\overline{\ 0\ }}\\
\end{array}
\begin{array}{c}
\\
\cdot\\
\\
\cdot\\
\\
\end{array}\begin{array}{c}
\widehat{\underline{\ 0\ }}\\
\Phi^{\ast (n_1^\ast)}[u^\ast,v_2]\\
\Phi^{(n_1^\ast)}[u^\ast v_2\gamma/v_1,v_1\gamma^{-1}]\\
\\
\uwidehat{\overline{\ 0\ }}\\
\end{array}
\begin{array}{c}
\\
\\
\ket{v_1\gamma^{-1},\lambda_2}\\
\ket{v_1,\lambda_2}\\
\\
\end{array}\nn\\
=\sum_{\lambda_1,\lambda_2}\mathfrak{q}_1^{|\lambda_1|}\mathfrak{q}_2^{|\lambda_2|}\frac{\prod_{x\in\lambda_1}\chi_x^{\kappa_1}\prod_{y\in\lambda_2}\chi_y^{\kappa_1}}{N_{\lambda_1\lambda_1}(1;q_1,q_2)N_{\lambda_2\lambda_2}(1;q_1,q_2)}{\cal G}^{-1}(v_1,v_2){\cal G}(v_1,v_2)\nn\\
\times Z_{bfd}(v_1,v_2;\lambda_2,\lambda_1|1)^{-1}Z_{bfd}(v_1,v_2;\lambda_2,\lambda_1|1)\nn\\
=\lt(\sum_{\lambda_1}\mathfrak{q}_1^{|\lambda_1|}\frac{\prod_{x\in\lambda_1}\chi_x^{\kappa_1}}{N_{\lambda_1\lambda_1}(1;q_1,q_2)}\rt)\lt(\sum_{\lambda_2}\mathfrak{q}_2^{|\lambda_2|}\frac{\prod_{y\in\lambda_2}\chi_y^{\kappa_2}}{N_{\lambda_2\lambda_2}(1;q_1,q_2)}\rt),\label{partition-D2}
\ea
where $\mathfrak{q}_1=u_1u_1^\ast v_2\gamma^{n_1+n_1^\ast +1}/v_1$, $\kappa_1=-n_1-n_1^\ast$, $\mathfrak{q}_2=-\frac{u}{u^\ast}\gamma^{n_1-n_1^\ast-1}$ and $\kappa_2=n^\ast_1-n_1$, we can see the decoupling of two D5 branes in the brane web, which explicitly realizes the $A_1\times A_1$ quiver.

The fundamental qq-character of the $D_2$ quiver can be obtained by considering the insertion, 
\begin{align}
\begin{tikzpicture}[baseline=(current  bounding  box.center)]
\draw(0,0.8)--(0,-0.8);
\draw(0,0.3)--(1,0.3);
\draw(0,-0.5)--(1.1,-0.5);
\draw(1,0.8)--(1,-0.45);
\draw(1,-0.55)--(1,-0.8);
\draw(1.1,-0.5) to [out=0,in=0] (1.1,-0.4);
\draw(1.1,-0.4)--(1,-0.4);
\node at (0.5,0.3) [above] {$x^-_>$};
\node at (2,0) {$+$};
\draw(3,0.8)--(3,-0.8);
\draw(3,0.3)--(4,0.3);
\draw(3,-0.5)--(4.1,-0.5);
\draw(4,0.8)--(4,-0.45);
\draw(4,-0.55)--(4,-0.8);
\draw(4.1,-0.5) to [out=0,in=0] (4.1,-0.4);
\draw(4.1,-0.4)--(4,-0.4);
\node at (3.5,-0.5) [above] {$x^-_>$};
\node at (5,0) {$+$};
\draw(6,0.8)--(6,-0.8);
\draw(6,0.3)--(7,0.3);
\draw(6,-0.5)--(7.1,-0.5);
\draw(7,0.8)--(7,-0.45);
\draw(7,-0.55)--(7,-0.8);
\draw(7.1,-0.5) to [out=0,in=0] (7.1,-0.4);
\draw(7.1,-0.4)--(7,-0.4);
\node at (6.5,0.3) [above] {$x^+_>$};
\node at (8,0) {$+$};
\draw(9,0.8)--(9,-0.8);
\draw(9,0.3)--(10,0.3);
\draw(9,-0.5)--(10.1,-0.5);
\draw(10,0.8)--(10,-0.45);
\draw(10,-0.55)--(10,-0.8);
\draw(10.1,-0.5) to [out=0,in=0] (10.1,-0.4);
\draw(10.1,-0.4)--(10,-0.4);
\node at (9.5,-0.5) [above] {$x^+_>$};
\end{tikzpicture}
\end{align}
and after some computation, it is found to take the form\footnote{Refer to \eqref{eq:chi_bar} and \eqref{eq:Y_normalized} for definitions of $Y$ and $\bar{\chi}$.} 
\ba
\bar{\chi}_{(1,1)}(z)= \lt(Y_1(zq_3^{-1})+\frac{1}{Y_1(z)}\rt)\lt(Y_2(zq_3^{-1})+\frac{1}{Y_2(z)}\rt),
\ea
from which we also see the manifestation of the decoupling of two gauge nodes. This factorization property together with the map between qq-character and representation theory we established for $A$-type quiver in Appendix \ref{a:qq-character} give the map between $D$-type qq-characters and the representation theory of $D$-type Lie groups. We have to further check the prefactor of $S$-functions in the expression of qq-characters for the consistency of the brane construction, and we give a non-trivial examination for adjoint representation of $D_4$ quiver in Appendix \ref{a:qq-D4}. 

We further propose an alternative construction with the following newly introduced vertex\footnote{Refer to section \ref{s:AFS} to see that the prefactor $t_{n}(\lambda,u,v)$ is a natural and appropriate choice here.}
\ba
\bra{v,\lambda}\bar{\Phi}^{\ast(n)}[u,v]=:\bar{\Phi}^{\ast(n)}_{\lambda}[u,v]=t_{n}(\lambda,u,v/\gamma):\bar{\Phi}^\ast_\emptyset (v)\prod_{x\in\lambda}\bar{\xi}(\chi_x):,
\label{eq:reflection_vertex}
\ea
with
\ba
&&\bar{\Phi}^\ast_\emptyset(v):=\exp\lt(-\sum_{n=1}^\infty \frac{1}{n}\frac{\gamma^{-n}v^n}{1-q_1^{-n}} a_{-n}\rt)\exp\lt(\sum_{n=1}^\infty \frac{1}{n}\frac{\gamma^{3n}v^{-n}}{1-q_2^{-n}} a_n\rt),\\
&&\bar{\xi}(z):=\exp\lt(-\sum_{n=1}^\infty \frac{1}{n}q_1^n(1-q_2^{n})z^n\gamma^{-n} a_{-n}\rt)\exp\lt(-\sum_{n=1}^\infty \frac{1}{n}(1-q_1^{-n})z^{-n}\gamma^{3n} a_{n}\rt).
\ea
We graphically represent it as 
\begin{align}
\begin{tikzpicture}[baseline=(current  bounding  box.center)]
\draw (0,1)--(0,-1);
\draw (-1,0)--(0,0);
\draw [ultra thick] (-0.2,-0.2)--(0.2,0.2);
\draw [ultra thick] (-0.2,0.2)--(0.2,-0.2);
\end{tikzpicture}
\end{align}
It plays a similar role to the reflection state, while the proposal in \cite{D-type} works only for U(1) gauge group in the node built with the reflection state or if one wants to left up the rank of the gauge group, one will have to use the generalized vertex introduced in \cite{BFHMZ}. On the other hand, we can raise the rank of the gauge group in the ordinary way by simply adding more D5 branes by using $\bar{\Phi}^{\ast(n)}[u,v]$ instead. For example the $D_2$ quiver gauge theory with U(1) $\otimes$ SU(2) gauge group can be built as 
\begin{align}
\begin{tikzpicture}[baseline=(current  bounding  box.center)]
\draw (0,1)--(0,-1);
\draw (-1,0)--(0,0);
\draw [ultra thick] (-0.1,-0.1)--(0.1,0.1);
\draw [ultra thick] (-0.1,0.1)--(0.1,-0.1);
\draw [ultra thick] (-0.1,-0.6)--(0.1,-0.4);
\draw [ultra thick] (-0.1,-0.4)--(0.1,-0.6);
\draw (0,-0.5)--(-1,-0.5);
\draw (-1,1)--(-1,-1);
\draw (-1,0.5)--(0,0.5);
\end{tikzpicture}
\end{align}

\paragraph{Superconformal quiver theories} We define the notion of superconformal gauge theories for general quiver gauge theories (even without Lagrangian description) through the asymptotic behavior of qq-characters, $\chi_{\vec{w}}(z)$ as $z\rightarrow \infty,\ 0$. As it is easier to explain the criteria in the 4d limit, let us put $z=e^{R\zeta}$ and consider the asymptotic behavior $\zeta\sim \infty$ in the $R\rightarrow 0$ limit. The function $Y(z)$ associated to an SU($N$) gauge node behaves as 
\ba
Y(e^{R\zeta})\sim \zeta^{N},\quad \zeta \sim\infty,\ R\rightarrow 0.
\ea
By requiring that all terms in the qq-characters (normalized or not) of a superconformal quiver theory to behave as $\zeta^n$ around $\zeta\sim \infty$ for certain integer $n$, we can see for example from the expression of the (unnormalized) fundamental qq-character of $A_r$ quiver, 
\ba
\hat{\chi}_{(1,0,0,\dots,0)}(z)=\frac{Y_1(zq_3^{-1})}{P_1(z)}+\frac{Y_2(zq_3^{-1}\mu_{12})}{Y_1(z)}+\frac{Y_3(zq_3^{-1}\mu_{12}\mu_{23})}{Y_2(z\mu_{12})}+\dots+\frac{P_r(z)}{Y_r(z\mu_{12}\dots\mu_{(N-1)N})},
\ea
where the hat symbol $\hat{}$ here stands for a different normalization for the qq-character, $P_1(z)$ and $P_r(z)$ are respectively polynomials of $z^\pm$ with degree $N_{f1}$ and $N_{fr}$, and they reduce to polynomials of $\zeta$ of degree $N_{f1}$ and $N_{fr}$ in the limit $R\rightarrow 0$, that the superconformal quiver theory in the current case is solved with respect to $n=0$ to 
\ba
N_{f1}=N_1=N_2=\dots=N_r=N_{fr}=^\exists N,
\ea
which is a well-known superconformal quiver gauge theory from the $\beta$-function analysis of the following quiver diagram, 
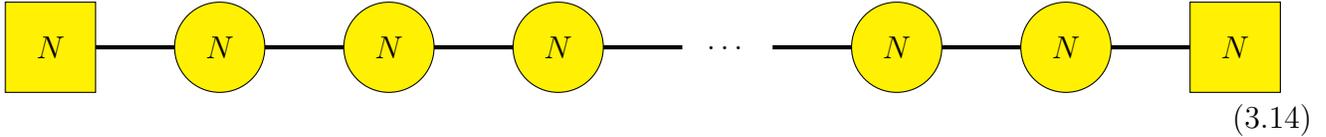
\begin{align}
\begin{tikzpicture}[scale=1.5, baseline=(current  bounding  box.center)]
\draw [ultra thick] (0,0)--(1.5,0);
\draw [ultra thick] (0,0)--(-1.5,0);
\draw [ultra thick] (1.5,0)--(3,0);
\draw [ultra thick] (3,0)--(4.1,0);
\draw [ultra thick] (4.9,0)--(6,0);
\draw [ultra thick] (6,0)--(7.5,0);
\draw [ultra thick] (7.5,0)--(9,0);
\draw [fill=yellow] (-1.9,0.4) rectangle (-1.1,-0.4);
\draw [fill=yellow] (0,0) circle (0.4);
\draw [fill=yellow] (1.5,0) circle (0.4);
\draw [fill=yellow] (3,0) circle (0.4);
\draw [fill=yellow] (6,0) circle (0.4);
\draw [fill=yellow] (7.5,0) circle (0.4);
\node at (4.5,0) {\dots};
\draw [fill=yellow] (8.6,0.4) rectangle (9.4,-0.4);
\node at (-1.5,0) {$N$};
\node at (0,0) {$N$};
\node at (1.5,0) {$N$};
\node at (3,0) {$N$};
\node at (6,0) {$N$};
\node at (7.5,0) {$N$};
\node at (9,0) {$N$};
\end{tikzpicture}
\end{align}
and for $n=1$ we have another example solution, 
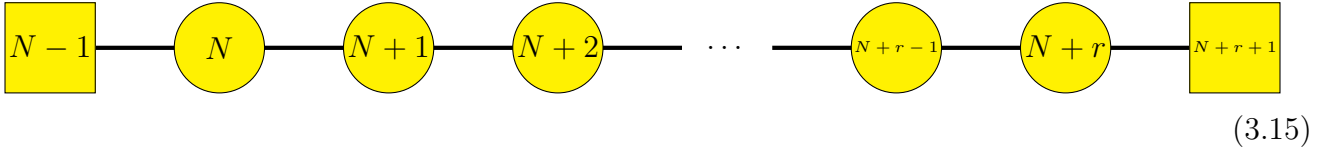
\begin{align}
\begin{tikzpicture}[scale=1.5]
\draw [ultra thick] (0,0)--(1.5,0);
\draw [ultra thick] (0,0)--(-1.5,0);
\draw [ultra thick] (1.5,0)--(3,0);
\draw [ultra thick] (3,0)--(4.1,0);
\draw [ultra thick] (4.9,0)--(6,0);
\draw [ultra thick] (6,0)--(7.5,0);
\draw [ultra thick] (7.5,0)--(9,0);
\draw [fill=yellow] (-1.9,0.4) rectangle (-1.1,-0.4);
\draw [fill=yellow] (0,0) circle (0.4);
\draw [fill=yellow] (1.5,0) circle (0.4);
\draw [fill=yellow] (3,0) circle (0.4);
\draw [fill=yellow] (6,0) circle (0.4);
\draw [fill=yellow] (7.5,0) circle (0.4);
\node at (4.5,0) {\dots};
\draw [fill=yellow] (8.6,0.4) rectangle (9.4,-0.4);
\node at (-1.5,0) {$N-1$};
\node at (0,0) {$N$};
\node at (1.5,0) {$N+1$};
\node at (3,0) {$N+2$};
\node at (6,0) {{\tiny $N+r-1$}};
\node at (7.5,0) {$N+r$};
\node at (9,0) {{\tiny $N+r+1$}};
\end{tikzpicture}
\end{align}
When $N=1$ in the above quiver diagram, this corresponds to the famous 5d version of $T_{r-1}$ theory. We note that the corresponding brane web of a superconformal quiver gauge theory appeared so far always has a realization with its all external legs always either in the vertical direction or in the horizontal direction\footnote{For example, such a web diagram can be obtained after some Hanany-Witten transition performed to the usual web diagram for $T_{r-1}$ theory (see \cite{T_N-jump} for this kind of discussion and how to deal with the jumps in the web diagram after Hanany-Witten transition).} . 

When we apply the same argument to the $D_2\simeq A_1\times A_1$ quiver, since two gauge nodes are decoupled, the superconformal quiver theory is simply a theory with $N_f=2N$ matters attached to each SU($N$) gauge node. Let us depict the brane web with orientifold for $N=1$ in the following. 
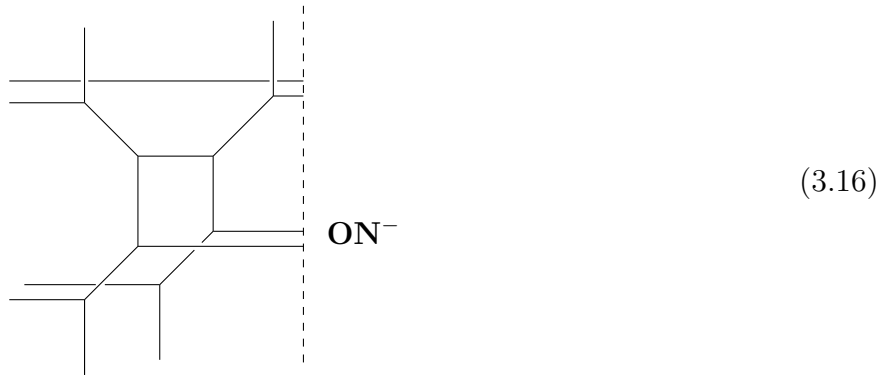
\begin{align}
\begin{tikzpicture}[baseline=(current  bounding  box.center)]
	\draw (-0.71,0.71)--(-0.71,1.71);
	\draw (-0.71,0.71)--(-1.71,0.71);
	\draw (0,0)--(-0.71,0.71);
	\draw (0,0)--(0,-1.2);
	\draw (0,-1.2)--(-0.71,-1.91);
	\draw (0,-1.2)--(2.2,-1.2);
	\draw (2.2,0.8)--(1.8,0.8);
	\draw (0,0)--(1,0);
	\draw (1,0)--(1.8,0.8);
	\draw (1.8,0.8)--(1.8,1.8);
	\draw (2.2,1)--(1.85,1);
	\draw (1.75,1)--(-0.66,1);
	\draw (-0.76,1)--(-1.71,1);
	\draw (1,-1)--(2.2,-1);
	\draw (-0.71,-1.91)--(-0.71,-2.91);
	\draw (-0.71,-1.91)--(-0.71,-2.91);
	\draw (-0.71,-1.91)--(-1.71,-1.91);
	\draw (1,-1)--(0.85,-1.15);
	\draw (0.75,-1.25)--(0.29,-1.71);
	\draw (1,0)--(1,-1);
	\draw[dashed] (2.2,2)--(2.2,-2.8);
	\draw (0.29,-1.71)--(-0.44,-1.71);
	\draw (-0.56,-1.71)--(-1.51,-1.71);
	\node at (3,-1) { ${\bf ON}^-$};
	\draw (0.29,-1.71)--(0.29,-2.71);
\end{tikzpicture}
\end{align}
Again we have all external legs either in the horizontal direction or in the vertical direction in the above diagram. This fact allows one to speculate the direction of each leg of new vertices such as $\tilde{\Phi}^\ast$ that we introduce in this article via vertex operators \eqref{eq:reflection_vertex}. For higher rank $D$-type quivers, the terms that appear in the (unnormalized) fundamental qq-character associated to the following substructure in the quiver diagram, 
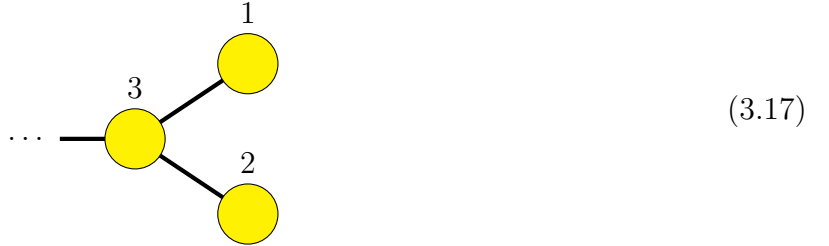
\begin{align}
\begin{tikzpicture}[baseline=(current  bounding  box.center)]
\draw [ultra thick] (0,0)--(1.5,1);
\draw [ultra thick] (0,0)--(1.5,-1);
\draw [ultra thick] (0,0)--(-1,0);
\draw [fill=yellow] (0,0) circle (0.4);
\draw [fill=yellow] (1.5,1) circle (0.4);
\draw [fill=yellow] (1.5,-1) circle (0.4);
\node at (-1,0) [left] {\dots};
\node at (0,0.4) [above] {3};
\node at (1.5,1.4) [above] {1};
\node at (1.5,-0.6) [above] {2};
\end{tikzpicture}
\end{align}
where we labeled three nodes by 1, 2, and 3, are given by 
\ba
\hat{\chi}_{0,0,\ldots,0,1}(z)=\dots+ \frac{Y_1(zq_3^{-1}\mu_{31})Y_2(zq_3^{-1}\mu_{32})}{Y_3(z)}+\frac{Y_2(zq_3^{-1}\mu_{32})}{Y_1(z\mu_{31})}+\frac{Y_1(zq_3^{-1}\mu_{31})}{Y_2(z\mu_{32})}+ \frac{Y_3(zq_3^{-1})}{Y_1(z\mu_{31})Y_2(z\mu_{32})}+\cdots.\nn\\
\ea
Without additional matter multiplets attached to node 1 or 2, the superconformal quiver theory condition ($\hat{\chi}_{0,0,\ldots,0,1}\sim \zeta^0$) requires 
\ba
N_3=2N_1=2N_2=2N.
\ea
Some examples of superconformal quiver theories can be found as 
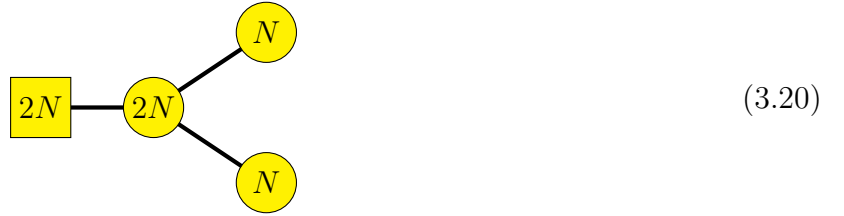
\begin{align}
\begin{tikzpicture}[baseline=(current  bounding  box.center)]
\draw [ultra thick] (0,0)--(1.5,1);
\draw [ultra thick] (0,0)--(1.5,-1);
\draw [ultra thick] (0,0)--(-1.5,0);
\draw [fill=yellow] (-1.9,0.4) rectangle (-1.1,-0.4);
\draw [fill=yellow] (0,0) circle (0.4);
\draw [fill=yellow] (1.5,1) circle (0.4);
\draw [fill=yellow] (1.5,-1) circle (0.4);
\node at (-1.5,0) {$2N$};
\node at (0,0) {$2N$};
\node at (1.5,1) {$N$};
\node at (1.5,-1) {$N$};
\end{tikzpicture}
\end{align}
and 
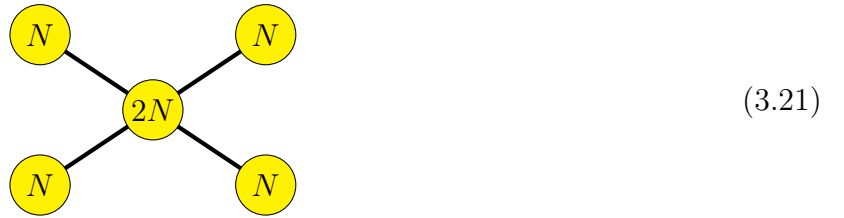
\begin{align}
\begin{tikzpicture}[baseline=(current  bounding  box.center)]
\draw [ultra thick] (0,0)--(1.5,1);
\draw [ultra thick] (0,0)--(1.5,-1);
\draw [ultra thick] (0,0)--(-1.5,1);
\draw [ultra thick] (0,0)--(-1.5,-1);
\draw [fill=yellow] (-1.5,1) circle (0.4);
\draw [fill=yellow] (-1.5,-1) circle (0.4);
\draw [fill=yellow] (0,0) circle (0.4);
\draw [fill=yellow] (1.5,1) circle (0.4);
\draw [fill=yellow] (1.5,-1) circle (0.4);
\node at (-1.5,1) {$N$};
\node at (-1.5,-1) {$N$};
\node at (0,0) {$2N$};
\node at (1.5,1) {$N$};
\node at (1.5,-1) {$N$};
\end{tikzpicture}
\end{align}
whose corresponding brane webs (for $N=1$) are respectively given by 
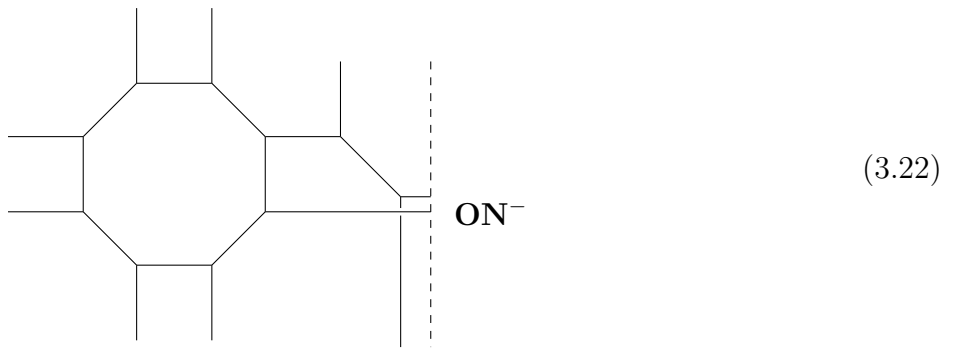
\begin{align}
	\begin{tikzpicture}[baseline=(current  bounding  box.center)]
	\draw (0,0)--(-0.71,0.71);
	\draw (0,0)--(0,-1);
	\draw (0,-1)--(-0.71,-1.71);
	\draw (0,-1)--(2.2,-1);
	\draw (2.2,-0.8)--(1.8,-0.8);
	\draw (0,0)--(1,0);
	\draw (1,0)--(1.8,-0.8);
	\draw (1.8,-0.8)--(1.8,-0.95);
	\draw (1.8,-1.05)--(1.8,-2.8);
	\draw (1,0)--(1,1);
	\draw[dashed] (2.2,1)--(2.2,-2.8);
	\node at (3,-1) { ${\bf ON}^-$};
	\draw (-0.71,0.71)--(-0.71,1.71);
	\draw (-0.71,0.71)--(-1.71,0.71);
	\draw (-1.71,0.71)--(-2.42,0);
	\draw (-1.71,0.71)--(-1.71,1.71);
	\draw (-0.71,-1.71)--(-0.71,-2.71);
	\draw (-0.71,-1.71)--(-1.71,-1.71);
	\draw (-1.71,-1.71)--(-1.71,-2.71);
	\draw (-1.71,-1.71)--(-2.42,-1);
	\draw (-2.42,-1)--(-2.42,0);
	\draw (-2.42,-1)--(-3.42,-1);
	\draw (-2.42,0)--(-3.42,-0);
	\end{tikzpicture}
\end{align}
and 
\begin{align}
	\begin{tikzpicture}[baseline=(current  bounding  box.center)]
	\draw (0,0)--(-0.71,0.71);
	\draw (0,0)--(0,-1);
	\draw (0,-1)--(-0.71,-1.71);
	\draw (0,-1)--(2.2,-1);
	\draw (2.2,-0.8)--(1.8,-0.8);
	\draw (0,0)--(1,0);
	\draw (1,0)--(1.8,-0.8);
	\draw (1.8,-0.8)--(1.8,-0.95);
	\draw (1.8,-1.05)--(1.8,-2.8);
	\draw (1,0)--(1,1);
	\draw[dashed] (2.2,1)--(2.2,-2.8);
	\node at (3,-1) { ${\bf ON}^-$};
	\draw (-0.71,0.71)--(-0.71,1.71);
	\draw (-0.71,0.71)--(-1.71,0.71);
	\draw (-1.71,0.71)--(-2.42,0);
	\draw (-1.71,0.71)--(-1.71,1.71);
	\draw (-0.71,-1.71)--(-0.71,-2.71);
	\draw (-0.71,-1.71)--(-1.71,-1.71);
	\draw (-1.71,-1.71)--(-1.71,-2.71);
	\draw (-1.71,-1.71)--(-2.42,-1);
	\draw (-2.42,-1)--(-2.42,0);
	\draw (-2.42,-1)--(-2.2-2.42,-1);
	\draw (-2.2-2.42,-0.8)--(-1.8-2.42,-0.8);
	\draw (-2.42,0)--(-1-2.42,0);
	\draw (-1-2.42,0)--(-1.8-2.42,-0.8);
	\draw (-1.8-2.42,-0.8)--(-1.8-2.42,-0.95);
	\draw (-1.8-2.42,-1.05)--(-1.8-2.42,-2.8);
	\draw (-1-2.42,0)--(-1-2.42,1);
	\draw[dashed] (-2.2-2.42,1)--(-2.2-2.42,-2.8);
	\node at (-3-2.42,-1) { ${\bf ON}^-$};
	\end{tikzpicture}
\end{align}
We will apply the same analysis to $BCFG$-type quivers later. 

We note that we can similarly introduce the following twisted vertex as a dual to \eqref{eq:reflection_vertex}, 
\ba
\bar{\Phi}^{(n)}[u,v]\ket{v,\lambda}=:\bar{\Phi}^{(n)}_{\lambda}[u,v]=t^\ast_{n}(\lambda,u,v/\gamma):\bar{\Phi}_\emptyset (v)\prod_{x\in\lambda}\bar{\eta}(\chi_x):,
\ea
with
\ba
&&\bar{\Phi}_\emptyset(v):=\exp\lt(\sum_{n=1}^\infty \frac{1}{n}\frac{q_3^{-n}v^n}{1-q_1^{-n}} a_{-n}\rt)\exp\lt(\sum_{n=1}^\infty \frac{1}{n}\frac{q_3^{n}v^{-n}}{1-q_2^{-n}} a_n\rt),\\
&&\bar{\eta}(z):=\exp\lt(-\sum_{n=1}^\infty \frac{1}{n}q_1^n(1-q_2^{n})z^nq_3^{-n} a_{-n}\rt)\exp\lt(-\sum_{n=1}^\infty \frac{1}{n}(1-q_1^{-n})z^{-n}q_3^{n} a_{n}\rt),
\ea
which is essentially the (inverse of the) usual topological vertex with a shift of Coulombmoduli $v$, to realize the substructure 
\begin{align}
\begin{tikzpicture}[xscale=-1, baseline=(current  bounding  box.center)]
\draw [ultra thick] (0,0)--(1.5,1);
\draw [ultra thick] (0,0)--(1.5,-1);
\draw [ultra thick] (0,0)--(-1,0);
\draw [fill=yellow] (0,0) circle (0.4);
\draw [fill=yellow] (1.5,1) circle (0.4);
\draw [fill=yellow] (1.5,-1) circle (0.4);
\node at (-1,0) [right] {\dots};
\end{tikzpicture}
\end{align}
in the quiver diagram. We will denote it as 
\begin{align}
\begin{tikzpicture}[xscale=-1, baseline=(current  bounding  box.center)]
\draw (0,1)--(0,-1);
\draw (-1,0)--(0,0);
\draw [ultra thick] (-0.2,-0.2)--(0.2,0.2);
\draw [ultra thick] (-0.2,0.2)--(0.2,-0.2);
\end{tikzpicture}
\end{align}
in the web diagram in this article.

\section{Partition Function of Fractional Quivers}\label{s:Frac-Part}

Now we turn to the construction of fractional quivers, focusing on quivers corresponding to non-simply-laced Lie algebras and their affine versions, with webs of vertex operators. To do so, let us first give a review on the instanton partition function of these theories. 

In \cite{Fractional-KP}, gauge theories associated to quivers of all Lie algebras of type $ABCDEFG$ (together with their affine and hyperbolic versions) are constructed. To write down the partition function of theory of this class, we assign an integer $d_i=(\alpha_i,\alpha_i)$ (c.f. Cartan matrix $c_{ij}=(\alpha^\vee_i,\alpha_j) = (\alpha_i, \alpha_j)/(\alpha_i,\alpha_i)$) to each node of the quiver.  

The vector multiplet contribution from the $i$-th node with Coulomb modulus $v_i$ is given by 
\ba
Z^{\rm vec}_i[\cX_i,d_i]=\mathbb{I}[\mathbf{V}_i]=\prod_{(x,x')\in \cX_i^2}\frac{\lt(q_1^{d_i}q_2x/x';q_2\rt)_\infty}{\lt(q_2x/x';q_2\rt)_\infty},
\ea
where $\cX_i=\{v_i q_1^{d_i(k-1)}q_2^{\lambda_k}\}^\infty_{k=1}$, the vector multiplet character
\ba
\mathbf{V}_i=\frac{1-q_1^{-d_i}}{1-q_2}\sum_{(x,x')\in \cX_i \times \cX_i} \frac{x'}{x},
\ea
and the index
\ba
\mathbb{I}\lt[\sum_k x_k\rt]=\prod_k(1-x_k^{-1}).
\ea
The bifundamental contribution associated to the link $e:i\rightarrow j$ with bifundamental mass $\mu_e$ is given by 
\ba
Z^{{\rm bf}}_{e:i\rightarrow j}[\cX_i,\cX_j,d_i,d_j;\mu_e]=\mathbb{I}\lt[\mathbf{H}_{e:i\rightarrow j}\rt]=\prod_{(x,x')\in\cX_i\times \cX_j}\prod_{r=0}^{d_j/d_{ij}-1}\frac{\lt(\mu_e^{-1}q_1^{-rd_{ij}}q_2 x/x';q_2\rt)_\infty}{\lt(\mu_e^{-1}q_1^{-rd_{ij}}q_1^{d_i}q_2 x/x';q_2\rt)_\infty},
\ea
where $d_{ij}={\rm gcd}(d_i,d_j)$ is the greatest common divisor of $d_i$ and $d_j$, and 
\ba
\mathbf{H}_{e:i\rightarrow j}=-\mu_e\frac{(1-q_1^{-d_i})(1-q_1^{d_j})}{(1-q_1^{d_{ij}})(1-q_2)}\sum_{(x,x')\in \cX_i\times \cX_j} \frac{x'}{x}.
\ea
When all $d_i=d_j=d_{ij}=1$,\footnote{To be more precise, as long as we have $d_i=d_0$ for some integer $d_0$ and all $^\forall i$, we recover the partition function for simply-laced quivers with $q_1$ replaced by $q_1^{d_0}$.} the quiver is simply-laced and the partition function reduces to that of an ADE quiver well-known from the localization calculation \cite{bifundInstanton}. 

In this article, we focus on the case with $d_j=d_{ij}$, which includes two important classes of theories: i) $d_j=1$ and ii) $d_i=d_j$. In this specific case, the bifundamental contribution is simplified to 
\ba
Z^{{\rm bf}}_{e:i\rightarrow j}[\cX_i,\cX_j,d_i,d_j;\mu_e]=\prod_{(x,x')\in\cX_i\times \cX_j}\frac{\lt(\mu_e^{-1}q_2 x/x';q_2\rt)_\infty}{\lt(\mu_e^{-1}q_1^{d_i}q_2 x/x';q_2\rt)_\infty}.
\ea

For each box $s=(j,k)$ in the ($j$-th row and $k$-th column of the) Young diagram $\lambda^{(i)}$ which labels $\cX_i$, we assign a coordinate to it, $\chi^{(d_i)}_s=v_iq_1^{d_i(j-1)}q_2^{k-1}$, where $v_i$ again is the Coulomb modulus associated to the $i$-th node. We further define
\ba
\cY^{(d)}_{\cX_i}(z):=\prod_{k=0}^{d/d_i-1}(1-v_iq_1^{kd_i}/z)\prod_{s\in\lambda^{(i)}}S_{d}(\chi_s^{(d_i)}/z),\label{def-frac-Y}
\ea
with 
\ba
S_d(z):=\frac{(1-q_1^dz)(1-q_2z)}{(1-z)(1-q_1^dq_2z)}.
\ea
In the special case we consider, $d_{ij}=d_j$, note that since 
\ba
S_{d_i}(z)=\prod_{k=0}^{d_i/d_j-1}S_{d_j}(q_1^{kd_j}z),\label{split-S}
\ea
we have 
\ba
\cY^{(d)}_{\cX_i}(z)=\prod_{k=0}^{d/d_i-1}\cY_{\lambda^{(i)}}(q_1^{-kd_i}z),\label{decompose-Y}
\ea
where we introduce a new notation $\cY_{\lambda^{(i)}}(z):=\cY^{(d_i)}_{\cX_i}(z)$. 

The recursive relation for $Z^{\rm bf}_{e:i\rightarrow j}$ can be computed to 
\ba
\frac{Z^{\rm bf}_{e:i\rightarrow j}[\cX_{i+s},\cX_j,d_i,d_j;\mu_e]}{Z^{\rm bf}_{e:i\rightarrow j}[\cX_{i},\cX_j,d_i,d_j;\mu_e]}=\prod_{x'\in \cX_j}\frac{1-\mu_e^{-1}q_1^{d_i}x_s/x'}{1-\mu_e^{-1}x_s/x'}=\tilde{\cY}_{\cX_j}^{(d_i)}(\chi^{(d_i)}_s\mu_e^{-1}q_1^{d_i}q_2),\label{frac-Z-1}\\
\frac{Z^{\rm bf}_{e:i\rightarrow j}[\cX_{i},\cX_{j+s},d_i,d_j;\mu_e]}{Z^{\rm bf}_{e:i\rightarrow j}[\cX_{i},\cX_j,d_i,d_j;\mu_e]}=\prod_{x'\in \cX_i}\frac{1-\mu^{-1}q_2x'/x_s}{1-\mu^{-1}q^{d_i}_1q_2x'/x_s}=\cY_{\cX_i}^{(d_i)}(\chi^{(d_j)}_s\mu_e),\label{frac-Z-2}
\ea
where $\tilde{\cY}_{\cX_j}^{(d_i)}(z)$ is defined as 
\ba
\tilde{\cY}_{\cX_j}^{(d_i)}(z):=\prod_{x'\in \cX_j}\frac{1-z/x'}{1-q_1^{-d_i}z/x'},
\ea
which is related to $\cY^{(d_i)}_{\cX_j}(z)$ through 
\ba
\tilde{\cY}_{\cX_j}^{(d_i)}(z)=\lt[\prod_{k=0}^{d_i/d_j-1}(-zq_1^{-k}/v_j)\rt] \cY_{\cX_j}^{(d_i)}(z).\label{tY-Y}
\ea
The equality for $\cY$-functions in (\ref{frac-Z-1}), (\ref{frac-Z-2}), (\ref{tY-Y}), can be derived from the recursive relations, 
\ba
\frac{\cY^{(d)}_{\cX_i+s}(z)}{\cY^{(d)}_{\cX_i}(z)}&=&S_d(\chi_s^{(d_i)}/z)=\frac{1-q_2\chi^{(d_i)}_s/z}{1-q^{d_i}_1q_2\chi^{(d_i)}_s/z}\frac{1-q^{d_i}_1\chi^{(d_i)}_s/z}{1-\chi^{(d_i)}_s/z}\nn\\
&=&\frac{1-x_s/z}{1-q^{d}_1x_s/z}\frac{1-q^{d}_1q_2^{-1}x_s/z}{1-q_2^{-1}x_s/z}=\frac{1-z/x_s}{1-z/(q^{d}_1x_s)}\frac{1-z/(q^{d}_1q_2^{-1}x_s)}{1-z/(q_2^{-1}x_s)}=\frac{\tilde{\cY}^{(d)}_{\cX_i+s}(z)}{\tilde{\cY}^{(d)}_{\cX_i}(z)},
\ea
and the initial condition, 
\ba
\cY^{(d)}_{\cX_i^\emptyset}(z)=\prod_{k=0}^{d/d_i-1}(1-v_iq_1^{kd_i}/z)=\prod_{x\in\cX_i^\emptyset}\frac{1-x/z}{1-q^{d}_1x/z},
\ea
\ba
\tilde{\cY}_{\cX_i}^{(d)}(z)=\prod_{x'\in \cX^\emptyset_i}\frac{1-z/x'}{1-q_1^{-d}z/x'}=\prod_{k=0}^{d/d_i-1}(1-z/(v_iq_1^{kd_i})).
\ea

Solving the recursive relation in a similar way by using the expressions of $\cY^{(d)}_{\cX_i}(z)$ and $\tilde{\cY}_{\cX_i}^{(d)}(z)$ in terms of $S$-functions, we obtain the following alternative expression for the instanton part of $Z^{\rm bf}_{e:i\rightarrow j}$, 
\ba
\frac{Z^{\rm bf}_{e:i\rightarrow j}[\cX_{i},\cX_j,d_i,d_j;\mu_e]}{Z^{\rm bf}_{e:i\rightarrow j}[v_i^\emptyset,v_j^\emptyset,d_i,d_j;\mu_e]}
=\prod_{x\in\lambda}\prod_{k=0}^{d_i/d_j-1}(1-\chi^{(d_i)}_xq_1^{d_i-kd_j}q_2\mu_e^{-1}/v_j)\prod_{y\in\nu}(1-v_i \mu_e^{-1}/\chi_y^{(d_j)})\nn\\
\times\prod_{x\in\lambda,y\in \nu}S_{d_i}(\mu_e^{-1}\chi_x^{(d_i)}/\chi_y^{(d_j)}),
\ea
where $\cX_i$ and $\cX_j$ are respectively the characteristic sets associated to Young diagram $\lambda$ and $\nu$. The above alternative expression will be a key rewriting formula in the construction with web of vertex operators presented in the following section. 

\section{Web Construction for BCFG Quivers and Half-blood Vertex}\label{s:BCFG}

We are now ready to construct the web of vertex operators that realizes non-simply-laced quivers of $BCFG$-type. 

 Let us define the following vertex operators, 
\ba
\tilde{\xi}^{(d_i)}_j(z):=\exp\lt(-\sum_{n=1}^\infty \frac{1}{n}\frac{1-q_1^{nd_j}}{1-q_1^{-nd_i}}(1-q_2^{n})z^n(\gamma_j^2/\gamma_i)^n a_{-n}\rt)\exp\lt(\sum_{n=1}^\infty \frac{1}{n}(1-q_1^{-nd_i})z^{-n}\gamma_i^n a_{n}\rt),
\ea
\ba
\tilde{\Phi}^{\ast(d_i)}_\emptyset[v_j]:=\exp\lt(\sum_{n=1}^\infty \frac{1}{n}\frac{(\gamma_j^2/\gamma_i)^nv_j^n}{1-q_1^{-nd_i}}a_{-n}\rt)\exp\lt(-\sum_{n=1}^\infty \frac{1}{n}\frac{1-q_1^{-nd_i}}{1-q_1^{-nd_j}}\frac{\gamma_i^nv_j^{-n}}{1-q_2^{-n}}a_n\rt),
\ea
where $\gamma_i:=q_1^{-d_i/2}q_2^{-1/2}$ is the refinement parameter associated to the $i$-th node. 
We further define a new vertex with the above two vertex operators, 
\ba
\tilde{\Phi}^{\ast(d_i)}_{\cX_j}[v_j]:=:\tilde{\Phi}^{\ast(d_i)}_\emptyset[v_j]\prod_{s\in\lambda^{(j)}}\tilde{\xi}_j^{(d_i)}(\chi^{(d_j)}_s):,
\ea
which depends on the information of two adjacent nodes, and therefore we would like to call it a half-blood vertex. One can easily check that it satisfies the following contraction rules, 
\ba
&&\contraction{}{\tilde{\Phi}^{\ast(d_i)}_{\cX^{(2)}_j}[v^{(2)}_j]}{}{\tilde{\Phi}^{\ast(d_i)}_{\cX^{(1)}_j}[v^{(1)}_j]}\tilde{\Phi}^{\ast(d_i)}_{\cX^{(2)}_j}[v^{(2)}_j]\tilde{\Phi}^{\ast(d_i)}_{\cX^{(1)}_j}[v^{(1)}_j]=\frac{1}{{\cal G}^{(d_j)}(v_j^{(1)},v^{(2)}_j\gamma_j^{-2})}\frac{Z^{\rm bf}_{e:i\rightarrow j}[v_j^{(1)\emptyset},v_j^{(2)\emptyset} \gamma_j^{-2},d_j,d_j;1]}{Z^{\rm bf}_{e:i\rightarrow j}[\cX^{(1)}_{j},\cX^{(2)}_j\gamma_j^{-2},d_j,d_j;1]}:\tilde{\Phi}^{\ast(d_i)}_{\cX^{(2)}_j}[v^{(2)}_j]\tilde{\Phi}^{\ast(d_i)}_{\cX^{(1)}_j}[v^{(1)}_j]:,\nn\\\label{contra-t2}
\ea
between two ``D5-branes'' with Coulomb moduli $v_j^{(1)}$ and $v_j^{(2)}$ belonging to the same node $j$, where ${\cal G}^{(d_j)}$ is the same ${\cal G}$ function with $q_1$ replaced by $q_1^{d_j}$, i.e. 
\ba
{\cal G}^{(d_j)}(v_1,v_2)=\exp\lt(\sum_{n>0}\frac{1}{n}\frac{(v_1/v_2)^n}{(1-q_1^{-d_jn})(1-q_2^{-n})}\rt),
\ea
and 
\ba
&&\contraction{}{\tilde{\Phi}^{\ast(d_i)}_{\cX_j}[v_j]}{}{\Phi^{(d_i)}_{\cX_i}[v_i]}\tilde{\Phi}^{\ast(d_i)}_{\cX_j}[v_j]\Phi^{(d_i)}_{\cX_i}[v_i]={\cal G}^{(d_j)}(v_i,v_j\gamma_i^{-1})\frac{Z^{\rm bf}_{e:i\rightarrow j}[\cX_{i},\cX_j\gamma_i^{-1},d_i,d_j;1]}{Z^{\rm bf}_{e:i\rightarrow j}[v_i^{\emptyset},v_j^{\emptyset} \gamma_j^{-1},d_i,d_j;1]}:\tilde{\Phi}^{\ast(d_i)}_{\cX_j}[v_j]\Phi^{(d_i)}_{\cX_i}[v_i]:,
\ea
where 
\ba
\Phi^{(d_i)}_{\cX_i}[v_i]:=\lt.\Phi_{\emptyset}[v_i]\prod_{x\in\lambda_i}\eta(\chi_x)\rt|_{q_1\rightarrow q_1^{d_i}}.
\ea
The above contraction relations all reduce to the familiar ones of the usual topological vertex when $d_i=d_j=1$. We remark that we did not include the prefactor, i.e. $t^\ast_n(\lambda,u,v)$ in $\Phi^{\ast}[u,v]$, in the above definition of half-blood vertex, as the quiver structure can be analyzed from the bifundamental contributions in the instanton partition function. We will include and fix this prefactor later when we discuss the AFS property of the half-blood vertex in section~\ref{s:AFS}.

Let us graphically represent the half-blood vertex as 
\begin{align}
\begin{tikzpicture}[scale=2, baseline=(current  bounding  box.center)]
\draw[ultra thick](0,0)--(0,-0.4);
\draw[ultra thick] (0,-0.6)--(0,-1);
\draw (0,-0.5) circle (0.1);
\node at (0.1,-0.5) [right] {$\tilde{\Phi}^{\ast(d_i)}_{\cX_j}[v_j]$};
\draw (-0.1,-0.5)--(-0.5,-0.5);
\end{tikzpicture}
\end{align}
then, for example, the web construction for $BC_2$ quiver is given by 
\begin{align}
\begin{tikzpicture}[scale=2,yscale=-1, baseline=(current  bounding  box.center)]
\draw[ultra thick](0,1)--(0,0.6);
\draw[ultra thick](0,0.4)--(0,-0.4);
\draw[ultra thick] (0,-0.6)--(0,-1);
\draw (0,0.5) circle (0.1);
\draw[ultra thick] (0.1,0.5)--(0.9,0.5);
\draw (1,0.5) circle (0.1);
\draw[ultra thick] (1,1)--(1,0.6);
\draw[ultra thick] (1,0.4)--(1,-0.5);
\draw (0,-0.5) circle (0.1);
\draw (-0.1,-0.5)--(-1,-0.5);
\draw (-1,0.5)--(-1,-1);
\node at (-0.1,0.5) [left] {$\Phi^{(d_2)}_{\cX_2}$};
\node at (0.5,0.5) [above] {$d_2=2$};
\node at (-0.5,-0.5) [below] {$d_1=1$};
\node at (0.1,-0.5) [right] {$\tilde{\Phi}^{\ast(d_2)}_{\cX_1}$};
\end{tikzpicture}
\end{align}
We represented the representation spaces of DIM$_{q_1^{d_i},q_2}$ (for $d_i>1$) with thick lines and those of DIM$_{q_1,q_2}$ with normal lines. The quiver structure in the above web diagram can be read (from the computation of the partition function) as 
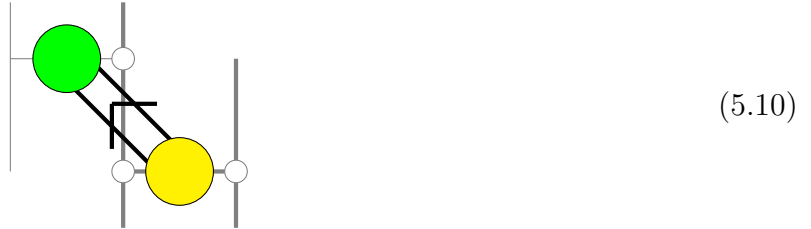
\begin{align}
\begin{tikzpicture}[scale=1.5,yscale=-1, baseline=(current  bounding  box.center)]
\draw[ultra thick, gray](0,1)--(0,0.6);
\draw[ultra thick, gray](0,0.4)--(0,-0.4);
\draw[ultra thick, gray] (0,-0.6)--(0,-1);
\draw[gray] (0,0.5) circle (0.1);
\draw[ultra thick, gray] (0.1,0.5)--(0.9,0.5);
\draw[gray] (1,0.5) circle (0.1);
\draw[ultra thick, gray] (1,1)--(1,0.6);
\draw[ultra thick, gray] (1,0.4)--(1,-0.5);
\draw[gray] (0,-0.5) circle (0.1);
\draw[gray] (-0.1,-0.5)--(-1,-0.5);
\draw[gray] (-1,0.5)--(-1,-1);
\draw [ultra thick] (-0.4,-0.6)--(0.6,0.4);
\draw [ultra thick] (-0.6,-0.4)--(0.4,0.6);
\draw [ultra thick] (-0.1,-0.1)--(0.3,-0.1);
\draw [ultra thick] (-0.1,-0.1)--(-0.1,0.3);
\draw[fill=green] (-0.5,-0.5) circle (0.3);
\draw[fill=yellow] (0.5,0.5) circle (0.3);
\end{tikzpicture}
\end{align}
where the green node is used to represent a node corresponding to a short root. We remark again that $\Phi^{(d_2)}_{\cX_2}$ is nothing but the usual refined topological vertex $\Phi_{\lambda_2}$ simply with $q_1$ replaced by $q_1^{d_2}$.

We can replace the vertices with generalized vertices as in \cite{BFHMZ} to increase the rank of gauge groups in the theory, but one nice feature of the half-blood vertex is that it is defined so that gluing several half-blood vertices together raises the rank of the gauge group: the following web diagram, 
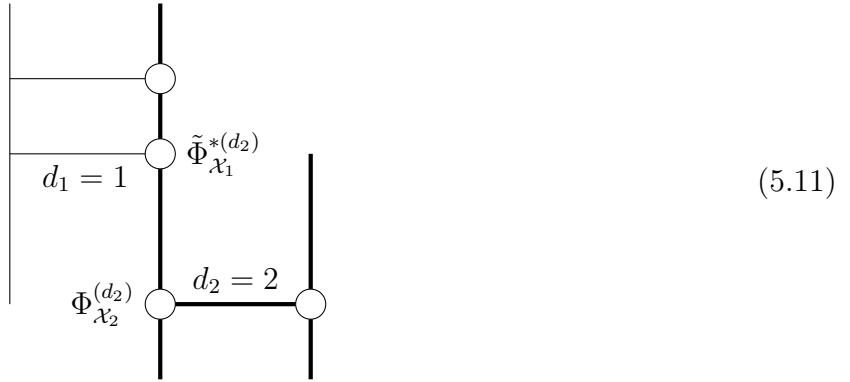
\begin{align}
\begin{tikzpicture}[scale=2, yscale=-1, baseline=(current  bounding  box.center)]
\draw[ultra thick](0,1)--(0,0.6);
\draw[ultra thick](0,0.4)--(0,-0.4);
\draw[ultra thick] (0,-0.6)--(0,-0.9);
\draw (0,0.5) circle (0.1);
\draw[ultra thick] (0.1,0.5)--(0.9,0.5);
\draw (1,0.5) circle (0.1);
\draw[ultra thick] (1,1)--(1,0.6);
\draw[ultra thick] (1,0.4)--(1,-0.5);
\draw (0,-0.5) circle (0.1);
\draw (-0.1,-0.5)--(-1,-0.5);
\draw (-1,0.5)--(-1,-1.5);
\node at (-0.1,0.5) [left] {$\Phi^{(d_2)}_{\cX_2}$};
\node at (0.5,0.5) [above] {$d_2=2$};
\node at (-0.5,-0.5) [below] {$d_1=1$};
\node at (0.1,-0.5) [right] {$\tilde{\Phi}^{\ast(d_2)}_{\cX_1}$};
\draw (0,-1) circle (0.1);
\draw[ultra thick] (0,-1.1)--(0,-1.5);
\draw (-0.1,-1)--(-1,-1);
\end{tikzpicture}
\end{align}
realizes the quiver, 
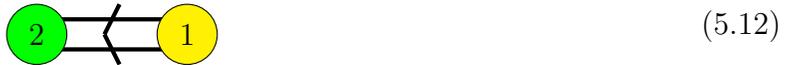
\begin{align}
\begin{tikzpicture}[baseline=(current  bounding  box.center)]
\draw [ultra thick] (0,0.2)--(2,0.2);
\draw [ultra thick] (0,-0.2)--(2,-0.2);
\draw[ultra thick] (0.9,0)--(1.1,0.4);
\draw[ultra thick] (0.9,0)--(1.1,-0.4);
\draw[fill=green] (0,0) circle (0.4);
\draw[fill=yellow] (2,0) circle (0.4);
\node at (0,0) {$2$};
\node at (2,0) {$1$};
\end{tikzpicture}
\end{align}
By replacing $d_2\rightarrow 3$, the quiver structure is lifted to $G_2$.

Adding more branes on the left side, we obtain higher-rank $C$-type quivers, 
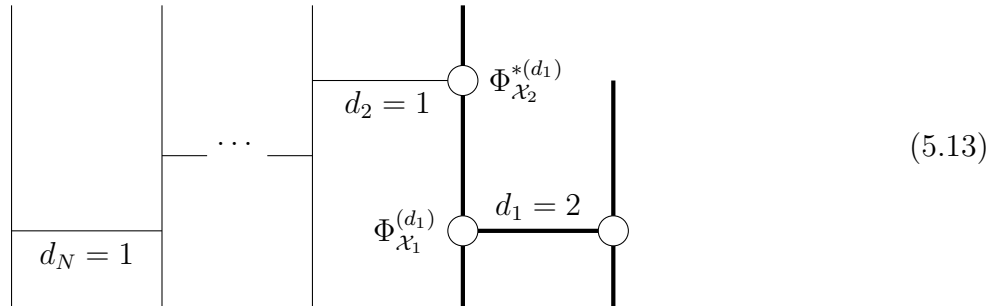
\begin{align}
\begin{tikzpicture}[scale=2,yscale=-1, baseline=(current  bounding  box.center)]
\draw[ultra thick](0,1)--(0,0.6);
\draw[ultra thick](0,0.4)--(0,-0.4);
\draw[ultra thick] (0,-0.6)--(0,-1);
\draw (0,0.5) circle (0.1);
\draw[ultra thick] (0.1,0.5)--(0.9,0.5);
\draw (1,0.5) circle (0.1);
\draw[ultra thick] (1,1)--(1,0.6);
\draw[ultra thick] (1,0.4)--(1,-0.5);
\draw (0,-0.5) circle (0.1);
\draw (-0.1,-0.5)--(-1,-0.5);
\draw (-1,1)--(-1,-1);
\node at (-0.1,0.5) [left] {$\Phi^{(d_1)}_{\cX_1}$};
\node at (0.5,0.5) [above] {$d_1=2$};
\node at (-0.5,-0.5) [below] {$d_2=1$};
\node at (0.1,-0.5) [right] {$\Phi^{\ast(d_1)}_{\cX_2}$};
\draw (-1,0)--(-1.3,0);
\draw (-1.7,0)--(-2,0);
\draw (-2,-1)--(-2,1);
\draw (-3,-1)--(-3,1);
\draw (-2,0.5)--(-3,0.5);
\node at (-1.5,0) [above] {\dots};
\node at (-2.5,0.5) [below] {$d_N=1$};
\end{tikzpicture}
\end{align}
to which the corresponding quiver diagram can be read from the web diagram as 
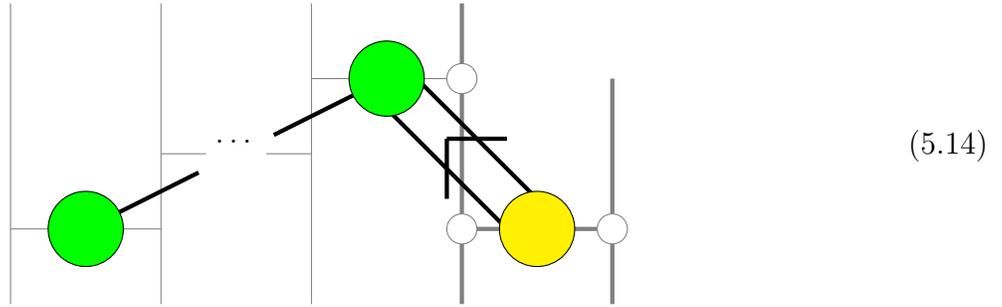
\begin{align}
\begin{tikzpicture}[scale=2,yscale=-1, baseline=(current  bounding  box.center)]
\draw[ultra thick, gray](0,1)--(0,0.6);
\draw[ultra thick, gray](0,0.4)--(0,-0.4);
\draw[ultra thick, gray] (0,-0.6)--(0,-1);
\draw[gray] (0,0.5) circle (0.1);
\draw[ultra thick,gray] (0.1,0.5)--(0.9,0.5);
\draw[gray] (1,0.5) circle (0.1);
\draw[ultra thick,gray] (1,1)--(1,0.6);
\draw[ultra thick,gray] (1,0.4)--(1,-0.5);
\draw[gray] (0,-0.5) circle (0.1);
\draw[gray] (-0.1,-0.5)--(-1,-0.5);
\draw[gray] (-1,1)--(-1,-1);
\draw[gray] (-1,0)--(-1.3,0);
\draw[gray] (-1.7,0)--(-2,0);
\draw[gray] (-2,-1)--(-2,1);
\draw[gray] (-3,-1)--(-3,1);
\draw[gray] (-2,0.5)--(-3,0.5);
\node at (-1.5,0) [above] {\dots};
\draw[ultra thick] (0.6,0.4)--(-0.4,-0.6);
\draw[ultra thick] (0.4,0.6)--(-0.6,-0.4);
\draw[ultra thick] (-0.1,-0.1)--(0.3,-0.1);
\draw[ultra thick] (-0.1,-0.1)--(-0.1,0.3);
\draw[ultra thick] (-0.5,-0.5)--(-1.25,-0.25/2);
\draw[ultra thick] (-2.5,0.5)--(-1.75,0.25/2);
\draw [fill=yellow] (0.5,0.5) circle (0.25);
\draw [fill=green] (-0.5,-0.5) circle (0.25);
\draw[fill=green] (-2.5,0.5) circle (0.25);
\end{tikzpicture}
\end{align}
and adding branes on the right side gives rise to higher-rank $B$-type quivers, 
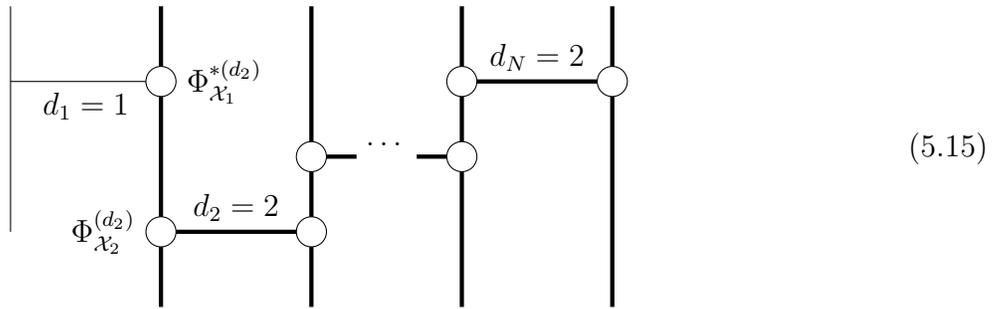
\begin{align}
\begin{tikzpicture}[scale=2,yscale=-1, baseline=(current  bounding  box.center)]
\draw[ultra thick](0,1)--(0,0.6);
\draw[ultra thick](0,0.4)--(0,-0.4);
\draw[ultra thick] (0,-0.6)--(0,-1);
\draw (0,0.5) circle (0.1);
\draw[ultra thick] (0.1,0.5)--(0.9,0.5);
\draw (1,0.5) circle (0.1);
\draw[ultra thick] (1,1)--(1,0.6);
\draw[ultra thick] (1,0.4)--(1,0.1);
\draw[ultra thick] (1,-0.1)--(1,-1);
\draw (0,-0.5) circle (0.1);
\draw (-0.1,-0.5)--(-1,-0.5);
\draw (-1,0.5)--(-1,-1);
\node at (-0.1,0.5) [left] {$\Phi^{(d_2)}_{\cX_2}$};
\node at (0.5,0.5) [above] {$d_2=2$};
\node at (-0.5,-0.5) [below] {$d_1=1$};
\node at (0.1,-0.5) [right] {$\Phi^{\ast(d_2)}_{\cX_1}$};
\draw (1,0) circle (0.1);
\draw[ultra thick] (1.1,0)--(1.3,0);
\draw[ultra thick] (1.7,0)--(1.9,0);
\draw[ultra thick] (2,1)--(2,0.1);
\draw[ultra thick] (2,-0.1)--(2,-0.4);
\draw[ultra thick] (2,-0.6)--(2,-1);
\draw (2.,0) circle (0.1);
\node at (1.5,0) [above] {\dots};
\draw (2,-0.5) circle (0.1);
\draw [ultra thick] (2.1,-0.5)--(2.9,-0.5);
\draw (3,-0.5) circle (0.1);
\draw[ultra thick] (3,1)--(3,-0.4);
\draw[ultra thick] (3,-0.6)--(3,-1);
\node at (2.5,-0.5) [above] {$d_N=2$};
\end{tikzpicture}
\end{align}
whose quiver structure is given by 
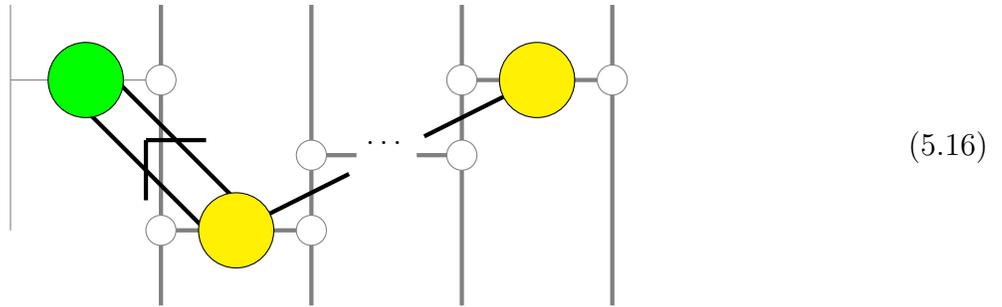
\begin{align}
\begin{tikzpicture}[scale=2,yscale=-1, baseline=(current  bounding  box.center)]
\draw[ultra thick,gray](0,1)--(0,0.6);
\draw[ultra thick,gray](0,0.4)--(0,-0.4);
\draw[ultra thick,gray] (0,-0.6)--(0,-1);
\draw[gray] (0,0.5) circle (0.1);
\draw[ultra thick,gray] (0.1,0.5)--(0.9,0.5);
\draw[gray] (1,0.5) circle (0.1);
\draw[ultra thick,gray] (1,1)--(1,0.6);
\draw[ultra thick,gray] (1,0.4)--(1,0.1);
\draw[ultra thick,gray] (1,-0.1)--(1,-1);
\draw[gray] (0,-0.5) circle (0.1);
\draw[gray] (-0.1,-0.5)--(-1,-0.5);
\draw[gray] (-1,0.5)--(-1,-1);
\draw[gray] (1,0) circle (0.1);
\draw[ultra thick,gray] (1.1,0)--(1.3,0);
\draw[ultra thick,gray] (1.7,0)--(1.9,0);
\draw[ultra thick,gray] (2,1)--(2,0.1);
\draw[ultra thick,gray] (2,-0.1)--(2,-0.4);
\draw[ultra thick,gray] (2,-0.6)--(2,-1);
\draw[gray] (2.,0) circle (0.1);
\node at (1.5,0) [above] {\dots};
\draw[gray] (2,-0.5) circle (0.1);
\draw [ultra thick,gray] (2.1,-0.5)--(2.9,-0.5);
\draw[gray] (3,-0.5) circle (0.1);
\draw[ultra thick,gray] (3,1)--(3,-0.4);
\draw[ultra thick,gray] (3,-0.6)--(3,-1);
\draw[ultra thick] (0.6,0.4)--(-0.4,-0.6);
\draw[ultra thick] (0.4,0.6)--(-0.6,-0.4);
\draw[ultra thick] (-0.1,-0.1)--(0.3,-0.1);
\draw[ultra thick] (-0.1,-0.1)--(-0.1,0.3);
\draw[ultra thick] (0.5,0.5)--(1.25,0.25/2);
\draw[ultra thick] (2.5,-0.5)--(1.75,-0.25/2);
\draw [fill=yellow] (0.5,0.5) circle (0.25);
\draw [fill=green] (-0.5,-0.5) circle (0.25);
\draw [fill=yellow] (2.5,-0.5) circle (0.25);
\end{tikzpicture}
\end{align}
Extending the web in two directions leads to $F_4$ quiver in a similar way: 
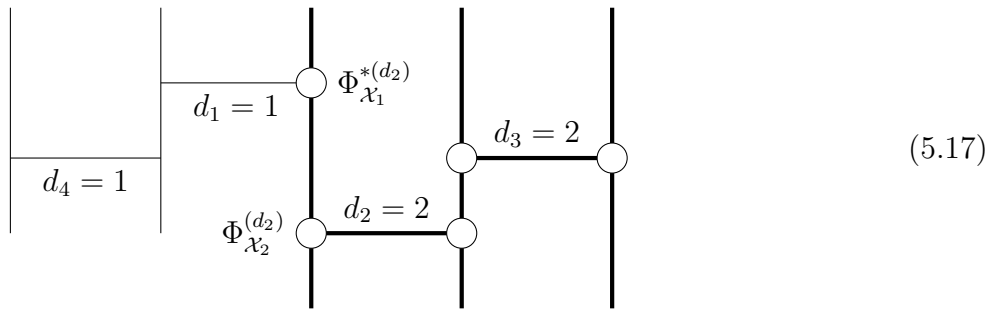
\begin{align}
\begin{tikzpicture}[scale=2,yscale=-1, baseline=(current  bounding  box.center)]
\draw[ultra thick](0,1)--(0,0.6);
\draw[ultra thick](0,0.4)--(0,-0.4);
\draw[ultra thick] (0,-0.6)--(0,-1);
\draw (0,0.5) circle (0.1);
\draw[ultra thick] (0.1,0.5)--(0.9,0.5);
\draw (1,0.5) circle (0.1);
\draw[ultra thick] (1,1)--(1,0.6);
\draw[ultra thick] (1,0.4)--(1,0.1);
\draw[ultra thick] (1,-0.1)--(1,-1);
\draw (0,-0.5) circle (0.1);
\draw (-0.1,-0.5)--(-1,-0.5);
\draw (-1,0.5)--(-1,-1);
\node at (-0.1,0.5) [left] {$\Phi^{(d_2)}_{\cX_2}$};
\node at (0.5,0.5) [above] {$d_2=2$};
\node at (-0.5,-0.5) [below] {$d_1=1$};
\node at (0.1,-0.5) [right] {$\Phi^{\ast(d_2)}_{\cX_1}$};
\draw (1,0) circle (0.1);
\draw[ultra thick] (1.1,0)--(1.9,0);
\draw[ultra thick] (2,1)--(2,0.1);
\draw[ultra thick] (2,-0.1)--(2,-1);
\draw (2.,0) circle (0.1);
\node at (1.5,0) [above] {$d_3=2$};
\draw (-2,0.5)--(-2,-1);
\draw (-2,0)--(-1,0);
\node at (-1.5,0) [below] {$d_4=1$};
\end{tikzpicture}
\end{align}
with the quiver structure, 
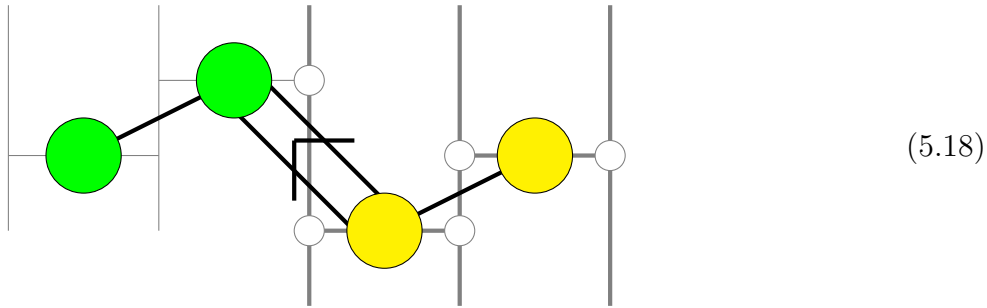
\begin{align}
\begin{tikzpicture}[scale=2,yscale=-1, baseline=(current  bounding  box.center)]
\draw[ultra thick,gray](0,1)--(0,0.6);
\draw[ultra thick,gray](0,0.4)--(0,-0.4);
\draw[ultra thick,gray] (0,-0.6)--(0,-1);
\draw[gray] (0,0.5) circle (0.1);
\draw[ultra thick,gray] (0.1,0.5)--(0.9,0.5);
\draw[gray] (1,0.5) circle (0.1);
\draw[ultra thick,gray] (1,1)--(1,0.6);
\draw[ultra thick,gray] (1,0.4)--(1,0.1);
\draw[ultra thick,gray] (1,-0.1)--(1,-1);
\draw[gray] (0,-0.5) circle (0.1);
\draw[gray] (-0.1,-0.5)--(-1,-0.5);
\draw[gray] (-1,0.5)--(-1,-1);
\draw[gray] (1,0) circle (0.1);
\draw[ultra thick,gray] (1.1,0)--(1.3,0);
\draw[ultra thick,gray] (1.7,0)--(1.9,0);
\draw[ultra thick,gray] (2,1)--(2,0.1);
\draw[ultra thick,gray] (2,-0.1)--(2,-1);
\draw[gray] (2.,0) circle (0.1);
\node at (1.5,0) [above] {\dots};
\draw[gray] (-2,-1)--(-2,0.5);
\draw[gray] (-2,0)--(-1,0);
\draw[ultra thick] (0.6,0.4)--(-0.4,-0.6);
\draw[ultra thick] (0.4,0.6)--(-0.6,-0.4);
\draw[ultra thick] (-0.1,-0.1)--(0.3,-0.1);
\draw[ultra thick] (-0.1,-0.1)--(-0.1,0.3);
\draw[ultra thick] (0.5,0.5)--(1.5,0);
\draw[ultra thick] (-0.5,-0.5)--(-1.5,0);
\draw [fill=yellow] (0.5,0.5) circle (0.25);
\draw [fill=green] (-0.5,-0.5) circle (0.25);
\draw [fill=yellow] (1.5,0) circle (0.25);
\draw [fill=green] (-1.5,0) circle (0.25);
\end{tikzpicture}
\end{align}

\paragraph{Superconformal quiver theories}

The qq-characters in several simple examples of $BCFG$-type quivers are carried out in Appendix \ref{a:qq-BCFG} via the Ward identity formalism. Now we use the result presented there to analyze the condition for superconformal quiver theories with non-simply-laced quivers. We again have the following asymptotic behavior for the $Y$-function associated to the $i$-th gauge node, 
\ba
Y^{(i)}(e^{R\zeta})\sim \zeta^{N},\quad \zeta \sim\infty,\ R\rightarrow 0.
\ea

By solving the superconformal conditions for $\chi^{{\rm BC}_2}_{(1,0)}$ and $\chi^{{\rm BC}_2}_{(0,1)}$, whose expressions (without matter contributions) can be found in (\ref{qq-BC2-10}) and (\ref{qq-BC2-01}), with the field contents of the theory labeled as 
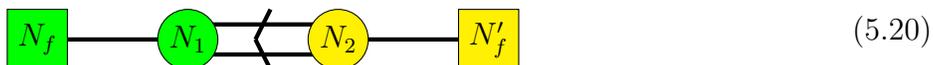
\begin{align}
\begin{tikzpicture}[baseline=(current  bounding  box.center)]
\draw [ultra thick] (0,0.2)--(2,0.2);
\draw [ultra thick] (0,-0.2)--(2,-0.2);
\draw[ultra thick] (0.9,0)--(1.1,0.4);
\draw[ultra thick] (0.9,0)--(1.1,-0.4);
\draw[ultra thick] (0,0)--(-2,0);
\draw[ultra thick] (4,0)--(2,0);
\draw[fill=green] (0,0) circle (0.4);
\draw[fill=yellow] (2,0) circle (0.4);
\draw[fill=green] (-2-0.4,0.4) rectangle (-2+0.4,-0.4);
\draw[fill=yellow] (4-0.4,0.4) rectangle (4+0.4,-0.4);
\node at (0,0) {$N_1$};
\node at (2,0) {$N_2$};
\node at (4,0) {$N'_f$};
\node at (-2,0) {$N_f$};
\end{tikzpicture}
\end{align}
we obtain the superconformal conditions, 
\ba
2N_1=N_2+N_f,\\
2N_2=2N_1+N'_f.
\ea
These equations can be rephrased as 
\ba
\sum_jc_{ij}N_j=N_{f,i},
\ea
with $N_{f,i}$ denoting the number of flavors attached to the $i$-th node. We note that this condition essentially comes from the ${\bf i}$-Weyl reflection relating a pair of terms in the qq-character, and thus it does note depend on which qq-character we consider to extract out the superconformal condition. 

Focus on two special solutions to the superconformal condition for $BC_2$, one with 
\ba
N'_f=0,\quad N_f=N_1=N_2,
\ea
and another with 
\ba
N_f=0,\quad N_2=2N_1=N'_f,\label{sol-B2}
\ea
we see that we cannot draw a consistent web diagram with the external legs of the middle brane in the unpreferred direction parallel to each other, which seems to contradict with the observation we had for $A$-type and $D$-type quivers that all external legs of a superconformal quiver theory can be put in either the preferred direction or the orthogonal direction. This suggests that there either exists certain kind of Hanany-Witten-like transiotion we do not understand for non-simply-laced quivers or there is a better (and more physical) construction for these quivers. 

We remark, however, based on the AFS property formulated in section \ref{s:AFS} for the half-blood vertex (\ref{half-blood-map}), we can draw a web diagram (with directions) shown in Figure \ref{BC2-SC} for the solution (\ref{sol-B2}), where by neglecting the leftmost Fock space all the external branes  are either in the horizontal or in the vertical directions, as in the brane webs for $A$-type and $D$-type superconformal quiver theories. 

\begin{figure}[H]
\begin{center}
\begin{tikzpicture}[scale=2,yscale=-1]
\draw[ultra thick](0+0.089,-0.5+0.045)--(0+0.89*0.4,-0.5+0.45*0.4);
\draw[ultra thick] (0,-0.6)--(0,-1);
\draw (0.89/2,-0.5+0.45/2) circle (0.1);
\draw[ultra thick] (0.89/2+0.1,-0.5+0.45/2)--(0.89/2+0.9,-0.5+0.45/2);
\draw (0.89/2+1,-0.5+0.45/2) circle (0.1);
\draw[ultra thick] (0.89/2+1+0.071,-0.5+0.45/2-0.071)--(0.89/2+1+0.71*0.4,-0.5+0.45/2-0.71*0.4);
\draw (0.89/2+1+0.71*0.5,-0.5+0.45/2-0.71*0.5) circle (0.1);
\draw[ultra thick] (0.89/2+1+0.71*0.5,-0.5+0.45/2-0.71*0.5-0.1)--(0.89/2+1+0.71*0.5,-0.5+0.45/2-0.71*0.5-0.5);
\draw[ultra thick] (0.89/2+1+0.71*0.5+0.1,-0.5+0.45/2-0.71*0.5)--(0.89/2+1+0.71*0.5+0.5,-0.5+0.45/2-0.71*0.5);
\draw[ultra thick] (0.89/2+1,-0.5+0.45/2+0.1)--(0.89/2+1,-0.5+0.45/2+0.4);
\draw (0.89/2+1,-0.5+0.45/2+0.5) circle (0.1);
\draw (0,-0.5) circle (0.1);
\draw (-0.1,-0.5)--(-1,-0.5);
\draw (-1,-0.5)--(-1,-1);
\draw (-1,-0.5)--(-1-0.71/2,-0.5+0.71/2);
\node at (0.89/2+0.5,-0.5+0.45/2) [above] {$d_2=2$};
\node at (-0.5,-0.5) [below] {$d_1=1$};
\draw[ultra thick] (0.89/2+0.071,-0.5+0.45/2+0.071)--(0.89/2+0.5-0.071,-0.5+0.45/2+0.5-0.071);
\draw (0.89/2+0.5,-0.5+0.45/2+0.5) circle (0.1);
\draw [ultra thick] (0.89/2+0.5+0.1,-0.5+0.45/2+0.5)--(0.89/2+0.9,-0.5+0.45/2+0.5);
\draw [ultra thick] (0.89/2+1+0.071,-0.5+0.45/2+0.5+0.071)--(0.89/2+1+0.4*0.71,-0.5+0.45/2+0.5+0.4*0.71);
\draw (0.89/2+1+0.5*0.71,-0.5+0.45/2+0.5+0.5*0.71) circle (0.1);
\draw [ultra thick] (0.89/2+1+0.5*0.71+0.1,-0.5+0.45/2+0.5+0.5*0.71)--(0.89/2+1.5+0.5*0.71,-0.5+0.45/2+0.5+0.5*0.71);
\draw [ultra thick] (0.89/2+1+0.5*0.71,-0.4+0.45/2+0.5+0.5*0.71)--(0.89/2+1+0.5*0.71,0.45/2+0.5+0.5*0.71);
\draw[ultra thick] (0.89/2+0.5,0.1+0.45/2)--(0.89/2+0.5,0.45/2+0.5+0.5*0.71);
\end{tikzpicture}
\caption{The web diagram for $BC_2$ quiver with $N_f=0$, $N_2=2N_1=N'_f=2$.}
\label{BC2-SC}
\end{center}
\end{figure}

The superconformal condition for $B_3$ quiver, with its field contents labeled as 
\begin{align}
\begin{tikzpicture}[baseline=(current  bounding  box.center)]
\draw [ultra thick] (0,0.2)--(2,0.2);
\draw [ultra thick] (0,-0.2)--(2,-0.2);
\draw[ultra thick] (0.9,0)--(1.1,0.4);
\draw[ultra thick] (0.9,0)--(1.1,-0.4);
\draw[ultra thick] (0,0)--(-2,0);
\draw[ultra thick] (4,0)--(2,0);
\draw[ultra thick] (4,0)--(6,0);
\draw[fill=green] (0,0) circle (0.4);
\draw[fill=yellow] (2,0) circle (0.4);
\draw[fill=green] (-2-0.4,0.4) rectangle (-2+0.4,-0.4);
\draw[fill=yellow] (4,0) circle (0.4);
\draw[fill=yellow] (6-0.4,0.4) rectangle (6+0.4,-0.4);
\node at (0,0) {$N_1$};
\node at (2,0) {$N_2$};
\node at (4,0) {$N_3$};
\node at (-2,0) {$N_f$};
\node at (6,0) {$N'_f$};
\end{tikzpicture}
\end{align}
can be read from the expression of the qq-character 
\ba
2N_1=N_f+N_2,\\
2N_2=2N_1+N_3,\\
2N_3=N_2+N'_f,
\ea
which is again equivalent to 
\ba
\sum_jc_{ij}N_j=N_{f,i}.
\ea
We again have the solution 
\ba
N_f=0,\quad N_2=N_3=N'_f=2N_1,
\ea
for which we can draw a similar diagram as Figure \ref{BC2-SC} with all external thick lines either in the horizontal or in the vertical directions. This class of solutions can always be found for general $B$-type quivers. 

The superconformal condition for $G_2$, whose field contents are again labeled by 
\begin{align}
\begin{tikzpicture}[baseline=(current  bounding  box.center)]
\draw [ultra thick] (0,0.3)--(2,0.3);
\draw [ultra thick] (0,0)--(2,0);
\draw [ultra thick] (0,-0.3)--(2,-0.3);
\draw[ultra thick] (0.9,0)--(1.1,0.4);
\draw[ultra thick] (0.9,0)--(1.1,-0.4);
\draw[ultra thick] (0,0)--(-2,0);
\draw[ultra thick] (4,0)--(2,0);
\draw[fill=green] (0,0) circle (0.4);
\draw[fill=yellow] (2,0) circle (0.4);
\draw[fill=green] (-2-0.4,0.4) rectangle (-2+0.4,-0.4);
\draw[fill=yellow] (4-0.4,0.4) rectangle (4+0.4,-0.4);
\node at (0,0) {$N_1$};
\node at (2,0) {$N_2$};
\node at (4,0) {$N'_f$};
\node at (-2,0) {$N_f$};
\end{tikzpicture}
\end{align}
is given by 
\ba
2N_1=N_2+N_f,\\
2N_2=3N_1+N'_f.
\ea
We again have two special solutions, 
\ba
N'_f=0,\quad N_1=2N_f,\ N_2=3N_f,
\ea
and 
\ba
N_f=0,\quad N_2=2N_1=2N'_f,
\ea
indicating different directions of the middle brane (in the unpreferred direction) of these two superconformal quiver theories. 

\section{$E$-type Quivers and ``Square-root'' Vertices}\label{s:square-root}

$E$-type quivers are more tricky to deal with at the level of the brane web. In the case of quivers corresponding to classical Lie algebras, one can see an implicit relation between the web diagram and the set of simple roots. It is most clear for $A$-type quivers: for the $A_n$ quiver, we have $n+1$ Fock spaces in unpreferred directions to express $n$ gauge nodes, and each gauge node is sandwiched by two topological vertices, $\Phi$ and $\Phi^\ast$. This resembles the construction of $n$ simple roots with $n+1$ independent unit vectors $e_i$, 
\ba
\alpha_i=e_i-e_{i+1}, \quad \text{s.t.} \ \alpha_i\cdot \alpha_i=2,\ \alpha_i\cdot\alpha_{i+1}=-1.
\ea
In the case of $D_n$ quiver, $n-1$ of $n$ simple roots are expressed in the same way, i.e. $\alpha_{i}=e_i-e_{i+1}$ for $i=1,\dots n-1$, while the last simple root is given by the sum of two unit vectors, 
\ba
\alpha_n=e_{n}+e_{n+1}.
\ea
The different form of the last simple root $\alpha_n$ exactly corresponds to the realization of a gauge node by gluing two topological vertices of type $\Phi$ with a reflection state, or one $\Phi$-vertex and one $\bar{\Phi}^\ast$-vertex, which essentially behaves like a $\Phi$-vertex. 

This correspondence can be found also in $BC$-type quivers, but in a more complicated (and subtle) way. In the case of $C$-type quiver, as the positive-mode part of the half-blood vertex $\tilde{\Phi}^{\ast(d_i)}_{\cX_j}[v_j]$ can be factorized into $d_i/d_j$ vertex operators, which appear in the topological vertex defined in DIM$_{q_1^{d_j},q_2}$, 
\ba
\exp\lt(\sum_{n=1}^\infty \frac{1}{n}(1-q_1^{-nd_i})z^{-n}\gamma_i^n a_{n}\rt)=\prod_{k=0}^{d_i/d_j-1}\exp\lt(\sum_{n=1}^\infty \frac{1}{n}q_1^{-nkd_j}(1-q_1^{-nd_j})z^{-n}\gamma_i^n a_{n}\rt),
\ea
\ba
\exp\lt(-\sum_{n=1}^\infty \frac{1}{n}\frac{1-q_1^{-nd_i}}{1-q_1^{-nd_j}}\frac{\gamma_i^nv_j^{-n}}{1-q_2^{-n}} a_n\rt)=\prod_{k=0}^{d_i/d_j-1}\exp\lt(-\sum_{n=1}^\infty \frac{1}{n}q_1^{-nkd_j}\frac{\gamma_i^nv_j^{-n}}{1-q_2^{-n}} a_n\rt),
\ea
a simple root of a different form, $\alpha_n=d_i/d_j e_n$, will be induced in the root system\footnote{We in fact absorbed $d_i/d_j-1$ vertex operators into the usual topological vertex $\Phi$ of the node with $d=d_i$ here.}. We note that we need to neglect the rightmost Fock space in the web construction using half-blood vertices described in the previous section to establish the correspondence with simple roots in the above discussion, and this neglect can only be done to $B$-type quivers. For $B$-type quivers, the half-blood vertex can only be viewed as one single topological vertex defined in DIM$_{q^{d_i},q_2}$, and thus we have a simple root of the form, $\alpha_1=-e_1$ (be neglecting the leftmost Fock space in the web construction given in the previous section). We note that this neglect indeed agrees with the discussion on the web diagram of superconformal quiver theories shown in Figure \ref{BC2-SC}.

We now turn to the construction of $E_8$ quiver from an analogy to the realization of its simple roots and obtain the whole $E$-series by removing some D5 branes from the web diagram. One beautiful realization of the simple-root system in 8 unit vectors is given by 
\ba
&&\alpha_1=e_1-e_2,\quad \alpha_2=e_2-e_3,\quad \alpha_3=e_3-e_4,\\
&&\alpha_4=e_4-e_5,\quad \alpha_5=e_5-e_6,\quad \alpha_6=e_6+e_7,\\
&&\alpha_7=-\frac{1}{2}e_1-\frac{1}{2}e_2-\frac{1}{2}e_3-\frac{1}{2}e_4-\frac{1}{2}e_5-\frac{1}{2}e_6-\frac{1}{2}e_7-\frac{1}{2}e_8,\quad \alpha_8=e_6-e_7,
\ea
where the label of the simple roots can be read from the following Dynkin diagram, 
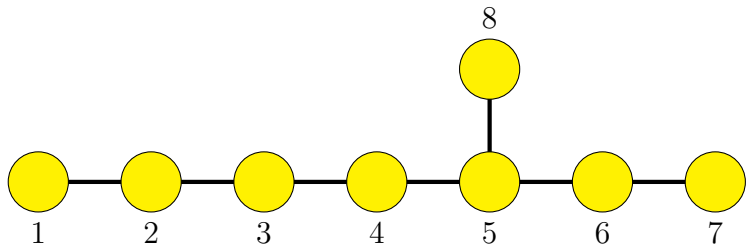
\begin{center}
\begin{tikzpicture}[baseline=(current  bounding  box.center)]
\draw[ultra thick] (0,0)--(1.5,0);
\draw[ultra thick] (0,0)--(0,1.5);
\draw[ultra thick] (0,0)--(-1.5,0);
\draw[ultra thick] (3,0)--(1.5,0);
\draw[ultra thick] (-3,0)--(-1.5,0);
\draw[ultra thick] (-3,0)--(-4.5,0);
\draw[ultra thick] (-6,0)--(-4.5,0);
\draw[fill=yellow] (0,0) circle (0.4);
\draw[fill=yellow] (1.5,0) circle (0.4);
\draw[fill=yellow] (-1.5,0) circle (0.4);
\draw[fill=yellow] (3,0) circle (0.4);
\draw[fill=yellow] (-3,0) circle (0.4);
\draw[fill=yellow] (0,1.5) circle (0.4);
\draw[fill=yellow] (-4.5,0) circle (0.4);
\draw[fill=yellow] (-6,0) circle (0.4);
\node at (0,1.9) [above] {8};
\node at (0,-0.4) [below] {5};
\node at (-1.5,-0.4) [below] {4};
\node at (-3,-0.4) [below] {3};
\node at (1.5,-0.4) [below] {6};
\node at (3,-0.4) [below] {7};
\node at (-4.5,-0.4) [below] {2};
\node at (-6,-0.4) [below] {1};
\end{tikzpicture}
\end{center}
The web construction can be found as 
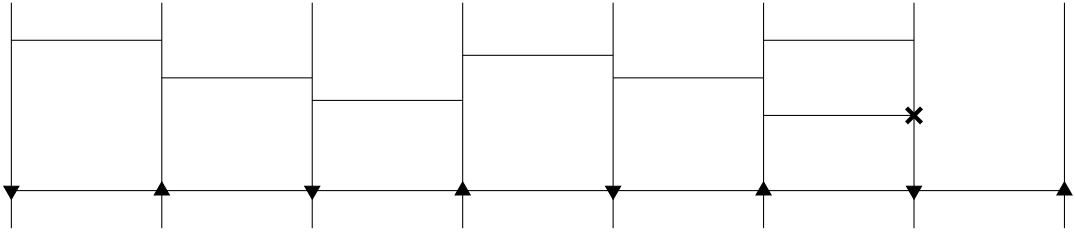
\begin{center}
\begin{tikzpicture}[baseline=(current  bounding  box.center)]
\draw (-1,1)--(-1,-2);
\draw (1,1)--(1,-2);
\draw (-3,1)--(-3,-2);
\draw (3,1)--(3,-2);
\draw (-5,1)--(-5,-2);
\draw (5,1)--(5,-2);
\draw (-7,1)--(-7,-2);
\draw (7,1)--(7,-2);
\draw (-7,0.5)--(-5,0.5);
\draw (-5,0)--(-3,0);
\draw (-3,-0.3)--(-1,-0.3);
\draw (-1,0.3)--(1,0.3);
\draw (1,0)--(3,0);
\draw (3,0.5)--(5,0.5);
\draw (3,-0.5)--(5,-0.5);
\draw [ultra thick] (5-0.1,-0.5+0.1)--(5+0.1,-0.5-0.1);
\draw [ultra thick] (5-0.1,-0.5-0.1)--(5+0.1,-0.5+0.1);
\draw (-7,-1.5)--(7,-1.5);
\draw[fill=black] (-7-0.1,-1.5+0.058)--(-7+0.1,-1.5+0.058)--(-7,-1.5-0.115)--cycle;
\draw[fill=black] (-5-0.1,-1.5-0.058)--(-5+0.1,-1.5-0.058)--(-5,-1.5+0.115)--cycle;
\draw[fill=black] (-3-0.1,-1.5+0.058)--(-3+0.1,-1.5+0.058)--(-3,-1.5-0.115)--cycle;
\draw[fill=black] (-1-0.1,-1.5-0.058)--(-1+0.1,-1.5-0.058)--(-1,-1.5+0.115)--cycle;
\draw[fill=black] (1-0.1,-1.5+0.058)--(1+0.1,-1.5+0.058)--(1,-1.5-0.115)--cycle;
\draw[fill=black] (3-0.1,-1.5-0.058)--(3+0.1,-1.5-0.058)--(3,-1.5+0.115)--cycle;
\draw[fill=black] (5-0.1,-1.5+0.058)--(5+0.1,-1.5+0.058)--(5,-1.5-0.115)--cycle;
\draw[fill=black] (7-0.1,-1.5-0.058)--(7+0.1,-1.5-0.058)--(7,-1.5+0.115)--cycle;
\end{tikzpicture}
\end{center}
where the long line (connecting 8 vertices) 
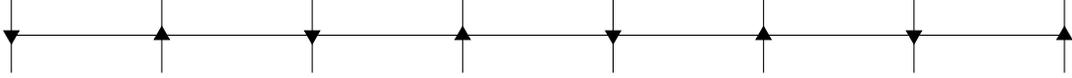
\begin{center}
\begin{tikzpicture}[baseline=(current  bounding  box.center)]
\draw (-7,-1.5)--(7,-1.5);
\draw[fill=black] (-7-0.1,-1.5+0.058)--(-7+0.1,-1.5+0.058)--(-7,-1.5-0.115)--cycle;
\draw[fill=black] (-5-0.1,-1.5-0.058)--(-5+0.1,-1.5-0.058)--(-5,-1.5+0.115)--cycle;
\draw[fill=black] (-3-0.1,-1.5+0.058)--(-3+0.1,-1.5+0.058)--(-3,-1.5-0.115)--cycle;
\draw[fill=black] (-1-0.1,-1.5-0.058)--(-1+0.1,-1.5-0.058)--(-1,-1.5+0.115)--cycle;
\draw[fill=black] (1-0.1,-1.5+0.058)--(1+0.1,-1.5+0.058)--(1,-1.5-0.115)--cycle;
\draw[fill=black] (3-0.1,-1.5-0.058)--(3+0.1,-1.5-0.058)--(3,-1.5+0.115)--cycle;
\draw[fill=black] (5-0.1,-1.5+0.058)--(5+0.1,-1.5+0.058)--(5,-1.5-0.115)--cycle;
\draw[fill=black] (7-0.1,-1.5-0.058)--(7+0.1,-1.5-0.058)--(7,-1.5+0.115)--cycle;
\draw (-1,-1)--(-1,-2);
\draw (1,-1)--(1,-2);
\draw (-3,-1)--(-3,-2);
\draw (3,-1)--(3,-2);
\draw (-5,-1)--(-5,-2);
\draw (5,-1)--(5,-2);
\draw (-7,-1)--(-7,-2);
\draw (7,-1)--(7,-2);
\end{tikzpicture}
\end{center}
is a short representation of the following computation, 
\begin{center}
\begin{align}
 \begin{tikzpicture}[baseline=(current  bounding  box.center)]
\draw[fill=black] (-7-0.1,-1.5+0.058)--(-7+0.1,-1.5+0.058)--(-7,-1.5-0.115)--cycle;
\draw[fill=black] (-5-0.1,-2.5-0.058)--(-5+0.1,-2.5-0.058)--(-5,-2.5+0.115)--cycle;
\draw[fill=black] (-3-0.1,-1.5+0.058)--(-3+0.1,-1.5+0.058)--(-3,-1.5-0.115)--cycle;
\draw[fill=black] (-1-0.1,-2.5-0.058)--(-1+0.1,-2.5-0.058)--(-1,-2.5+0.115)--cycle;
\draw[fill=black] (1-0.1,-1.5+0.058)--(1+0.1,-1.5+0.058)--(1,-1.5-0.115)--cycle;
\draw[fill=black] (3-0.1,-2.5-0.058)--(3+0.1,-2.5-0.058)--(3,-2.5+0.115)--cycle;
\draw[fill=black] (5-0.1,-1.5+0.058)--(5+0.1,-1.5+0.058)--(5,-1.5-0.115)--cycle;
\draw[fill=black] (7-0.1,-2.5-0.058)--(7+0.1,-2.5-0.058)--(7,-2.5+0.115)--cycle;
\draw (-1,-1)--(-1,-3);
\draw (1,-1)--(1,-3);
\draw (-3,-1)--(-3,-3);
\draw (3,-1)--(3,-3);
\draw (-5,-1)--(-5,-3);
\draw (5,-1)--(5,-3);
\draw (-7,-1)--(-7,-3);
\draw (7,-1)--(7,-3);
\draw (-7,-1.5)--(-6.7,-1.5);
\node at (-8,-2) {$\sum_\lambda a_\lambda$};
\node at (-6.7,-1.5) [right] {$\ket{v,\lambda}$};
\draw (-5,-2.5)--(-5.3,-2.5);
\node at (-5.3,-2.5) [left] {$\bra{v,\lambda}$};
\draw (-3,-1.5)--(-2.7,-1.5);
\node at (-2.7,-1.5) [right] {$\ket{v,\lambda}$};
\draw (-1,-2.5)--(-1.3,-2.5);
\node at (-1.3,-2.5) [left] {$\bra{v,\lambda}$};
\draw (1,-1.5)--(1.3,-1.5);
\node at (1.3,-1.5) [right] {$\ket{v,\lambda}$};
\draw (3,-2.5)--(2.7,-2.5);
\node at (2.7,-2.5) [left] {$\bra{v,\lambda}$};
\draw (5,-1.5)--(5.3,-1.5);
\node at (5.3,-1.5) [right] {$\ket{v,\lambda}$};
\draw (7,-2.5)--(6.7,-2.5);
\node at (6.7,-2.5) [left] {$\bra{v,\lambda}$};
\end{tikzpicture}\nn\\\label{long-line}
\end{align}
\end{center}
and we introduced two types of new vertices as the ``square-root'' of $\bar{\Phi}$ and $\Phi^\ast$: 
\begin{center}
\begin{align}
\begin{tikzpicture}[baseline=(current  bounding  box.center)]
\draw (0,1)--(0,-1);
\draw (0,0)--(1,0);
\draw[fill=black] (0.1,0.058)--(-0.1,0.058)--(0,-0.115)--cycle;
\draw [<->] (1.5,0)--(2,0);
\node at (3.5,0) {$\bar{\Phi}^{[1/2]}[v]$,};
\end{tikzpicture}
\end{align}
\end{center}
\ba
\bar{\Phi}^{[1/2]}[v]\ket{v,\lambda}=\bar{\Phi}^{\frac{1}{2}}_\emptyset[v]\prod_{x\in\lambda}\bar{\eta}^{\frac{1}{2}}(\chi_x),
\ea
\begin{center}
\begin{align}
\begin{tikzpicture}[baseline=(current  bounding  box.center)]
\draw (1,1)--(1,-1);
\draw (0,0)--(1,0);
\draw[fill=black] (1+0.1,-0.058)--(1-0.1,-0.058)--(1,0.115)--cycle;
\draw [<->] (1.5,0)--(2,0);
\node at (3.5,0) {$\Phi^{\ast[\frac{1}{2}]}[v]$,};
\end{tikzpicture}
\end{align}
\end{center}
\ba
\bra{v,\lambda}\Phi^{\ast[\frac{1}{2}]}[v]=\Phi^{\ast\frac{1}{2}}_\emptyset[v]\prod_{x\in\lambda}\xi^{\frac{1}{2}}(\chi_x).
\ea
We notice again that we omitted the prefactor that corresponds to $t_n(\lambda,u,v)$ in $\Phi[u,v]$ in the expression of new vertices here. As $\bar{\Phi}$ vertex behaves in a similar way to $\Phi^\ast$ as a vertex operator in the unpreferred direction, the construction of the 7-th node, i.e. the long line, indeed resembles the expression of the simple root $\alpha_7$. 

With the long line standing for the 7-th node in the quiver diagram, the quiver structure can be read from the web diagram (by computing the instanton partition function) as 
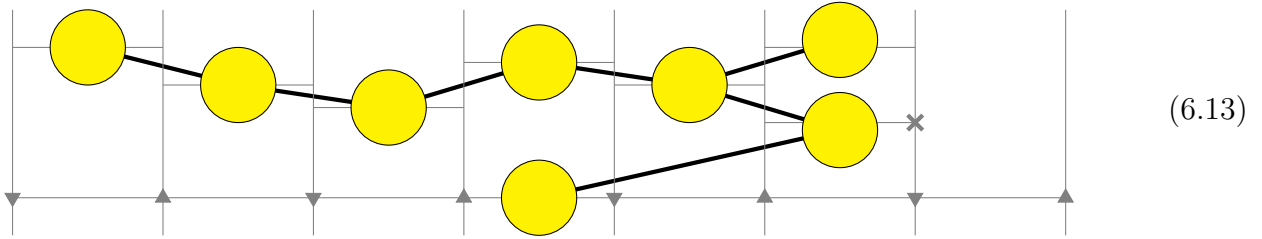
\begin{align}
\begin{tikzpicture}[baseline=(current  bounding  box.center)]
\draw[ultra thick] (-6,0.5)--(-4,0);
\draw[ultra thick] (-2,-0.3)--(-4,0);
\draw[ultra thick] (-2,-0.3)--(0,0.3);
\draw[ultra thick] (2,0)--(0,0.3);
\draw[ultra thick] (2,0)--(4,0.6);
\draw[ultra thick] (2,0)--(4,-0.6);
\draw[ultra thick] (0,-1.5)--(4,-0.6);
\draw[gray] (-1,1)--(-1,-2);
\draw[gray] (1,1)--(1,-2);
\draw[gray] (-3,1)--(-3,-2);
\draw[gray] (3,1)--(3,-2);
\draw[gray] (-5,1)--(-5,-2);
\draw[gray] (5,1)--(5,-2);
\draw[gray] (-7,1)--(-7,-2);
\draw[gray] (7,1)--(7,-2);
\draw[gray] (-7,0.5)--(-5,0.5);
\draw[gray] (-5,0)--(-3,0);
\draw[gray] (-3,-0.3)--(-1,-0.3);
\draw[gray] (-1,0.3)--(1,0.3);
\draw[gray] (1,0)--(3,0);
\draw[gray] (3,0.5)--(5,0.5);
\draw[gray] (3,-0.5)--(5,-0.5);
\draw [ultra thick, gray] (5-0.1,-0.5+0.1)--(5+0.1,-0.5-0.1);
\draw [ultra thick, gray] (5-0.1,-0.5-0.1)--(5+0.1,-0.5+0.1);
\draw[gray] (-7,-1.5)--(7,-1.5);
\draw[fill=gray, gray] (-7-0.1,-1.5+0.058)--(-7+0.1,-1.5+0.058)--(-7,-1.5-0.115)--cycle;
\draw[fill=gray, gray] (-5-0.1,-1.5-0.058)--(-5+0.1,-1.5-0.058)--(-5,-1.5+0.115)--cycle;
\draw[fill=gray,gray] (-3-0.1,-1.5+0.058)--(-3+0.1,-1.5+0.058)--(-3,-1.5-0.115)--cycle;
\draw[fill=gray,gray] (-1-0.1,-1.5-0.058)--(-1+0.1,-1.5-0.058)--(-1,-1.5+0.115)--cycle;
\draw[fill=gray,gray] (1-0.1,-1.5+0.058)--(1+0.1,-1.5+0.058)--(1,-1.5-0.115)--cycle;
\draw[fill=gray,gray] (3-0.1,-1.5-0.058)--(3+0.1,-1.5-0.058)--(3,-1.5+0.115)--cycle;
\draw[fill=gray,gray] (5-0.1,-1.5+0.058)--(5+0.1,-1.5+0.058)--(5,-1.5-0.115)--cycle;
\draw[fill=gray,gray] (7-0.1,-1.5-0.058)--(7+0.1,-1.5-0.058)--(7,-1.5+0.115)--cycle;
\draw[fill=yellow] (-6,0.5) circle (0.5);
\draw[fill=yellow] (-4,0) circle (0.5);
\draw[fill=yellow] (-2,-0.3) circle (0.5);
\draw[fill=yellow] (0,0.3) circle (0.5);
\draw[fill=yellow] (2,0) circle (0.5);
\draw[fill=yellow] (4,0.6) circle (0.5);
\draw[fill=yellow] (4,-0.6) circle (0.5);
\draw[fill=yellow] (0,-1.5) circle (0.5);
\end{tikzpicture}
\end{align}
For example, it is easy to see that the D5 brane corresponding to the first node and the long line completely decouple from each other (factors from the contraction in the left-most Fock space and those from the second left Fock space cancel with each other as in the computation for $D_2$ quiver, refer to (\ref{partition-D2})). 
We first remark that at the 7-th node simply piling up the long lines as 
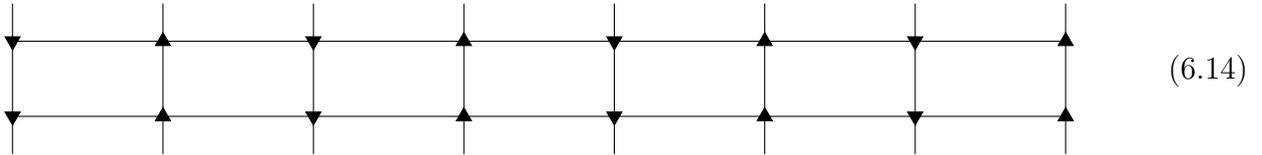
\begin{align}
\begin{tikzpicture}[baseline=(current  bounding  box.center)]
\draw (-7,-1.5)--(7,-1.5);
\draw (-7,-2.5)--(7,-2.5);
\draw[fill=black] (-7-0.1,-1.5+0.058)--(-7+0.1,-1.5+0.058)--(-7,-1.5-0.115)--cycle;
\draw[fill=black] (-5-0.1,-1.5-0.058)--(-5+0.1,-1.5-0.058)--(-5,-1.5+0.115)--cycle;
\draw[fill=black] (-3-0.1,-1.5+0.058)--(-3+0.1,-1.5+0.058)--(-3,-1.5-0.115)--cycle;
\draw[fill=black] (-1-0.1,-1.5-0.058)--(-1+0.1,-1.5-0.058)--(-1,-1.5+0.115)--cycle;
\draw[fill=black] (1-0.1,-1.5+0.058)--(1+0.1,-1.5+0.058)--(1,-1.5-0.115)--cycle;
\draw[fill=black] (3-0.1,-1.5-0.058)--(3+0.1,-1.5-0.058)--(3,-1.5+0.115)--cycle;
\draw[fill=black] (5-0.1,-1.5+0.058)--(5+0.1,-1.5+0.058)--(5,-1.5-0.115)--cycle;
\draw[fill=black] (7-0.1,-1.5-0.058)--(7+0.1,-1.5-0.058)--(7,-1.5+0.115)--cycle;
\draw[fill=black] (-7-0.1,-2.5+0.058)--(-7+0.1,-2.5+0.058)--(-7,-2.5-0.115)--cycle;
\draw[fill=black] (-5-0.1,-2.5-0.058)--(-5+0.1,-2.5-0.058)--(-5,-2.5+0.115)--cycle;
\draw[fill=black] (-3-0.1,-2.5+0.058)--(-3+0.1,-2.5+0.058)--(-3,-2.5-0.115)--cycle;
\draw[fill=black] (-1-0.1,-2.5-0.058)--(-1+0.1,-2.5-0.058)--(-1,-2.5+0.115)--cycle;
\draw[fill=black] (1-0.1,-2.5+0.058)--(1+0.1,-2.5+0.058)--(1,-2.5-0.115)--cycle;
\draw[fill=black] (3-0.1,-2.5-0.058)--(3+0.1,-2.5-0.058)--(3,-2.5+0.115)--cycle;
\draw[fill=black] (5-0.1,-2.5+0.058)--(5+0.1,-2.5+0.058)--(5,-2.5-0.115)--cycle;
\draw[fill=black] (7-0.1,-2.5-0.058)--(7+0.1,-2.5-0.058)--(7,-2.5+0.115)--cycle;
\draw (-1,-1)--(-1,-3);
\draw (1,-1)--(1,-3);
\draw (-3,-1)--(-3,-3);
\draw (3,-1)--(3,-3);
\draw (-5,-1)--(-5,-3);
\draw (5,-1)--(5,-3);
\draw (-7,-1)--(-7,-3);
\draw (7,-1)--(7,-3);
\end{tikzpicture}
\end{align}
raises the rank of the gauge group of the 7-th node, and therefore one is allowed to raise or lower the rank of gauge group at each node freely. The partition function obtained from this construction is consistent with that written down in terms of the $E_8$ ($q$-)Cartan matrix \cite{Kimura-Pestun}.  Another remark is that the summation computation given in (\ref{long-line}) can be achieved by combining several pieces of trivalent vertices, whose definitions and graphical representations are given by 
\begin{align}
\begin{tikzpicture}[baseline=(current  bounding  box.center)]
\draw (0,0)--(1,0);
\draw[fill=black] (1,0) circle (0.1);
\draw (1,0)--(1,0.3);
\draw (1,0)--(1,-0.3);
\draw (1,0.3)--(2,0.3);
\draw (1,-0.3)--(2,-0.3);
\draw [<->] (2.5,0)--(3,0);
\node at (6.5,0) {$\sum_\lambda a_\lambda \ket{v,\lambda}\otimes \bra{v,\lambda}\otimes\bra{v,\lambda},$};
\end{tikzpicture}
\end{align}
\begin{align}
\begin{tikzpicture}[baseline=(current  bounding  box.center)]
\draw (0,0)--(-1,0);
\draw[fill=black] (-1,0) circle (0.1);
\draw (-1,0)--(-1,0.3);
\draw (-1,0)--(-1,-0.3);
\draw (-1,0.3)--(-2,0.3);
\draw (-1,-0.3)--(-2,-0.3);
\draw [<->] (0.5,0)--(1,0);
\node at (4.5,0) {$\sum_\lambda a_\lambda \ket{v,\lambda}\otimes \ket{v,\lambda}\otimes\bra{v,\lambda}.$};
\end{tikzpicture}
\end{align}
These trivalent vertices are natural generalizations of the identity operator ${\bf 1}=\sum_\lambda a_\lambda\ket{v,\lambda}\bra{v,\lambda}$ and the reflection state $\ket{\Omega, \alpha}\rangle=\sum_{\lambda}a_\lambda\ket{v,\lambda}\otimes \ket{v\alpha, \lambda}$, and they are the basic building blocks for such generalizations with more legs. We give the prescription to deal with these trivalent vertices in Appendix \ref{a:triv-vert} when deriving the qq-characters via the Ward identity approach. We leave the concrete computation of qq-characters of $E$-type quivers to future works. 

\section{$DE$-type Quiver Construction with Trivalent Vertex}\label{s:trivalent}

As we have already introduced the trivalent vertex $\sum_\lambda a_\lambda \ket{v,\lambda}\otimes \bra{v,\lambda}\otimes\bra{v,\lambda}$ in the previous section to realize $E$-type quiver gauge theories, it can in fact be used to realize an alternative (brute-force) way for $D$ and $E$-type quivers. An S-dual version of this realization has already been proposed in~\cite{Ohmori-Hayashi} to perform instanton counting for $D$ and $E$-type gauge groups. The brane web diagram of $D_4$ quiver gauge theory, for example, built with this trivalent vertex is given by
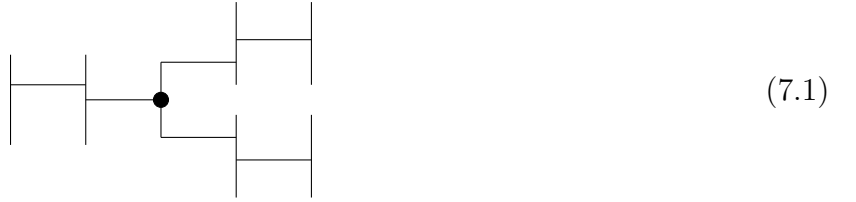
\begin{align}
\begin{tikzpicture}[baseline=(current  bounding  box.center)]
\draw (0,0)--(1,0);
\draw[fill=black] (1,0) circle (0.1);
\draw (1,0)--(1,0.5);
\draw (1,0)--(1,-0.5);
\draw (1,0.5)--(2,0.5);
\draw (1,-0.5)--(2,-0.5);
\draw (2,1.3)--(2,0.2);
\draw (2,0.8)--(3,0.8);
\draw (3,1.3)--(3,0.2);
\draw (2,-1.3)--(2,-0.2);
\draw (2,-0.8)--(3,-0.8);
\draw (3,-1.3)--(3,-0.2);
\draw (0,0.6)--(0,-0.6);
\draw (-1,0.6)--(-1,-0.6);
\draw (-1,0.2)--(0,0.2);
\end{tikzpicture}
\end{align}
and one can read off the information of the quiver from the diagram as
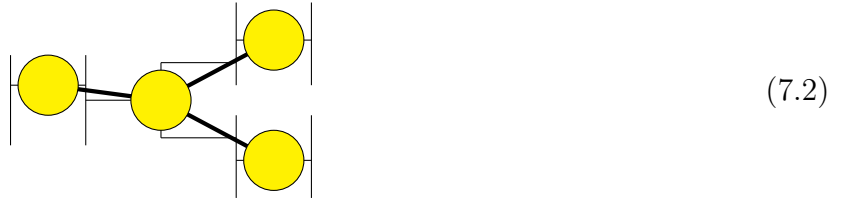
\begin{align}
\begin{tikzpicture}[baseline=(current  bounding  box.center)]
\draw (0,0)--(1,0);
\draw[fill=black] (1,0) circle (0.1);
\draw (1,0)--(1,0.5);
\draw (1,0)--(1,-0.5);
\draw (1,0.5)--(2,0.5);
\draw (1,-0.5)--(2,-0.5);
\draw (2,1.3)--(2,0.2);
\draw (2,0.8)--(3,0.8);
\draw (3,1.3)--(3,0.2);
\draw (2,-1.3)--(2,-0.2);
\draw (2,-0.8)--(3,-0.8);
\draw (3,-1.3)--(3,-0.2);
\draw (0,0.6)--(0,-0.6);
\draw (-1,0.6)--(-1,-0.6);
\draw (-1,0.2)--(0,0.2);
\draw [ultra thick] (1,0)--(2.5,0.8);
\draw [ultra thick] (1,0)--(2.5,-0.8);
\draw [ultra thick] (1,0)--(-0.5,0.2);
\draw[fill=yellow] (1,0) circle (0.4);
\draw[fill=yellow] (2.5,0.8) circle (0.4);
\draw[fill=yellow] (2.5,-0.8) circle (0.4);
\draw[fill=yellow] (-0.5,0.2) circle (0.4);
\end{tikzpicture}
\end{align}

We remark on a disadvantage of this construction: to raise the rank of the gauge group in the central node makes the web diagram extremely complicated. Even in the unrefined limit, where $\Phi^\ast$ is simplified to $\Phi^{-1}$, the web diagram for an SU(2) gauge group in the problem node looks like  
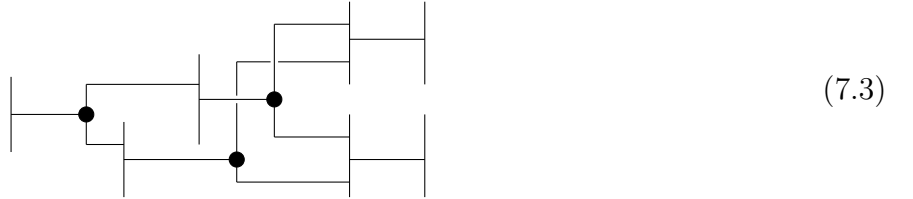
\begin{align}
\begin{tikzpicture}[baseline=(current  bounding  box.center)]
\draw (0,0)--(1,0);
\draw[fill=black] (1,0) circle (0.1);
\draw (1,0)--(1,1);
\draw (1,0)--(1,-0.5);
\draw (1,1)--(2,1);
\draw (1,-0.5)--(2,-0.5);
\draw (2,1.3)--(2,0.2);
\draw (2,0.8)--(3,0.8);
\draw (3,1.3)--(3,0.2);
\draw (2,-1.3)--(2,-0.2);
\draw (2,-0.8)--(3,-0.8);
\draw (3,-1.3)--(3,-0.2);
\draw (0,0.6)--(0,-0.6);
\draw (-1.5,0.2)--(0,0.2);
\draw[fill=black] (0.5,-0.8) circle (0.1);
\draw (0.5,-0.8)--(0.5,-0.05); 
\draw (0.5,0.05)--(0.5,0.5);
\draw (0.5,0.5)--(0.95,0.5);
\draw (1.05,0.5)--(2,0.5);
\draw (0.5,-0.8)--(0.5,-1.1);
\draw (0.5,-1.1)--(2,-1.1);
\draw (0.5,-0.8)--(-1,-0.8);
\draw (-1,-0.3)--(-1,-1.3);
\draw[fill=black] (-1.5,-0.2) circle (0.1);
\draw (-1.5,-0.2)--(-1.5,0.2);
\draw (-1.5,-0.2)--(-1.5,-0.6);
\draw (-1.5,-0.6)--(-1,-0.6);
\draw (-1.5,-0.2)--(-2.5,-0.2);
\draw (-2.5,0.3)--(-2.5,-0.7);
\end{tikzpicture}
\end{align}
and it is more complex and almost not tractable in the refined case. We can bypass this problem by introducing generalized trivalent vertices as how generalized topological vertices are defined in~\cite{BFHMZ}. Essentially it was also the generalized trivalent vertex associated with the SU(2) gauge group that helped in \cite{Ohmori-Hayashi} build the $D$, $E$-type quiver gauge theories with SU(2) gauge groups, which can be S-dualized to ($A_1$ quiver) gauge theories with $D$, $E$-type gauge groups. In the current context, this brute-force construction does not seem to be very interesting to us. 

\section{Affine Quivers}\label{s:affine}

There is almost no difficulty to proceed to affine quivers (including the twisted ones), with all the new vertices we introduced in this article. For example, the untwisted affine $D$-type quivers, $D^{(1)}_n$ has already appeared as an example of superconformal quiver theories before. The untwisted affine $E$-type quivers, $E^{(1)}_{6,7,8}$, are the only tricky ones to construct. Let us describe in this section how to build $E^{(1)}_{6}$, $E^{(1)}_{7}$ and $E^{(1)}_{8}$ quivers in the language of web diagram by only using the topological vertices introduced in this article and their ``square-roots''.

The Dynkin diagram for $E^{(1)}_7$ reads 
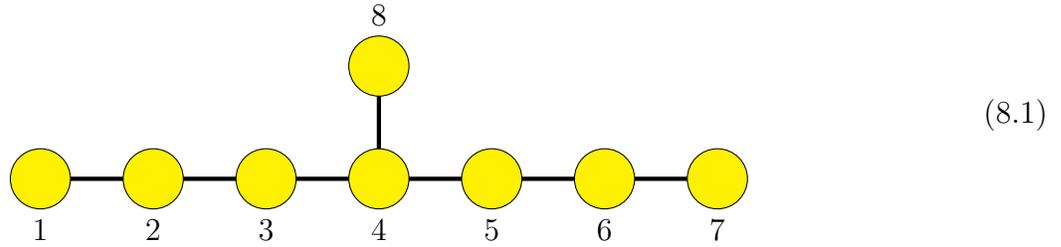
\begin{align}
\begin{tikzpicture}[baseline=(current  bounding  box.center)]
\draw[ultra thick] (0,0)--(1.5,0);
\draw[ultra thick] (0,0)--(0,1.5);
\draw[ultra thick] (0,0)--(-1.5,0);
\draw[ultra thick] (3,0)--(1.5,0);
\draw[ultra thick] (-3,0)--(-1.5,0);
\draw[ultra thick] (-3,0)--(-4.5,0);
\draw[ultra thick] (3,0)--(4.5,0);
\draw[fill=yellow] (0,0) circle (0.4);
\draw[fill=yellow] (1.5,0) circle (0.4);
\draw[fill=yellow] (-1.5,0) circle (0.4);
\draw[fill=yellow] (3,0) circle (0.4);
\draw[fill=yellow] (-3,0) circle (0.4);
\draw[fill=yellow] (0,1.5) circle (0.4);
\draw[fill=yellow] (-4.5,0) circle (0.4);
\draw[fill=yellow] (4.5,0) circle (0.4);
\node at (0,1.9) [above] {8};
\node at (0,-0.4) [below] {4};
\node at (-1.5,-0.4) [below] {3};
\node at (-3,-0.4) [below] {2};
\node at (1.5,-0.4) [below] {5};
\node at (3,-0.4) [below] {6};
\node at (-4.5,-0.4) [below] {1};
\node at (4.5,-0.4) [below] {7};
\end{tikzpicture}
\end{align}
The construction for the $E^{(1)}_7$ quiver, which is actually simpler, is given by 
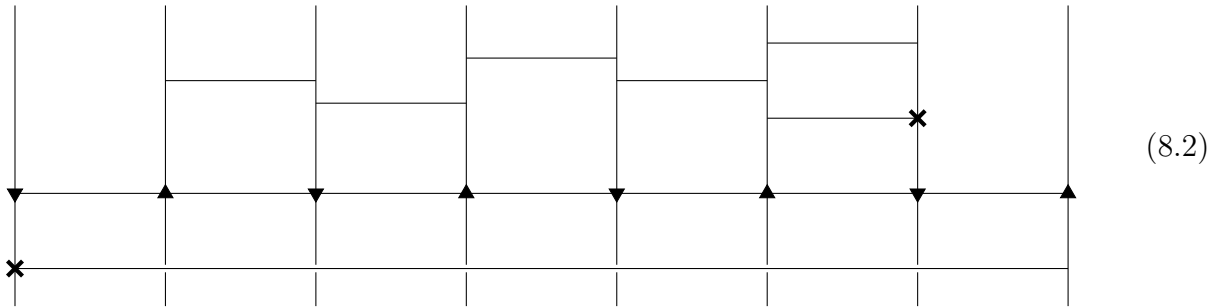
\begin{align}
\begin{tikzpicture}[baseline=(current  bounding  box.center)]
\draw (-1,1)--(-1,-2.45);
\draw (-1,-2.55)--(-1,-3);
\draw (1,1)--(1,-2.45);
\draw (1,-2.55)--(1,-3);
\draw (-3,1)--(-3,-2.45);
\draw (-3,-2.55)--(-3,-3);
\draw (3,1)--(3,-2.45);
\draw (3,-2.55)--(3,-3);
\draw (-5,1)--(-5,-2.45);
\draw (-5,-2.55)--(-5,-3);
\draw (5,1)--(5,-2.45);
\draw (5,-2.55)--(5,-3);
\draw (-7,1)--(-7,-3);
\draw (7,1)--(7,-3);
\draw (-5,0)--(-3,0);
\draw (-3,-0.3)--(-1,-0.3);
\draw (-1,0.3)--(1,0.3);
\draw (1,0)--(3,0);
\draw (3,0.5)--(5,0.5);
\draw (3,-0.5)--(5,-0.5);
\draw [ultra thick] (5-0.1,-0.5+0.1)--(5+0.1,-0.5-0.1);
\draw [ultra thick] (5-0.1,-0.5-0.1)--(5+0.1,-0.5+0.1);
\draw (-7,-1.5)--(7,-1.5);
\draw[fill=black] (-7-0.1,-1.5+0.058)--(-7+0.1,-1.5+0.058)--(-7,-1.5-0.115)--cycle;
\draw[fill=black] (-5-0.1,-1.5-0.058)--(-5+0.1,-1.5-0.058)--(-5,-1.5+0.115)--cycle;
\draw[fill=black] (-3-0.1,-1.5+0.058)--(-3+0.1,-1.5+0.058)--(-3,-1.5-0.115)--cycle;
\draw[fill=black] (-1-0.1,-1.5-0.058)--(-1+0.1,-1.5-0.058)--(-1,-1.5+0.115)--cycle;
\draw[fill=black] (1-0.1,-1.5+0.058)--(1+0.1,-1.5+0.058)--(1,-1.5-0.115)--cycle;
\draw[fill=black] (3-0.1,-1.5-0.058)--(3+0.1,-1.5-0.058)--(3,-1.5+0.115)--cycle;
\draw[fill=black] (5-0.1,-1.5+0.058)--(5+0.1,-1.5+0.058)--(5,-1.5-0.115)--cycle;
\draw[fill=black] (7-0.1,-1.5-0.058)--(7+0.1,-1.5-0.058)--(7,-1.5+0.115)--cycle;
\draw (-7,-2.5)--(7,-2.5);
\draw[ultra thick] (-7-0.1,-2.5+0.1)--(-7+0.1,-2.5-0.1);
\draw[ultra thick] (-7-0.1,-2.5-0.1)--(-7+0.1,-2.5+0.1);
\end{tikzpicture}
\end{align}
where the horizontal line (corresponding to the 7-th node) at the bottom only intersects with the Fock space on the most left and that on the most right. The quiver structure is again contained in the web diagram as 
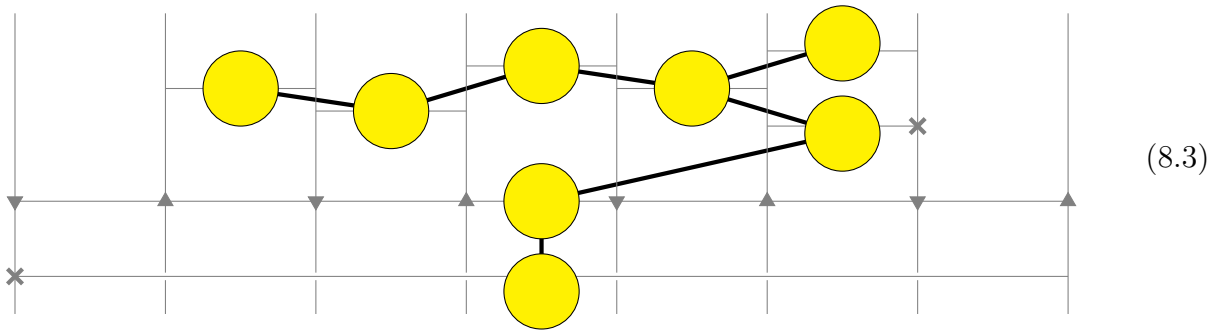
\begin{align}
\begin{tikzpicture}[baseline=(current  bounding  box.center)]
\draw[ultra thick] (-2,-0.3)--(-4,0);
\draw[ultra thick] (-2,-0.3)--(0,0.3);
\draw[ultra thick] (2,0)--(0,0.3);
\draw[ultra thick] (2,0)--(4,0.6);
\draw[ultra thick] (2,0)--(4,-0.6);
\draw[ultra thick] (0,-1.5)--(4,-0.6);
\draw[ultra thick] (0,-1.5)--(0,-2.7);
\draw[gray] (-1,1)--(-1,-2.45);
\draw[gray] (-1,-2.55)--(-1,-3);
\draw[gray] (1,1)--(1,-2.45);
\draw[gray] (1,-2.55)--(1,-3);
\draw[gray] (-3,1)--(-3,-2.45);
\draw[gray] (-3,-2.55)--(-3,-3);
\draw[gray] (3,1)--(3,-2.45);
\draw[gray] (3,-2.55)--(3,-3);
\draw[gray] (-5,1)--(-5,-2.45);
\draw[gray] (-5,-2.55)--(-5,-3);
\draw[gray] (5,1)--(5,-2.45);
\draw[gray] (5,-2.55)--(5,-3);
\draw[gray] (-7,1)--(-7,-3);
\draw[gray] (7,1)--(7,-3);
\draw[gray] (-5,0)--(-3,0);
\draw[gray] (-3,-0.3)--(-1,-0.3);
\draw[gray] (-1,0.3)--(1,0.3);
\draw[gray] (1,0)--(3,0);
\draw[gray] (3,0.5)--(5,0.5);
\draw[gray] (3,-0.5)--(5,-0.5);
\draw [ultra thick, gray] (5-0.1,-0.5+0.1)--(5+0.1,-0.5-0.1);
\draw [ultra thick, gray] (5-0.1,-0.5-0.1)--(5+0.1,-0.5+0.1);
\draw[gray] (-7,-1.5)--(7,-1.5);
\draw[fill=gray, gray] (-7-0.1,-1.5+0.058)--(-7+0.1,-1.5+0.058)--(-7,-1.5-0.115)--cycle;
\draw[fill=gray, gray] (-5-0.1,-1.5-0.058)--(-5+0.1,-1.5-0.058)--(-5,-1.5+0.115)--cycle;
\draw[fill=gray,gray] (-3-0.1,-1.5+0.058)--(-3+0.1,-1.5+0.058)--(-3,-1.5-0.115)--cycle;
\draw[fill=gray,gray] (-1-0.1,-1.5-0.058)--(-1+0.1,-1.5-0.058)--(-1,-1.5+0.115)--cycle;
\draw[fill=gray,gray] (1-0.1,-1.5+0.058)--(1+0.1,-1.5+0.058)--(1,-1.5-0.115)--cycle;
\draw[fill=gray,gray] (3-0.1,-1.5-0.058)--(3+0.1,-1.5-0.058)--(3,-1.5+0.115)--cycle;
\draw[fill=gray,gray] (5-0.1,-1.5+0.058)--(5+0.1,-1.5+0.058)--(5,-1.5-0.115)--cycle;
\draw[fill=gray,gray] (7-0.1,-1.5-0.058)--(7+0.1,-1.5-0.058)--(7,-1.5+0.115)--cycle;
\draw[fill=yellow] (-4,0) circle (0.5);
\draw[fill=yellow] (-2,-0.3) circle (0.5);
\draw[fill=yellow] (0,0.3) circle (0.5);
\draw[fill=yellow] (2,0) circle (0.5);
\draw[fill=yellow] (4,0.6) circle (0.5);
\draw[fill=yellow] (4,-0.6) circle (0.5);
\draw[fill=yellow] (0,-1.5) circle (0.5);
\draw[gray] (-7,-2.5)--(7,-2.5);
\draw[ultra thick,gray] (-7-0.1,-2.5+0.1)--(-7+0.1,-2.5-0.1);
\draw[ultra thick,gray] (-7-0.1,-2.5-0.1)--(-7+0.1,-2.5+0.1);
\draw[fill=yellow] (0,-2.7) circle (0.5);
\end{tikzpicture}
\end{align}
Note that the effective Coulomb modulus of the 7-th node is shifted by $\gamma$ in this construction. The realization of the 7-th node is designed so that piling up the same structure raises the rank of the gauge group there. This will be true for all constructions presented in this section. 

The Dynkin diagram for $E^{(1)}_8$ is given by 
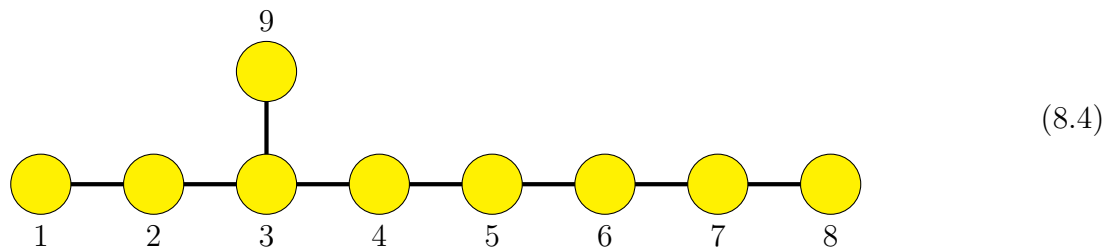
\begin{align}
\begin{tikzpicture}[baseline=(current  bounding  box.center)]
\draw[ultra thick] (0,0)--(1.5,0);
\draw[ultra thick] (0,0)--(0,1.5);
\draw[ultra thick] (0,0)--(-1.5,0);
\draw[ultra thick] (3,0)--(1.5,0);
\draw[ultra thick] (-3,0)--(-1.5,0);
\draw[ultra thick] (3,0)--(4.5,0);
\draw[ultra thick] (4.5,0)--(6,0);
\draw[ultra thick] (7.5,0)--(6,0);
\draw[fill=yellow] (0,0) circle (0.4);
\draw[fill=yellow] (1.5,0) circle (0.4);
\draw[fill=yellow] (-1.5,0) circle (0.4);
\draw[fill=yellow] (3,0) circle (0.4);
\draw[fill=yellow] (-3,0) circle (0.4);
\draw[fill=yellow] (0,1.5) circle (0.4);
\draw[fill=yellow] (4.5,0) circle (0.4);
\draw[fill=yellow] (6,0) circle (0.4);
\draw[fill=yellow] (7.5,0) circle (0.4);
\node at (0,1.9) [above] {9};
\node at (0,-0.4) [below] {3};
\node at (-1.5,-0.4) [below] {2};
\node at (-3,-0.4) [below] {1};
\node at (1.5,-0.4) [below] {4};
\node at (3,-0.4) [below] {5};
\node at (4.5,-0.4) [below] {6};
\node at (6,-0.4) [below] {7};
\node at (7.5,-0.4) [below] {8};
\end{tikzpicture}
\end{align}
and can be constructed as 
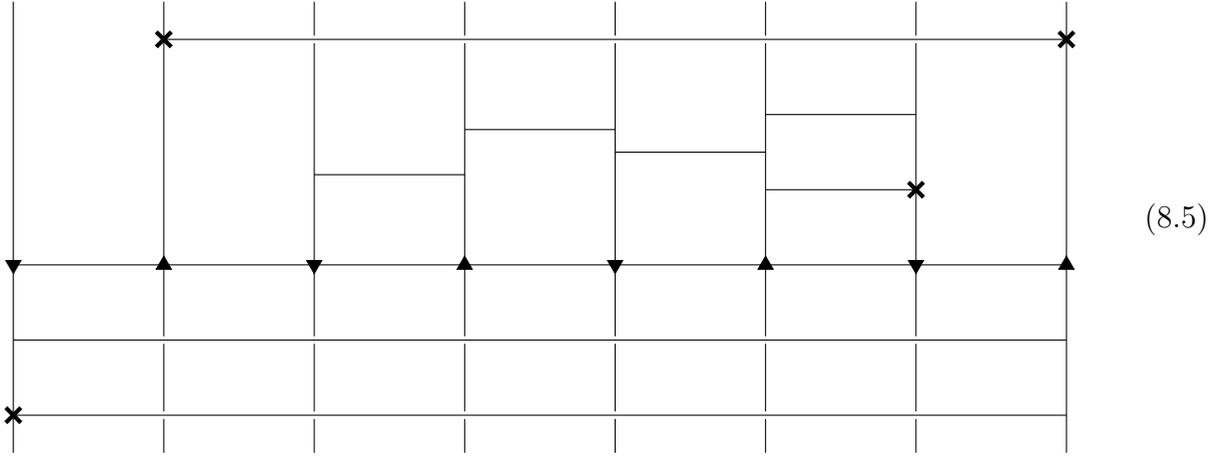
\begin{align}
\begin{tikzpicture}[baseline=(current  bounding  box.center)]
\draw (-1,1.45)--(-1,-2.45);
\draw (-1,-2.55)--(-1,-3.45);
\draw (-1,1.55)--(-1,2);
\draw (-1,-3.55)--(-1,-4);
\draw (1,1.45)--(1,-2.45);
\draw (1,-2.55)--(1,-3.45);
\draw (1,1.55)--(1,2);
\draw (1,-3.55)--(1,-4);
\draw (-3,1.45)--(-3,-2.45);
\draw (-3,-2.55)--(-3,-3.45);
\draw (-3,1.55)--(-3,2);
\draw (-3,-3.55)--(-3,-4);
\draw (3,1.45)--(3,-2.45);
\draw (3,-2.55)--(3,-3.45);
\draw (3,1.55)--(3,2);
\draw (3,-3.55)--(3,-4);
\draw (-5,2)--(-5,-2.45);
\draw (-5,-2.55)--(-5,-3.45);
\draw (-5,-3.55)--(-5,-4);
\draw (5,1.45)--(5,-2.45);
\draw (5,-2.55)--(5,-3.45);
\draw (5,1.55)--(5,2);
\draw (5,-3.55)--(5,-4);
\draw (-7,2)--(-7,-4);
\draw (7,2)--(7,-4);
\draw (-3,-0.3)--(-1,-0.3);
\draw (-1,0.3)--(1,0.3);
\draw (1,0)--(3,0);
\draw (3,0.5)--(5,0.5);
\draw (3,-0.5)--(5,-0.5);
\draw [ultra thick] (5-0.1,-0.5+0.1)--(5+0.1,-0.5-0.1);
\draw [ultra thick] (5-0.1,-0.5-0.1)--(5+0.1,-0.5+0.1);
\draw (-7,-1.5)--(7,-1.5);
\draw[fill=black] (-7-0.1,-1.5+0.058)--(-7+0.1,-1.5+0.058)--(-7,-1.5-0.115)--cycle;
\draw[fill=black] (-5-0.1,-1.5-0.058)--(-5+0.1,-1.5-0.058)--(-5,-1.5+0.115)--cycle;
\draw[fill=black] (-3-0.1,-1.5+0.058)--(-3+0.1,-1.5+0.058)--(-3,-1.5-0.115)--cycle;
\draw[fill=black] (-1-0.1,-1.5-0.058)--(-1+0.1,-1.5-0.058)--(-1,-1.5+0.115)--cycle;
\draw[fill=black] (1-0.1,-1.5+0.058)--(1+0.1,-1.5+0.058)--(1,-1.5-0.115)--cycle;
\draw[fill=black] (3-0.1,-1.5-0.058)--(3+0.1,-1.5-0.058)--(3,-1.5+0.115)--cycle;
\draw[fill=black] (5-0.1,-1.5+0.058)--(5+0.1,-1.5+0.058)--(5,-1.5-0.115)--cycle;
\draw[fill=black] (7-0.1,-1.5-0.058)--(7+0.1,-1.5-0.058)--(7,-1.5+0.115)--cycle;
\draw (-7,-2.5)--(7,-2.5);
\draw (-7,-3.5)--(7,-3.5);
\draw[ultra thick] (-7-0.1,-3.5+0.1)--(-7+0.1,-3.5-0.1);
\draw[ultra thick] (-7-0.1,-3.5-0.1)--(-7+0.1,-3.5+0.1);
\draw(-5,1.5)--(7,1.5);
\draw[ultra thick] (-5-0.1,1.5+0.1)--(-5+0.1,1.5-0.1);
\draw[ultra thick] (-5-0.1,1.5-0.1)--(-5+0.1,1.5+0.1);
\draw[ultra thick] (7-0.1,1.5+0.1)--(7+0.1,1.5-0.1);
\draw[ultra thick] (7-0.1,1.5-0.1)--(7+0.1,1.5+0.1);
\end{tikzpicture}
\end{align}
We note that the top line only intersects the second left Fock space and the most right one, and two bottom lines, as in the case of $E^{(1)}_7$, intersect only with the Fock spaces at the most left and right. The quiver structure, which is more non-trivial compared to previous examples in this article, in this web diagram, is given as 
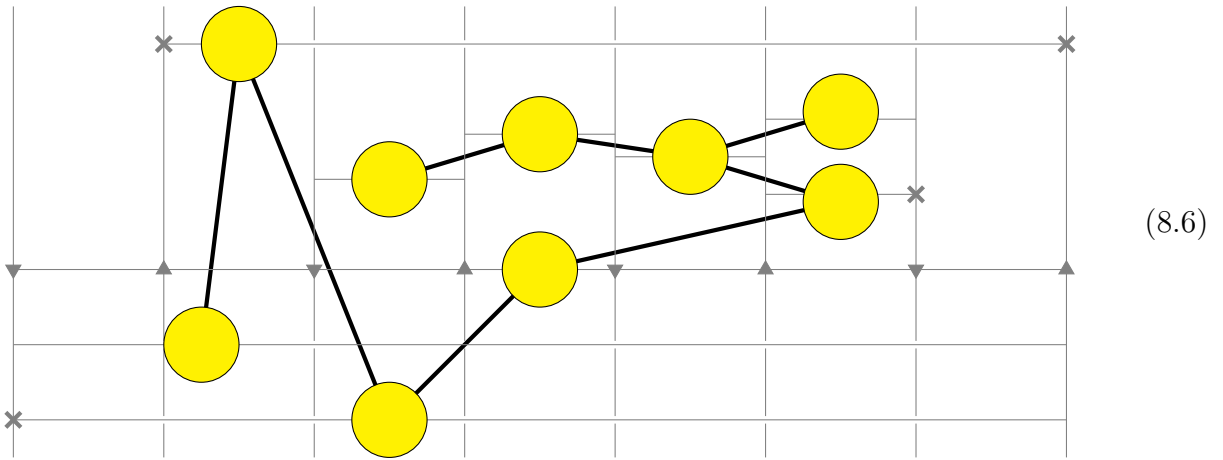
\begin{align}
\begin{tikzpicture}[baseline=(current  bounding  box.center)]
\draw[ultra thick] (-2,-0.3)--(0,0.3);
\draw[ultra thick] (2,0)--(0,0.3);
\draw[ultra thick] (2,0)--(4,0.6);
\draw[ultra thick] (2,0)--(4,-0.6);
\draw[ultra thick] (0,-1.5)--(4,-0.6);
\draw[ultra thick] (0,-1.5)--(-2,-3.5);
\draw[ultra thick] (-4,1.5)--(-2,-3.5);
\draw[ultra thick] (-4,1.5)--(-4.5,-2.5);
\draw[gray] (-1,1.45)--(-1,-2.45);
\draw[gray] (-1,-2.55)--(-1,-3.45);
\draw[gray] (-1,1.55)--(-1,2);
\draw[gray] (-1,-3.55)--(-1,-4);
\draw[gray] (1,1.45)--(1,-2.45);
\draw[gray] (1,-2.55)--(1,-3.45);
\draw[gray] (1,1.55)--(1,2);
\draw[gray] (1,-3.55)--(1,-4);
\draw[gray] (-3,1.45)--(-3,-2.45);
\draw[gray] (-3,-2.55)--(-3,-3.45);
\draw[gray] (-3,1.55)--(-3,2);
\draw[gray] (-3,-3.55)--(-3,-4);
\draw[gray] (3,1.45)--(3,-2.45);
\draw[gray] (3,-2.55)--(3,-3.45);
\draw[gray] (3,1.55)--(3,2);
\draw[gray] (3,-3.55)--(3,-4);
\draw[gray] (-5,2)--(-5,-2.45);
\draw[gray] (-5,-2.55)--(-5,-3.45);
\draw[gray] (-5,-3.55)--(-5,-4);
\draw[gray] (5,1.45)--(5,-2.45);
\draw[gray] (5,-2.55)--(5,-3.45);
\draw[gray] (5,1.55)--(5,2);
\draw[gray] (5,-3.55)--(5,-4);
\draw[gray] (-7,2)--(-7,-4);
\draw[gray] (7,2)--(7,-4);
\draw[gray] (-3,-0.3)--(-1,-0.3);
\draw[gray] (-1,0.3)--(1,0.3);
\draw[gray] (1,0)--(3,0);
\draw[gray] (3,0.5)--(5,0.5);
\draw[gray] (3,-0.5)--(5,-0.5);
\draw [ultra thick,gray] (5-0.1,-0.5+0.1)--(5+0.1,-0.5-0.1);
\draw [ultra thick,gray] (5-0.1,-0.5-0.1)--(5+0.1,-0.5+0.1);
\draw[gray] (-7,-1.5)--(7,-1.5);
\draw[fill=gray,gray] (-7-0.1,-1.5+0.058)--(-7+0.1,-1.5+0.058)--(-7,-1.5-0.115)--cycle;
\draw[fill=gray,gray] (-5-0.1,-1.5-0.058)--(-5+0.1,-1.5-0.058)--(-5,-1.5+0.115)--cycle;
\draw[fill=gray,gray] (-3-0.1,-1.5+0.058)--(-3+0.1,-1.5+0.058)--(-3,-1.5-0.115)--cycle;
\draw[fill=gray,gray] (-1-0.1,-1.5-0.058)--(-1+0.1,-1.5-0.058)--(-1,-1.5+0.115)--cycle;
\draw[fill=gray,gray] (1-0.1,-1.5+0.058)--(1+0.1,-1.5+0.058)--(1,-1.5-0.115)--cycle;
\draw[fill=gray,gray] (3-0.1,-1.5-0.058)--(3+0.1,-1.5-0.058)--(3,-1.5+0.115)--cycle;
\draw[fill=gray,gray] (5-0.1,-1.5+0.058)--(5+0.1,-1.5+0.058)--(5,-1.5-0.115)--cycle;
\draw[fill=gray,gray] (7-0.1,-1.5-0.058)--(7+0.1,-1.5-0.058)--(7,-1.5+0.115)--cycle;
\draw[gray] (-7,-2.5)--(7,-2.5);
\draw[gray] (-7,-3.5)--(7,-3.5);
\draw[ultra thick,gray] (-7-0.1,-3.5+0.1)--(-7+0.1,-3.5-0.1);
\draw[ultra thick,gray] (-7-0.1,-3.5-0.1)--(-7+0.1,-3.5+0.1);
\draw[gray] (-5,1.5)--(7,1.5);
\draw[ultra thick,gray] (-5-0.1,1.5+0.1)--(-5+0.1,1.5-0.1);
\draw[ultra thick,gray] (-5-0.1,1.5-0.1)--(-5+0.1,1.5+0.1);
\draw[ultra thick,gray] (7-0.1,1.5+0.1)--(7+0.1,1.5-0.1);
\draw[ultra thick,gray] (7-0.1,1.5-0.1)--(7+0.1,1.5+0.1);
\draw[fill=yellow] (-2,-0.3) circle (0.5);
\draw[fill=yellow] (0,0.3) circle (0.5);
\draw[fill=yellow] (2,0) circle (0.5);
\draw[fill=yellow] (4,0.6) circle (0.5);
\draw[fill=yellow] (4,-0.6) circle (0.5);
\draw[fill=yellow] (0,-1.5) circle (0.5);
\draw[fill=yellow] (-2,-3.5) circle (0.5);
\draw[fill=yellow] (-4,1.5) circle (0.5);
\draw[fill=yellow] (-4.5,-2.5) circle (0.5);
\end{tikzpicture}
\end{align}
The effective Coulomb moduli of both the 6-th node and the 7-th node are shifted by $\gamma$. We remark that there is no space left in this approach to add more gauge nodes to the quiver diagram of $E^{(1)}_8$. 

The Dynkin diagram for $E^{(1)}_6$ reads 
\begin{align}
\begin{tikzpicture}[baseline=(current  bounding  box.center)]
\draw[ultra thick] (0,0)--(1.5,0);
\draw[ultra thick] (0,0)--(0,1.5);
\draw[ultra thick] (0,0)--(-1.5,0);
\draw[ultra thick] (3,0)--(1.5,0);
\draw[ultra thick] (-3,0)--(-1.5,0);
\draw[ultra thick] (0,1.5)--(0,3);
\draw[fill=yellow] (0,0) circle (0.4);
\draw[fill=yellow] (1.5,0) circle (0.4);
\draw[fill=yellow] (-1.5,0) circle (0.4);
\draw[fill=yellow] (3,0) circle (0.4);
\draw[fill=yellow] (-3,0) circle (0.4);
\draw[fill=yellow] (0,1.5) circle (0.4);
\draw[fill=yellow] (0,3) circle (0.4);
\node at (0.4,1.5) [right] {6};
\node at (0.4,3) [right] {7};
\node at (0,-0.4) [below] {3};
\node at (-1.5,-0.4) [below] {2};
\node at (-3,-0.4) [below] {1};
\node at (1.5,-0.4) [below] {4};
\node at (3,-0.4) [below] {5};
\end{tikzpicture}
\end{align}
We can realize it with the following web diagram 
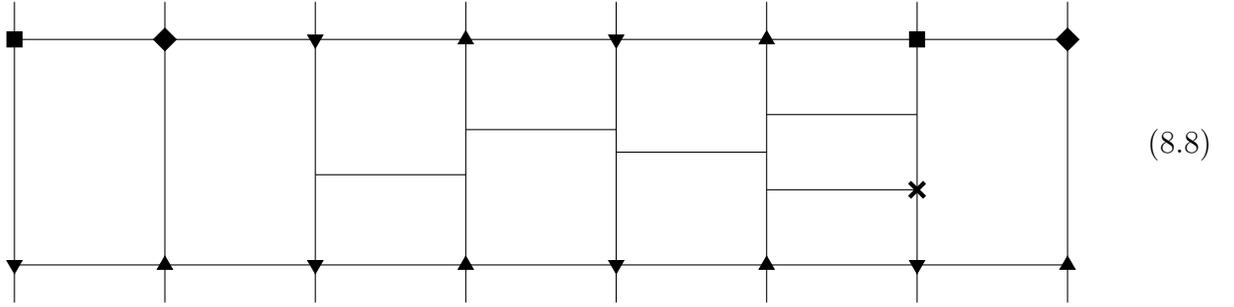
\begin{align}
\begin{tikzpicture}[baseline=(current  bounding  box.center)]
\draw (-1,2)--(-1,-2);
\draw (1,2)--(1,-2);
\draw (-3,2)--(-3,-2);
\draw (3,2)--(3,-2);
\draw (-5,2)--(-5,-2);
\draw (5,2)--(5,-2);
\draw (-7,2)--(-7,-2);
\draw (7,2)--(7,-2);
\draw (-3,-0.3)--(-1,-0.3);
\draw (-1,0.3)--(1,0.3);
\draw (1,0)--(3,0);
\draw (3,0.5)--(5,0.5);
\draw (3,-0.5)--(5,-0.5);
\draw [ultra thick] (5-0.1,-0.5+0.1)--(5+0.1,-0.5-0.1);
\draw [ultra thick] (5-0.1,-0.5-0.1)--(5+0.1,-0.5+0.1);
\draw (-7,-1.5)--(7,-1.5);
\draw[fill=black] (-7-0.1,-1.5+0.058)--(-7+0.1,-1.5+0.058)--(-7,-1.5-0.115)--cycle;
\draw[fill=black] (-5-0.1,-1.5-0.058)--(-5+0.1,-1.5-0.058)--(-5,-1.5+0.115)--cycle;
\draw[fill=black] (-3-0.1,-1.5+0.058)--(-3+0.1,-1.5+0.058)--(-3,-1.5-0.115)--cycle;
\draw[fill=black] (-1-0.1,-1.5-0.058)--(-1+0.1,-1.5-0.058)--(-1,-1.5+0.115)--cycle;
\draw[fill=black] (1-0.1,-1.5+0.058)--(1+0.1,-1.5+0.058)--(1,-1.5-0.115)--cycle;
\draw[fill=black] (3-0.1,-1.5-0.058)--(3+0.1,-1.5-0.058)--(3,-1.5+0.115)--cycle;
\draw[fill=black] (5-0.1,-1.5+0.058)--(5+0.1,-1.5+0.058)--(5,-1.5-0.115)--cycle;
\draw[fill=black] (7-0.1,-1.5-0.058)--(7+0.1,-1.5-0.058)--(7,-1.5+0.115)--cycle;
\draw (-7,1.5)--(7,1.5);
\draw[fill=black] (-7-0.1,1.5+0.1) rectangle (-7+0.1,1.5-0.1);
\draw[fill=black] (5-0.1,1.5+0.1) rectangle (5+0.1,1.5-0.1);
\draw[fill=black] (-5,1.5+0.15)--(-5+0.15,1.5)--(-5,1.5-0.15)--(-5-0.15,1.5)--cycle;
\draw[fill=black] (7,1.5+0.15)--(7+0.15,1.5)--(7,1.5-0.15)--(7-0.15,1.5)--cycle;
\draw[fill=black] (3+0.1,1.5-0.058)--(3-0.1,1.5-0.058)--(3,1.5+0.115)--cycle;
\draw[fill=black] (-1+0.1,1.5-0.058)--(-1-0.1,1.5-0.058)--(-1,1.5+0.115)--cycle;
\draw[fill=black] (-3+0.1,1.5+0.058)--(-3-0.1,1.5+0.058)--(-3,1.5-0.115)--cycle;
\draw[fill=black] (1+0.1,1.5+0.058)--(1-0.1,1.5+0.058)--(1,1.5-0.115)--cycle;
\end{tikzpicture}
\end{align}
where we further introduced the ``square-root'' of $\Phi$ and $\bar{\Phi}^\ast$ as 
\begin{align}
\begin{tikzpicture}[baseline=(current  bounding  box.center)]
\draw (0,1)--(0,-1);
\draw (0,0)--(1,0);
\draw[fill=black] (-0.1,0.1) rectangle (0.1,-0.1);
\draw [<->] (1.5,0)--(2,0);
\node at (3.5,0) {$\Phi^{[1/2]}[v]$,};
\end{tikzpicture}
\end{align}
\ba
\Phi^{[1/2]}[v]\ket{v,\lambda}=\Phi^{\frac{1}{2}}_\emptyset[v]\prod_{x\in\lambda}\eta^{\frac{1}{2}}(\chi_x),
\ea
\begin{align}
\begin{tikzpicture}[baseline=(current  bounding  box.center)]
\draw (1,1)--(1,-1);
\draw (0,0)--(1,0);
\draw[fill=black] (1,0.15)--(1.15,0)--(1,-0.15)--(0.85,0)--cycle;
\draw [<->] (1.5,0)--(2,0);
\node at (3.5,0) {$\bar{\Phi}^{\ast[\frac{1}{2}]}[v]$,};
\end{tikzpicture}
\end{align}
\ba
\bra{v,\lambda}\bar{\Phi}^{\ast[\frac{1}{2}]}[v]=\bar{\Phi}^{\ast\frac{1}{2}}_\emptyset[v]\prod_{x\in\lambda}\bar{\xi}^{\frac{1}{2}}(\chi_x).
\ea
The quiver structure can be read from the web diagram as 
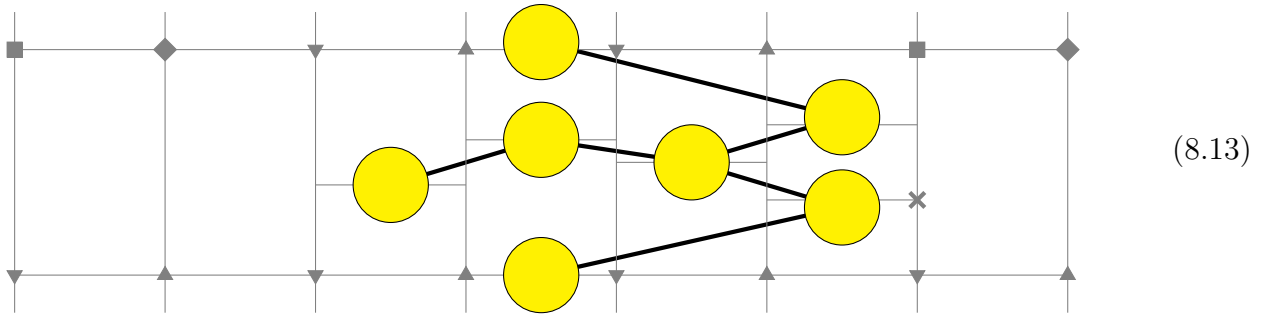
\begin{align}
\begin{tikzpicture}[baseline=(current  bounding  box.center)]
\draw[ultra thick] (-2,-0.3)--(0,0.3);
\draw[ultra thick] (2,0)--(0,0.3);
\draw[ultra thick] (2,0)--(4,0.6);
\draw[ultra thick] (2,0)--(4,-0.6);
\draw[ultra thick] (0,-1.5)--(4,-0.6);
\draw[ultra thick] (0,1.6)--(4,0.6);
\draw[gray] (-1,2)--(-1,-2);
\draw[gray] (1,2)--(1,-2);
\draw[gray] (-3,2)--(-3,-2);
\draw[gray] (3,2)--(3,-2);
\draw[gray] (-5,2)--(-5,-2);
\draw[gray] (5,2)--(5,-2);
\draw[gray] (-7,2)--(-7,-2);
\draw[gray] (7,2)--(7,-2);
\draw[gray] (-3,-0.3)--(-1,-0.3);
\draw[gray] (-1,0.3)--(1,0.3);
\draw[gray] (1,0)--(3,0);
\draw[gray] (3,0.5)--(5,0.5);
\draw[gray] (3,-0.5)--(5,-0.5);
\draw [ultra thick,gray] (5-0.1,-0.5+0.1)--(5+0.1,-0.5-0.1);
\draw [ultra thick,gray] (5-0.1,-0.5-0.1)--(5+0.1,-0.5+0.1);
\draw[gray] (-7,-1.5)--(7,-1.5);
\draw[fill=gray,gray] (-7-0.1,-1.5+0.058)--(-7+0.1,-1.5+0.058)--(-7,-1.5-0.115)--cycle;
\draw[fill=gray,gray] (-5-0.1,-1.5-0.058)--(-5+0.1,-1.5-0.058)--(-5,-1.5+0.115)--cycle;
\draw[fill=gray,gray] (-3-0.1,-1.5+0.058)--(-3+0.1,-1.5+0.058)--(-3,-1.5-0.115)--cycle;
\draw[fill=gray,gray] (-1-0.1,-1.5-0.058)--(-1+0.1,-1.5-0.058)--(-1,-1.5+0.115)--cycle;
\draw[fill=gray,gray] (1-0.1,-1.5+0.058)--(1+0.1,-1.5+0.058)--(1,-1.5-0.115)--cycle;
\draw[fill=gray,gray] (3-0.1,-1.5-0.058)--(3+0.1,-1.5-0.058)--(3,-1.5+0.115)--cycle;
\draw[fill=gray,gray] (5-0.1,-1.5+0.058)--(5+0.1,-1.5+0.058)--(5,-1.5-0.115)--cycle;
\draw[fill=gray,gray] (7-0.1,-1.5-0.058)--(7+0.1,-1.5-0.058)--(7,-1.5+0.115)--cycle;
\draw[gray] (-7,1.5)--(7,1.5);
\draw[fill=gray,gray] (-7-0.1,1.5+0.1) rectangle (-7+0.1,1.5-0.1);
\draw[fill=gray,gray] (5-0.1,1.5+0.1) rectangle (5+0.1,1.5-0.1);
\draw[fill=gray,gray] (-5,1.5+0.15)--(-5+0.15,1.5)--(-5,1.5-0.15)--(-5-0.15,1.5)--cycle;
\draw[fill=gray,gray] (7,1.5+0.15)--(7+0.15,1.5)--(7,1.5-0.15)--(7-0.15,1.5)--cycle;
\draw[fill=gray,gray] (3+0.1,1.5-0.058)--(3-0.1,1.5-0.058)--(3,1.5+0.115)--cycle;
\draw[fill=gray,gray] (-1+0.1,1.5-0.058)--(-1-0.1,1.5-0.058)--(-1,1.5+0.115)--cycle;
\draw[fill=gray,gray] (-3+0.1,1.5+0.058)--(-3-0.1,1.5+0.058)--(-3,1.5-0.115)--cycle;
\draw[fill=gray,gray] (1+0.1,1.5+0.058)--(1-0.1,1.5+0.058)--(1,1.5-0.115)--cycle;
\draw[fill=yellow] (-2,-0.3) circle (0.5);
\draw[fill=yellow] (0,0.3) circle (0.5);
\draw[fill=yellow] (2,0) circle (0.5);
\draw[fill=yellow] (4,0.6) circle (0.5);
\draw[fill=yellow] (4,-0.6) circle (0.5);
\draw[fill=yellow] (0,-1.5) circle (0.5);
\draw[fill=yellow] (0,1.6) circle (0.5);
\end{tikzpicture}
\end{align}

\paragraph{Superconformal quiver theories of $E$-type}

It is not difficult to find solutions to superconformal $E$-type quivers via the condition 
\ba
\sum_jc_{ij}N_j=N_{f,i}.
\ea 
For example we have superconformal theories associated with the $E_6$ quiver given by 
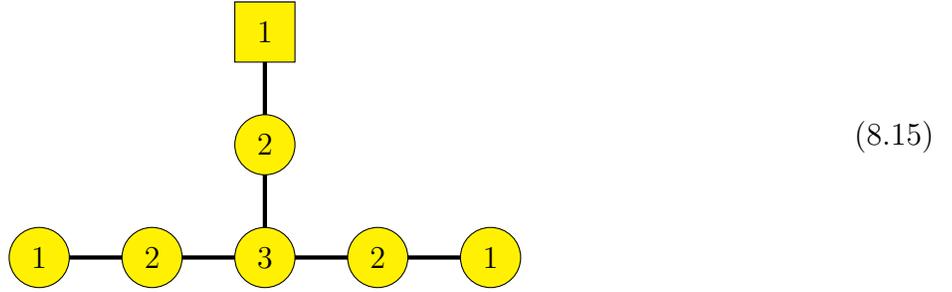
\begin{align}
\begin{tikzpicture}[baseline=(current  bounding  box.center)]
\draw[ultra thick] (0,0)--(1.5,0);
\draw[ultra thick] (0,0)--(0,1.5);
\draw[ultra thick] (0,0)--(-1.5,0);
\draw[ultra thick] (3,0)--(1.5,0);
\draw[ultra thick] (-3,0)--(-1.5,0);
\draw[ultra thick] (0,1.5)--(0,3);
\draw[fill=yellow] (0,0) circle (0.4);
\draw[fill=yellow] (1.5,0) circle (0.4);
\draw[fill=yellow] (-1.5,0) circle (0.4);
\draw[fill=yellow] (3,0) circle (0.4);
\draw[fill=yellow] (-3,0) circle (0.4);
\draw[fill=yellow] (0,1.5) circle (0.4);
\draw[fill=yellow] (-0.4,3.4) rectangle (0.4,2.6);
\node at (0,1.5) {2};
\node at (0,3) {1};
\node at (0,0) {3};
\node at (-1.5,0) {2};
\node at (-3,0) {1};
\node at (1.5,0) {2};
\node at (3,0) {1};
\end{tikzpicture}
\end{align}
and 
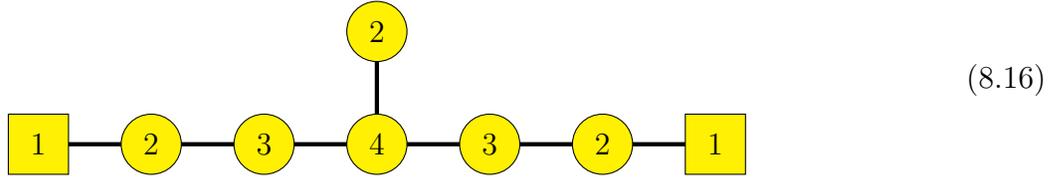
\begin{align}
\begin{tikzpicture}[baseline=(current  bounding  box.center)]
\draw[ultra thick] (0,0)--(1.5,0);
\draw[ultra thick] (0,0)--(0,1.5);
\draw[ultra thick] (0,0)--(-1.5,0);
\draw[ultra thick] (3,0)--(1.5,0);
\draw[ultra thick] (-3,0)--(-1.5,0);
\draw[ultra thick] (4.5,0)--(3,0);
\draw[ultra thick] (-4.5,0)--(-3,0);
\draw[fill=yellow] (0,0) circle (0.4);
\draw[fill=yellow] (1.5,0) circle (0.4);
\draw[fill=yellow] (-1.5,0) circle (0.4);
\draw[fill=yellow] (3,0) circle (0.4);
\draw[fill=yellow] (-3,0) circle (0.4);
\draw[fill=yellow] (0,1.5) circle (0.4);
\draw[fill=yellow] (4.1,0.4) rectangle (4.9,-0.4);
\draw[fill=yellow] (-4.1,0.4) rectangle (-4.9,-0.4);
\node at (0,1.5) {2};
\node at (0,0) {4};
\node at (-1.5,0) {3};
\node at (-3,0) {2};
\node at (1.5,0) {3};
\node at (3,0) {2};
\node at (4.5,0) {1};
\node at (-4.5,0) {1};
\end{tikzpicture}
\end{align}
These examples in principle can respectively be realized in terms of a brane web from the web construction of $E^{(1)}_6$ and $E^{(1)}_7$ quivers, by moving one NS5 brane of the extra gauge node(s) to the infinity, i.e. ungauging extra nodes. However, in the construction of affine $E$-type described above, we cannot move, for instance, any NS5 brane attached to the 7-th node without ungauging the 5-th node in $E^{(1)}_6$ (refer to the Dynkin diagram for the label of nodes). We remark that the fact we cannot add fundamental matter multiplets into our construction might be related to that we start from the expression of simple roots in $E_8$. It may help if we start from the following quiver, 
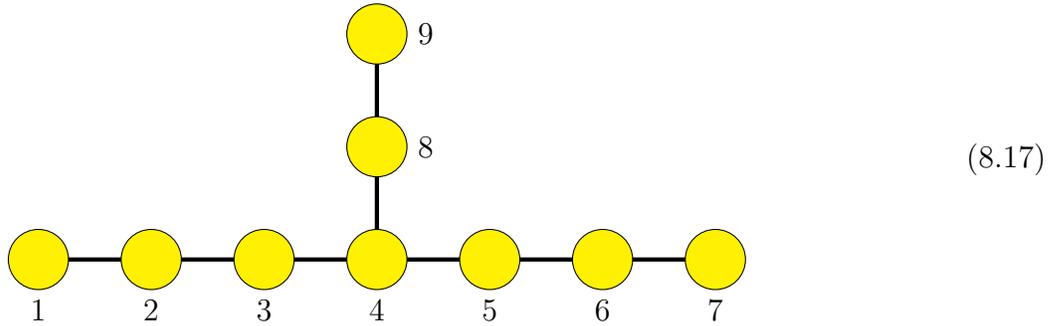
\begin{align}
\begin{tikzpicture}[baseline=(current  bounding  box.center)]
\draw[ultra thick] (0,0)--(1.5,0);
\draw[ultra thick] (0,0)--(0,1.5);
\draw[ultra thick] (0,0)--(-1.5,0);
\draw[ultra thick] (0,3)--(0,1.5);
\draw[ultra thick] (3,0)--(1.5,0);
\draw[ultra thick] (-3,0)--(-1.5,0);
\draw[ultra thick] (-3,0)--(-4.5,0);
\draw[ultra thick] (3,0)--(4.5,0);
\draw[fill=yellow] (0,0) circle (0.4);
\draw[fill=yellow] (1.5,0) circle (0.4);
\draw[fill=yellow] (-1.5,0) circle (0.4);
\draw[fill=yellow] (3,0) circle (0.4);
\draw[fill=yellow] (-3,0) circle (0.4);
\draw[fill=yellow] (0,1.5) circle (0.4);
\draw[fill=yellow] (0,3) circle (0.4);
\draw[fill=yellow] (-4.5,0) circle (0.4);
\draw[fill=yellow] (4.5,0) circle (0.4);
\node at (0.4,1.5) [right] {8};
\node at (0.4,3) [right] {9};
\node at (0,-0.4) [below] {4};
\node at (-1.5,-0.4) [below] {3};
\node at (-3,-0.4) [below] {2};
\node at (1.5,-0.4) [below] {5};
\node at (3,-0.4) [below] {6};
\node at (-4.5,-0.4) [below] {1};
\node at (4.5,-0.4) [below] {7};
\end{tikzpicture}
\end{align} 
but we leave this kind of exploration to a future work.

\section{AFS Property of New Vertices}\label{s:AFS}

In this section, we explore potential (more mathematical) definitions for those new topological vertices we introduced in this article to realize quivers beyond $A$-type. Let us first recall how the usual refined topological vertex can be formulated in the algebraic language, i.e. as an intertwiner mapping two representations into one in the DIM algebra. We will refer to this formulation as Awata-Feigin-Shiraishi (AFS) property of the topological vertex, and we intend to mimic it for all the other vertices. The AFS properties for $\Phi$ and $\Phi^\ast$ read 
\ba
(\rho^{(1,n)}_{u}\otimes \rho^{(0,1)}_{\vec{v}})\Delta(g(z))\Phi^{\ast(n)}_{\vec{\lambda}}[u,\vec{v}]=\Phi^{\ast(n)}_{\vec{\lambda}}[u,\vec{v}]\rho^{(1,n+1)}_{u'}(g(z)),\\
\Phi^{(n)}_{\vec{\lambda}}[u,\vec{v}](\rho^{(0,1)}_{\vec{v}}\otimes \rho^{(1,n)}_u)\Delta(g(z))=\rho^{(1,n+1)}_{u'}(g(z))\Phi^{(n)}_{\vec{\lambda}}[u,\vec{v}],
\ea
where $g=x^\pm,\ \psi^\pm$. The coproduct structure is given by 
\begin{align}
\begin{split}\label{AFS_coproduct}
&\Delta(x^+(z))=x^+(z)\otimes 1+\psi^-(\hat{\gamma}_{(1)}^{1/2}z)\otimes x^+(\hat{\gamma}_{(1)}z),\\
&\Delta(x^-(z))=x^-(\hat{\gamma}_{(2)} z)\otimes \psi^+(\hat{\gamma}_{(2)}^{1/2}z)+1\otimes x^-(z),\\
&\Delta(\psi^+(z))=\psi^+(\hat{\gamma}_{(2)}^{1/2}z)\otimes\psi^+(\hat{\gamma}_{(1)}^{-1/2}z),\\
&\Delta(\psi^-(z))=\psi^-(\hat{\gamma}_{(2)}^{-1/2}z)\otimes\psi^-(\hat{\gamma}_{(1)}^{1/2}z),
\end{split}
\end{align}
where $\hat{\gamma}_{(1)}=\hat{\gamma}\otimes 1$ and $\hat{\gamma}_{(2)}=1\otimes \hat{\gamma}$. Diagrammatically, the AFS property can be represented as 
\begin{align}
\begin{tikzpicture}[yscale=-1, baseline=(current  bounding  box.center)]
\draw (0,0.5)--(0,-0.5);
\draw (0,0)--(0.5,0);
\node at (0.5,0) [right] {$x^+(z)$};
\draw (2.4,0)--(2.6,0);
\draw (2.5,0.1)--(2.5,-0.1);
\draw (3.5,0.5)--(3.5,-0.5);
\draw (3.5,0)--(4,0);
\node at (4,0) [right] {$\psi^-(z)$};
\node at (3.5,0.5) [below] {$x^+(z)$};
\draw (5.4,0.05)--(5.6,0.05);
\draw (5.4,-0.05)--(5.6,-0.05);
\draw (6.5,0.5)--(6.5,-0.5);
\draw (6.5,0)--(7,0);
\node at (6.5,-0.5) [above] {$x^+(z)$};
\end{tikzpicture}
\end{align}
\begin{align}
\begin{tikzpicture}[yscale=-1, baseline=(current  bounding  box.center)]
\draw (0,0.5)--(0,-0.5);
\draw (0,0)--(0.5,0);
\node at (0.5,0) [right] {$x^-(\gamma z)$};
\node at (0,0.5) [below] {$\psi^+(\gamma^{1/2}z)$};
\draw (2.4,0)--(2.6,0);
\draw (2.5,0.1)--(2.5,-0.1);
\draw (3.5,0.5)--(3.5,-0.5);
\draw (3.5,0)--(4,0);
\node at (3.5,0.5) [below] {$x^-(z)$};
\draw (4.9,0.05)--(5.1,0.05);
\draw (4.9,-0.05)--(5.1,-0.05);
\draw (6,0.5)--(6,-0.5);
\draw (6,0)--(6.5,0);
\node at (6,-0.5) [above] {$x^-(z)$};
\end{tikzpicture}
\end{align}
\begin{align}
\begin{tikzpicture}[yscale=-1, baseline=(current  bounding  box.center)]
\draw (0,0.5)--(0,-0.5);
\draw (0,0)--(-0.5,0);
\node at (0,-0.5) [above] {$x^+(z)$};
\draw (0.4,0)--(0.6,0);
\draw (0.5,0.1)--(0.5,-0.1);
\draw (3,0.5)--(3,-0.5);
\draw (2.5,0)--(3,0);
\node at (2.5,-0.5) [above] {$\psi^-(\gamma^{1/2} z)$};
\node at (2.5,0) [left] {$x^+(\gamma z)$};
\draw (3.9,0.05)--(4.1,0.05);
\draw (3.9,-0.05)--(4.1,-0.05);
\draw (6,0.5)--(6,-0.5);
\draw (6,0)--(5.5,0);
\node at (6,0.5) [below] {$x^+(z)$};
\end{tikzpicture}
\end{align}
\begin{align}
\begin{tikzpicture}[yscale=-1, baseline=(current  bounding  box.center)]
\draw (0,0.5)--(0,-0.5);
\draw (0,0)--(-0.5,0);
\node at (0,-0.5) [above] {$x^-(z)$};
\node at (-0.5,0) [left] {$\psi^+(z)$};
\draw (0.4,0)--(0.6,0);
\draw (0.5,0.1)--(0.5,-0.1);
\draw (3,0.5)--(3,-0.5);
\draw (2.5,0)--(3,0);
\node at (2.5,0) [left] {$x^-(z)$};
\draw (3.9,0.05)--(4.1,0.05);
\draw (3.9,-0.05)--(4.1,-0.05);
\draw (6,0.5)--(6,-0.5);
\draw (6,0)--(5.5,0);
\node at (6,0.8) {$x^-(z)$};
\end{tikzpicture}
\end{align}
We remark that what essentially happens when we confirm the validness of the AFS property is that operators acting from both sides of the topological vertex in the unpreferred direction give the same result, but these functions are respectively expanded around $z\sim 0$ and $z\sim \infty$, so their difference will be given by the summation of a $\delta$-function times a residue over all the other poles of the function. 
\ba
\lt.f(z)\rt|_{z\sim \infty}-\lt.f(z)\rt|_{z\sim 0}=\sum_{a:\ {\rm pole}}\delta(x/a)\Res_{x'\rightarrow a}x^{\prime -1}f(x').
\ea

Let us start from $\bar{\Phi}^\ast$ vertex. For simplicity, we focus on the explicit calculation of acting $x^+(z)$ in the unpreferred direction. We have 
\ba
\contraction{}{\eta(z)}{}{\bar{\Phi}^{\ast}_\lambda[u,v]}\eta(z)\bar{\Phi}^{\ast}_\lambda[u,v]&=&\frac{1}{1-v/(z\gamma)}\prod_{x\in\lambda}S(\chi_x/(z\gamma))^{-1}:\eta(z)\bar{\Phi}^{\ast}_\lambda[u,v]:\nn\\
&=&\frac{1}{\cY_\lambda(z\gamma)}:\eta(z)\bar{\Phi}^{\ast}_\lambda[u,v]:,
\ea
\ba
\contraction{}{\bar{\Phi}^{\ast}_\lambda[u,v]}{}{\eta(z)}\bar{\Phi}^{\ast}_\lambda[u,v]\eta(z)&=&\frac{1}{1-z\gamma/v}\prod_{x\in\lambda}S(\chi_x/(z\gamma))^{-1}:\eta(z)\bar{\Phi}^{\ast}_\lambda[u,v]:\nn\\
&=&-\frac{v}{z\gamma}\frac{1}{\cY_\lambda(z\gamma)}:\eta(z)\bar{\Phi}^{\ast}_\lambda[u,v]:,
\ea
and thus by formulating $\bar{\Phi}^\ast$ as a map\footnote{The axio-dilaton charge (or the center of DIM) is not preserved under this map due to the effective existence of an orientifold plane.}, 
\ba
\bar{\Phi}^{(p)\ast}[u,v]:\ (p,1)_{u}\rightarrow (p+1,1)_{-uv}\otimes (1,0)_v,
\ea
we find the AFS property for it given by 
\begin{align}
\begin{tikzpicture}[baseline=(current  bounding  box.center)]
\draw (0,1)--(0,-1);
\draw (-1,0)--(0,0);
\draw[ultra thick] (-0.1,0.1)--(0.1,-0.1);
\draw[ultra thick] (0.1,0.1)--(-0.1,-0.1);
\node at (0,1) [above] {$x^+(z)$};
\draw (1.4,0)--(1.6,0);
\draw (3.5,1)--(3.5,-1);
\draw (2.5,0)--(3.5,0);
\draw[ultra thick] (3.5-0.1,0.1)--(3.5+0.1,-0.1);
\draw[ultra thick] (3.5+0.1,0.1)--(3.5-0.1,-0.1);
\node at (3.5,-1) [below] {$q_3x^+(z)$};
\draw (4.9,0.05)--(5.1,0.05);
\draw (4.9,-0.05)--(5.1,-0.05);
\draw (9,1)--(9,-1);
\draw (8,0)--(9,0);
\draw[ultra thick] (9-0.1,0.1)--(9+0.1,-0.1);
\draw[ultra thick] (9+0.1,0.1)--(9-0.1,-0.1);
\node at (8,0) [left] {$-x^-(z\gamma)$};
\node at (9,-1) [below] {$:\eta(z)\bar{\xi}^{-1}(z\gamma):$}; 
\end{tikzpicture}
\end{align}
We note that 
\ba
:\eta(z)\bar{\xi}^{-1}(z\gamma): \ =\exp\lt(-\sum_{n=1}^\infty \frac{1}{n}(1-q^{-n}_1)z^{-n}(1-q_3^n) a_n\rt),
\ea
only has positive modes (as $\psi^+(z)$). It is natural that $\bar{\Phi}^\ast$ reflects $x^+$ in the unpreferred direction to $x^-$ in the preferred direction, which is also expected from the construction with a reflection state. 

Similarly, we can write down the AFS property for the $\bar{\Phi}$ vertex as 
\begin{align}
\begin{tikzpicture}[baseline=(current  bounding  box.center)]
\draw (0,1)--(0,-1);
\draw (0,0)--(1,0);
\draw[ultra thick] (-0.1,0.1)--(0.1,-0.1);
\draw[ultra thick] (-0.1,-0.1)--(0.1,0.1);
\node at (0,1) [above] {$q_3x^+(z)$};
\node at (1,0) [right] {$\psi^-(zq_3)$};
\draw (2.9,0)--(3.1,0);
\draw (4,1)--(4,-1);
\draw (4,0)--(5,0);
\draw[ultra thick] (4-0.1,0.1)--(4+0.1,-0.1);
\draw[ultra thick] (4-0.1,-0.1)--(4+0.1,0.1);
\node at (4,-1) [below] {$x^+(z)$};
\draw (5.9,0.05)--(6.1,0.05);
\draw (5.9,-0.05)--(6.1,-0.05);
\draw (7,1)--(7,-1);
\draw (7,0)--(8,0);
\draw[ultra thick] (7-0.1,0.1)--(7+0.1,-0.1);
\draw[ultra thick] (7-0.1,-0.1)--(7+0.1,0.1);
\node at (8,0) [right] {$x^-(zq_3)$};
\node at (7,1) [above] {$\gammah^p$};
\end{tikzpicture}
\end{align}
by thinking of $\bar{\Phi}$ as a map, 
\ba
\bar{\Phi}^{(p)}[u,v]:\ (p+1,1)_{-uv}\otimes (1,0)_v\rightarrow (p,1)_u.
\ea

For the half-blood vertex $\tilde{\Phi}^{\ast(d_i)}_{j}$ (with ${\rm gcd}(d_i,d_j)=d_j$), 
\begin{align}
\begin{tikzpicture}[scale=2, baseline=(current  bounding  box.center)]
\draw[ultra thick](0,0)--(0,-0.4);
\draw[ultra thick] (0,-0.6)--(0,-1);
\draw (0,-0.5) circle (0.1);
\draw (-0.1,-0.5)--(-0.5,-0.5);
\end{tikzpicture}
\end{align}
we can write down a similar relation, 
\begin{align}
\begin{tikzpicture}[scale=2,yscale=-1, baseline=(current  bounding  box.center)]
\draw[ultra thick](0,0)--(0,-0.4);
\draw[ultra thick] (0,-0.6)--(0,-1);
\draw (0,-0.5) circle (0.1);
\draw (-0.1,-0.5)--(-0.5,-0.5);
\node at (0,-1) [above] {$:\prod_{k=0}^{d_i/d_j-1}x^+(zq_1^{-kd_j}\gamma_j^2/\gamma_i^2):$};
\draw (0.9,-0.5)--(1.1,-0.5);
\draw (1,-0.6)--(1,-0.4);
\draw[ultra thick](4.5,0)--(4.5,-0.4);
\draw[ultra thick] (4.5,-0.6)--(4.5,-1);
\draw (4.5,-0.5) circle (0.1);
\draw (4.4,-0.5)--(4,-0.5);
\node at (4,-0.5) [left] {$\sum_{l=0}^{d_i/d_j-1}x^+(zq_1^{-ld_j}\gamma_i)$};
\node at (4.5,-1) [above] {$:\tilde{\varphi}^{-(d_i)}_j(zq_1^{-ld_j}\gamma_i^{1/2})\prod_{\substack{k=0\\k\neq l}}^{d_i/d_j-1}x^+(zq_1^{-kd_j}\gamma_j^2/\gamma_i^2):$};
\draw (5.4,-0.475)--(5.6,-0.475);
\draw (5.4,-0.525)--(5.6,-0.525);
\draw[ultra thick](7,0)--(7,-0.4);
\draw[ultra thick] (7,-0.6)--(7,-1);
\draw (7,-0.5) circle (0.1);
\draw (6.9,-0.5)--(6.5,-0.5);
\node at (7,0) [below] {$x^+(z)$};
\end{tikzpicture}
\end{align}
by considering $\tilde{\Phi}^{\ast(d_i)}_{j}$ as a map, 
\ba
\tilde{\Phi}^{(p)\ast(d_i)}_{j}[u,v_j]:\ (d_i/d_j \cdot (p+1),1)^{(d_i)}_{u'}\rightarrow (p,1)^{(d_i)}_{u}\otimes (1,0)^{(d_j)}_{v_j},\label{half-blood-map}
\ea
where $p$ might be fractional, $(p,q)^{(d_i)}$ is a $(p,q)$ representation of the algebra DIM$_{q_1^{d_i},q_2}$, and 
\ba
u'=(-uv_j)^{d_i/d_j}q_1^{-\frac{(3p+1)}{2}(d_i/d_j-1)d_i},
\ea
\ba
&&\tilde{\varphi}^{-(d_i)}_j(z):= \ :\eta^{(d_i)}(z\gamma_i^{-1/2})\tilde{\xi}_j(z\gamma_i^{1/2}):,\\
&&\tilde{\psi}^{-(d_i)}_j(z)\mapsto \gamma_i^p\tilde{\varphi}^{-(d_i)}_j(z)\quad {\rm under\ the\ representation}\ (p,1)^{(d_i)}.
\ea
The prefactor $\tilde{t}_j^{(p)\ast(d_i)}(u,v_j)$ in the full expression of the half-blood topological vertex 
\ba
\tilde{\Phi}^{\ast(d_i)}_{\cX_j}[v_j]=\tilde{t}_j^{(p)\ast(d_i)}(u,v_j):\tilde{\Phi}^{\ast(d_i)}_\emptyset[v_j]\prod_{s\in\lambda^{(j)}}\tilde{\xi}_j^{(d_i)}(\chi^{(d_j)}_s):,
\ea
is found to take the form 
\ba
\tilde{t}_j^{(p)\ast(d_i)}(u,v_j)=(u\gamma_j)^{-|\lambda|}\prod_{x\in\lambda^{(j)}}(\chi_x\gamma_j^{2}/\gamma_i^3)^p,
\ea
to satisfy the AFS property proposed here. We note that the AFS property of $\tilde{\Phi}^{\ast(d_i)}_{j}$ certainly reduces to that of $\Phi^\ast$ defined in DIM$_{q_1^{d_j},q_2}$ when we take $d_i=d_j$. 

The AFS property for a ``square-root'' vertex can be worked out in the same spirit. Let us first consider the vertex $\Phi^{[1/2]}[u,v]$, which can be formulated as a map, 
\ba
\Phi^{[1/2](p)}[u,v]:\ (p,1)_u\otimes(1,0)_v\rightarrow (p+\frac{1}{2},1)_{u(-v)^{\frac{1}{2}}}.
\ea
Then the AFS property can be found as 
\begin{align}
\begin{tikzpicture}[baseline=(current  bounding  box.center)]
\draw (-1,1)--(-1,-1);
\draw (-1,0)--(0,0);
\draw[fill=black] (-1-0.1,0.1) rectangle (-1+0.1,-0.1);
\node at (-1,1) [above] {$\eta^{\frac{3}{2}}(z)$};
\node at (0,0) [right] {$x^+(z)\cY^{\frac{3}{2}}(z)$};
\draw (2.4,0)--(2.6,0);
\draw (2.5,0.1)--(2.5,-0.1);
\draw (3.5,1)--(3.5,-1);
\draw (3.5,0)--(4.5,0);
\draw[fill=black] (3.5-0.1,0.1) rectangle (3.5+0.1,-0.1);
\node at (4.5,0) [right] {$\psi^-(z)$};
\node at (3.5,-1) [below] {$:x^+(z)x^+(z):$};
\draw (5.9+0.5,0.05)--(6.1+0.5,0.05);
\draw (5.9+0.5,-0.05)--(6.1+0.5,-0.05);
\draw (7.5,1)--(7.5,-1);
\draw (7.5,0)--(8.5,0);
\draw[fill=black] (7.5-0.1,0.1) rectangle (7.5+0.1,-0.1);
\node at (7.5,1) [above] {$:x^+(z)x^+(z):$};
\end{tikzpicture}
\end{align}
where the prefactor $t^{[\frac{1}{2}](p)}(u,v)$ in 
\ba
\Phi^{[1/2](p)}_\lambda[u,v]=t^{[\frac{1}{2}]}_{p}(u,v)\Phi^{\frac{1}{2}}_\emptyset[v]\prod_{x\in\lambda}\eta^{\frac{1}{2}}(\chi_x),
\ea
is fixed to 
\ba
t^{[\frac{1}{2}]}_{p}(\lambda,u,v)=\lt(-u^2v\rt)^{|\lambda|}\prod_{x\in\lambda}(\gamma/\chi_x)^{2n+1},
\ea
and the operator $\cY(z)$ is a diagonal operator in the basis $\ket{v,\lambda}$ defined in (\ref{def-Y-op}). The vertex $\Phi^{\ast[\frac{1}{2}]}$ can be thought of as a map, 
\ba
\Phi^{\ast[\frac{1}{2}](p)}[u,v]:\ (p+\frac{1}{2},1)_{u(-v)^{\frac{1}{2}}}\rightarrow (p,1)_u\otimes (1,0)_v,
\ea
and the AFS property for it is given by 
\begin{align}
\begin{tikzpicture}[baseline=(current  bounding  box.center)]
\draw (-1,1)--(-1,-1);
\draw (-2,0)--(-1,0);
\draw[fill=black] (-1+0.1,-0.058)--(-1-0.1,-0.058)--(-1,0.115)--cycle;
\node at (-1,1) [above] {$:x^+(z)x^+(z):$};
\draw (0.4,0)--(0.6,0);
\draw (0.5,-0.1)--(0.5,0.1);
\draw (5.5,1)--(5.5,-1);
\draw (4.5,0)--(5.5,0);
\draw[fill=black] (5.5+0.1,-0.058)--(5.5-0.1,-0.058)--(5.5,0.115)--cycle;
\draw (6.9,0.05)--(7.1,0.05);
\draw (6.9,-0.05)--(7.1,-0.05);
\draw (9.5,1)--(9.5,-1);
\draw (8.5,0)--(9.5,0);
\draw[fill=black] (9.5+0.1,-0.058)--(9.5-0.1,-0.058)--(9.5,0.115)--cycle;
\node at (9.5,-1) [below] {$:x^+(z)x^+(z):$};
\node at (4.5,0) [left] {$\cY(z\gamma^{-1})^{-\frac{3}{2}}x^+(\gamma z)$};
\node at (5.5,1) [above] {$:\eta^{\frac{3}{2}}(z)\lt(\psi^-(\gamma^{\frac{1}{2}} z)\rt)^{\frac{1}{2}}:$};
\end{tikzpicture}
\end{align}
with the prefactor $t^{\ast[\frac{1}{2}]}_p[u,v]$, 
\ba
t^{\ast[\frac{1}{2}]}_p[u,v]=(\gamma u^2)^{-|\lambda|}\prod_{x\in\lambda}(\chi_x/\gamma)^{2p}.
\ea
The AFS properties formulated here for the square-root vertices are obviously not as elegant as the other vertices and it seems that the DIM algebra is not designed for them. 

We can write down the AFS property in parallel for 
\ba
\bar{\Phi}^{\ast[\frac{1}{2}](p)}[u,v]:\ (p,1)_{u}\otimes(1,0)_v\rightarrow (p+\frac{1}{2},1)_{u(-v)^{\frac{1}{2}}\gamma^{-1}},
\ea
and 
\ba
\bar{\Phi}^{[\frac{1}{2}](p)}[u,v]:\ (p+\frac{1}{2},1)_{u(-v)^{\frac{1}{2}}\gamma^{-\frac{3}{2}}}\rightarrow (p,1)_u\otimes (1,0)_v,
\ea
are respectively given by 
\begin{align}
\begin{tikzpicture}[baseline=(current  bounding  box.center)]
\draw (0,1)--(0,-1);
\draw (-1,0)--(0,0);
\draw[fill=black] (0,0.15)--(0.15,0)--(0,-0.15)--(-0.15,0)--cycle;
\node at (0,1) [above] {$:x^+(z)x^+(z):$};
\draw (1.4,0)--(1.6,0);
\draw (3.5,1)--(3.5,-1);
\draw (2.5,0)--(3.5,0);
\draw[fill=black] (3.5,0.15)--(3.5+0.15,0)--(3.5,-0.15)--(3.5-0.15,0)--cycle;
\node at (3.5,-1) [below] {$:x^+(z)x^+(z):$};
\draw (4.9,0.05)--(5.1,0.05);
\draw (4.9,-0.05)--(5.1,-0.05);
\draw (10,1)--(10,-1);
\draw (9,0)--(10,0);
\draw[fill=black] (10,0.15)--(10+0.15,0)--(10,-0.15)--(10-0.15,0)--cycle;
\node at (9,0) [left] {$-\cY^{\frac{3}{2}}(z\gamma)x^-(z\gamma)$};
\node at (10,1) [above] {$\eta^{\frac{3}{2}}(z)$};
\node at (10,-1) [below] {$:\eta^{\frac{1}{2}}(z)\bar{\xi}^{-\frac{1}{2}}(z\gamma):$}; 
\end{tikzpicture}
\end{align}
and 
\begin{align}
\begin{tikzpicture}[baseline=(current  bounding  box.center)]
\draw (0,1)--(0,-1);
\draw (0,0)--(1,0);
\draw[fill=black] (0.1,0.058)--(-0.1,0.058)--(0,-0.115)--cycle;
\node at (0,1) [above] {$:x^+(z)x^+(z):$};
\node at (1,0) [right] {$\psi^-(zq_3)$};
\draw (2.9,0)--(3.1,0);
\draw (4,1)--(4,-1);
\draw (4,0)--(5,0);
\draw[fill=black] (4+0.1,0.058)--(4-0.1,0.058)--(4,-0.115)--cycle;
\node at (4,-1) [below] {$:x^+(z)x^+(z):$};
\draw (5.9,0.05)--(6.1,0.05);
\draw (5.9,-0.05)--(6.1,-0.05);
\draw (7,1)--(7,-1);
\draw (7,0)--(8,0);
\draw[fill=black] (7+0.1,0.058)--(7-0.1,0.058)--(7,-0.115)--cycle;
\node at (7,1) [above] {$\eta^{\frac{3}{2}}(z)$};
\node at (8,0) [right] {$x^-(zq_3)\cY(zq_3)^{-\frac{3}{2}}$};
\end{tikzpicture}
\end{align}
with the prefactors respectively fixed as 
\ba
\bar{t}^{\ast[\frac{1}{2}]}_p[u,v]=t^{[\frac{1}{2}]}_{p}(\lambda,u,v/q_3),
\ea
and 
\ba
\bar{t}^{[\frac{1}{2}]}_{p}(\lambda,u,v)=t^{\ast[\frac{1}{2}]}_p[u/q_3,v].
\ea

We will list the AFS properties for some vertices associated to acting $x^-$('s) in the unpreferred direction in Appendix \ref{AFS-xm}. 

\section{Conclusion and Discussion}

In this article, we propose a web construction of all $ABCDEFG$-type and affine quiver gauge theories. It is based on a web of vertex operators defined in different Fock spaces glued together by the $(p,q)=(1,0)$ representations of the DIM algebra in the preferred direction. Unlike the construction for $A$-type and $D$-type quivers, which is nothing but the topological vertex formalism on the $(p,q)$-brane web (further with orientifolds for $D$-type cases), we do not have a clear physical meaning for that for non-simply-laced quivers and $E$-type ones. The results presented in this article might be a very preliminary form of the final answer, and our aim here is to reproduce the instanton partition functions of these quiver gauge theories on a full $\Omega$-background from the web construction. We always put a $(p,q)=(1,0)$ representation space of the DIM algebra in the preferred direction due to that we can utilize its nice property to reproduce the expressions of qq-characters via the Ward identity approach in the web diagram. In the unpreferred direction, in addition to the vertex operators corresponding to the well-known refined topological vertices, we introduced several new vertices, which can be divided into two classes. The first class contains vertices obtained by twisting of the usual topological vertices. They have the same contraction rule with themselves as the usual topological vertex before twisting, but can give an abundant structure when contracted with other vertices. The second class, roughly speaking, is obtained by taking fractional power of the first classes, and the ``square-root'' vertices seem to play a role in the construction of $E$-type quivers and their affine extensions. 

As the topological vertex formalism can be extended to 6d $\cN=(1,0)$ gauge theories on $\mathbb{C}^2_{q_1,q_2}\times T^2$ \cite{Zhu-elliptic, Foda-Zhu}, we can define the corresponding 6d versions of our new vertices to formulate 6d non-simply-laced quiver gauge theories. However, we do not know yet the relation with the classification of 6d $\cN=(1,0)$ theories via the geometric picture of the F-theory \cite{Morrison-Taylor}. 

The fiber-base duality is always a topic that cannot be avoided when we consider quivers structure or gauge groups beyond the $A$-type Lie algebra. We have proposed the AFS properties for new vertices introduced in this article, which can be used in principle to construct a fiber-base dual version of them to analyze the duality in full details. However, our AFS properties are formulated inside the DIM algebra, both in the preferred and unpreferred directions. From our experience to generalize the topological vertex formalism to 6d and to 5d theories on $\mathbb{C}^2/\mathbb{Z}_n\times S^1$ \cite{KZ-Zn, Foda-Zn, qq-Zn}, the underlying algebra seems to get modified whenever we go to a different theory (and is determined by the Nekrasov factor). This suggests that a potential modification of our construction might be necessary in the future study.

An additional guide line that has not been considered in this article for our construction is the integrability of the underlying gauge theories. The correspondence between gauge theories in the Nekrasov-Shatashivili (NS) limit and quantum integrable spin chain was first proposed in \cite{NS1, NS2, NPS}, and it was also worked out for non-simply-laced quivers in \cite{Kimura-Chen}. In the full $\Omega$-background, there is a universal R-matrix associated to the DIM algebra \cite{FJMM1}, which is a $q$-deformed version \cite{Harada1} of the celebrated Maulik-Okounkov's R-amtrix \cite{MO} found through a geometric approach to 4d $\cN=2$ gauge theories. The relation between these two integrable models are unclear at the moment, but it is natural to expect the quantum integrablity of non-simply laced quivers in the NS limit to be uplifted to an integrability described by an R-matrix embedded in some quantum algebra. Note that such an algebraic structure of the topological vertex formalism for $A$-type quivers can be very powerful to prove statements such as the fiber-base duality in a rigorous way \cite{S-FOS}. 

\section*{Acknowledgement}

The work of TK was supported in part by JSPS Grant-in-Aid for Scientific Research (No.~JP17K18090), the MEXT-Supported Program for the Strategic Research Foundation at Private Universities ``Topological Science'' (No.~S1511006), and JSPS Grant-in-Aid for Scientific Research on Innovative Areas ``Topological Materials Science'' (No.~JP15H05855), and ``Discrete Geometric Analysis for Materials Design'' (No.~JP17H06462). RZ was supported by JSPS fellowship for young scientist at the beginning stage of this work, and he would like to thank the hospitality of Simons center and Galileo Galilei Institute, where part of the results of this paper were presented. \\

\begin{flushleft}
{\LARGE {\bf Appendix}}
\end{flushleft}

\vspace{-2em}

\appendix

\section{qq-character derived from Ward identity}\label{a:qq-character}

The qq-character is an operator-valued quantity introduced by Nekrasov \cite{BPS/CFT} as a quantized object which accounts for the integrability structure of the gauge theory with eight supercharges.

It only depends on the quiver structure of the gauge theory, and its formula (for 5d theories) is given by (a 5d version of Nekrasov's integral formula in \cite{BPS/CFT})
\ba
\chi_{\vec{w}}(z;\vec{\nu}) =\sum_{\vec{a}}\prod_{i\in {\rm Vert}(\Gamma')}\frac{1}{a_i!}\lt(\frac{\mathfrak{q}_i(1-q_3^{-1})}{2\pi i(1-q_1)(1-q_2)}\rt)^{a_i}\prod_{j=1}^{w_i}\cY_i(zq_3^{-1}\nu_{i,j})\oint_{C_{w,\nu,a}}\Upsilon_{w,\nu,a}(z),
\ea
where 
\ba
\Upsilon_{w,\nu,a}(z)=\prod_{e\in {\rm Edge}(\Gamma')}\Upsilon^e_{w,\nu,a}(z)\prod_{i\in{\rm Vert}(\Gamma')}\Upsilon^i_{w,\nu,a}(z),
\ea
\ba
\Upsilon^e_{w,\nu,a}(z)=\prod_{k=1}^{a_{t(e)}}\cY_{s(e)}(zq_3^{-1}\mu_e^{-1}\phi_k^{(t(e))})\prod_{l=1}^{a_{s(e)}}\cY_{t(e)}(z\mu_e\phi_l^{(s(e))})S(\phi_k^{(t(e))}/\phi_l^{(s(e))}\mu_e),
\ea
\ba
\Upsilon_{w,\nu,a}^i(z)=\prod_{k=1}^{a_i}\frac{{\rm d}\phi^{(i)}_k P_i(z\phi_k^{(i)})}{\phi_k^{(i)}\cY_i(zq_3^{-1}\phi^{(i)}_k)\cY_i(z\phi^{(i)}_k)}\prod_{l\neq k}S(\phi^{(i)}_l/\phi^{(i)}_k)^{-1}\prod_{j=1}^{w_i}S(\nu_{i,j}/\phi^{(i)}_k),
\ea
and the integral contour picks up the poles in $\frac{1}{1-z}$ factors of $S(z)$'s. $z_{i,j}:=z\nu_{i,j}$ are independent parameters in the qq-character.  $\vec{w}$ is the highest weight of the representation of the quiver Lie algebra that characterizes the qq-character. $P_i(z)$ in the above formula is some fractional function which contains the information of matter contents and Chern-Simons levels of the gauge theory. The information of the gauge group is encoded in the expression of $\cY_i$, in the case of U(1) gauge theory,
\ba
\bra{v_i,\lambda}\cY_i(z)\ket{v_i,\lambda}=\cY_\lambda(z),\label{def-Y-op}
\ea
and for SU($N$), it is given by $N$ products of U(1) $\cY$-functions. In the $(1,0)$ representation (used in the preferred direction of the web diagram) of the DIM algebra, one can formulate $\cY_i(z)$ as a diagonal operator in the basis $\ket{v_i,\lambda}$. 

The qq-charcter can be derived from the web diagram following the prescription: 
\begin{itemize}
\item First we consider an insertion of operators in DIM algebra in the preferred direction, whose action on $\Phi$ gives the residue of the highest weight term in the qq-character. 

\item Using the Ward identity\footnote{The reason we call it a Ward identity is that it resembles the conformal Ward identity for 2-pt functions in 2d CFT, and the DIM algebra might indeed be a symmetry of the underlying gauge theory/topological string theory.} (i.e. the action of the operator considered in the first step stays the same either we consider it on $\Phi$ or $\Phi^\ast$) to convert the residue obtained from the previous step to the residue of some other term in the qq-character. 

\item Consider new insertions of operators in the web diagram with which all the poles of the form $\chi_x$ of the new term obtained from the second step are covered so that we can rewrite the sum of these residues of this new term into a contour integral around the pole at $z_{i,j}$, the origin and the infinity. 

\item Repeat from the second step to pick up all poles of all possible terms until we get a closed form of the qq-character in a contour integral form.
\end{itemize}

Let us describe the computation of the fundamental qq-character in pure U(1) gauge theory ($A_1$ quiver) with Chern-Simons level $\kappa$ as an example, and then use the above prescription to derive the spin-1 (3-dim) qq-character of the same theory in a brief manner. Let us consider the following insertion in the preferred direction of the web diagram, 
\begin{align}
\begin{tikzpicture}[baseline=(current  bounding  box.center)]
\draw (0,0.5)--(0,-0.5);
\draw (1,0.5)--(1,-0.5);
\draw (0,0)--(1,0);
\node at (0.5,0.3) {${\small x^-_{>}}$};
\node at (-0.5,0) [left] {$ \Phi^{(n)}[u,v]$};
\node at (1.5,0) [right] {$ \Phi^{(n^\ast)}[u^\ast,v]$};
\end{tikzpicture}
\end{align}
where $x^-_>=\sum_{k>0}x^-_kz^{-k}$ (see Appendix \ref{a:rep-DIM} for the meaning and how to evaluate this operator). The trivial identity equating the action of $x^-_>$ on $ \Phi^{(n)}[u,v]$ and the action on $ \Phi^{(n^\ast)}[u^\ast,v]$ is translated to 
\ba
z^{1-\kappa}u^{\ast}\sum_\lambda \mathfrak{q}^{|\lambda|}\frac{\prod_{y\in\lambda}\chi_y^\kappa}{N_{\lambda\lambda}(1;q_1,q_2)}\sum_{x\in R(\lambda)}\frac{1}{z-\chi_x}\Res_{z\rightarrow\chi_x}\cY_\lambda(zq_3^{-1})=uv\gamma\sum_\lambda \mathfrak{q}^{|\lambda|}\frac{\prod_{y\in\lambda}\chi_y^\kappa}{N_{\lambda\lambda}(1;q_1,q_2)}\sum_{x\in A(\lambda)}\frac{1}{z-\chi_x}\Res_{z\rightarrow\chi_x}\frac{1}{\cY_\lambda(z)},\nn
\ea
where the gauge coupling $\mathfrak{q}=-\frac{u\gamma^{n-n^\ast-1}}{u^\ast}$ and the Chern-Simons level $\kappa=n^\ast-n$. We can rewrite the above equation in the form of contour integral, 
\ba
\sum_\lambda \mathfrak{q}^{|\lambda|}\frac{\prod_{y\in\lambda}\chi_y^\kappa}{N_{\lambda\lambda}(1;q_1,q_2)}\oint_{C_\lambda}{\rm d}x\lt(\frac{1}{z-x}\cY_\lambda(xq_3^{-1})+\frac{\mathfrak{q}vq_3z^{\kappa-1}}{(z-x)\cY_\lambda(x)}\rt)=0,
\ea
where the contour $C_\lambda$ is around all finite poles in $\cY_\lambda(xq_3^{-1})$ and $1/\cY_\lambda(x)$, i.e. $\{\chi_x \mid x\in A(\lambda)\cup R(\lambda)\}$. One can then convert the contour integral to be a simple formula that allows us to evaluate the expectation value of the qq-character as 
\ba
\langle \chi_1(z)\rangle=\sum_\lambda \mathfrak{q}^{|\lambda|}\frac{\prod_{y\in\lambda}\chi_y^\kappa}{N_{\lambda\lambda}(1;q_1,q_2)}\oint_{x\sim 0,\infty}{\rm d}x\lt(\frac{1}{z-x}\cY_\lambda(xq_3^{-1})+\frac{\mathfrak{q}vq_3z^{\kappa-1}}{(z-x)\cY_\lambda(x)}\rt),
\ea
where the expectation value with respect to the instanton partition function $Z_{\mathrm{U}(1)}=\sum_\lambda \mathfrak{q}^{|\lambda|}\frac{\prod_{y\in\lambda}\chi_y^\kappa}{N_{\lambda\lambda}(1;q_1,q_2)}$ is given by 
\ba
\langle \cdots \rangle=\sum_\lambda \mathfrak{q}^{|\lambda|}\frac{\prod_{y\in\lambda}\chi_y^\kappa}{N_{\lambda\lambda}(1;q_1,q_2)}\bra{ v,\lambda}\cdots\ket{v,\lambda},
\ea
and one can extract out the operator-valued expression of the qq-character in terms of the $\cY$-operator as 
\ba
\chi_1(z)=\cY(zq_3^{-1})+\mathfrak{q}vq_3z^{\kappa-1}\cY(z)^{-1}.
\ea
One can redefine the $\cY$ operator, which will be denoted as $Y$ instead (and change the normalization of the qq-character), to schematically write the above qq-character as 
\ba
\bar{\chi}_1(z)=Y(z)+Y(zq_1^{-1}q_2^{-1})^{-1},
\label{eq:chi_bar}
\ea
where\footnote{The concrete form of this normalized $Y$-function depends on the Chern-Simons level and matter contents. Its existence is more important than the concrete expression of it in our computation. Such a normalized $Y$-function can also be defined for $\cY^d_{\cX_i}$ in (\ref{def-frac-Y}) in the same way, and we will not declare its existence before using it directly again.} 
\ba
Y(z)=z^{-\frac{1}{2}(\kappa-1)}\mathfrak{q}^{-\frac{1}{2}}v^{-\frac{1}{2}}\gamma^{\frac{\kappa-3}{2}}\cY(zq_3^{-1}).
\label{eq:Y_normalized}
\ea

The qq-character of the spin-1 representation of the $A_1$ quiver starts with the term $Y(z_1)Y(z_2)$ corresponding to the highest weight in the representation. One can complete the contour integral with the contour surrounding all its poles by considering the following  two insertions (for simplicity of notation, we omit the subscript $_>$ in $x^-_>$ inserted from now on), 
\begin{align}
\begin{tikzpicture}[baseline=(current  bounding  box.center)]
\draw (0,0.5)--(0,-0.5);
\draw (3,0.5)--(3,-0.5);
\draw (0,0)--(3,0);
\node at (1.5,0.3) {$x^-(z_1)\cY(z_2q_3^{-1})$};
\end{tikzpicture}
\end{align}
and 
\begin{align}
\begin{tikzpicture}[baseline=(current  bounding  box.center)]
\draw (0,0.5)--(0,-0.5);
\draw (3,0.5)--(3,-0.5);
\draw (0,0)--(3,0);
\node at (1.5,0.3) {$x^-(z_2)\cY(z_1q_3^{-1})$};
\end{tikzpicture}
\end{align}
The Ward identity for the first insertion reveals a new term 
\ba
\chi_2(z_1,z_2)\supset S(z_2/z_1)\frac{{\color{blue}Y(z_2)}}{Y(z_1q_1^{-1}q_2^{-1})},
\ea
in the qq-character, where we used the blue color in the above equation to show that the residues for poles in $Y(z_2)$ have not been included in the analysis so far. The $S$-function appears in the computation when we move $x^-$ to the right of the $\cY$-operator. To further complete the contour integral for this term, we need to consider the insertion of $x^-(z_2)\cY^{-1}(z_1)$ in the preferred direction, 
\begin{align}
\begin{tikzpicture}[baseline=(current  bounding  box.center)]
\draw (0,0.5)--(0,-0.5);
\draw (3,0.5)--(3,-0.5);
\draw (0,0)--(3,0);
\node at (1.5,0.3) {${\small \cY^{-1}(z_1)x^-(z_2)}$};
\end{tikzpicture}
\end{align}
which will give rise to a new term, 
\ba
\chi_2(z_1,z_2)\supset \frac{1}{{\color{blue}Y(z_1q_1^{-1}q_2^{-1})}Y(z_2q_1^{-1}q_2^{-1})}.\label{term-chi2}
\ea
Similarly, the Ward identity with respect to the insertion of $x^-(z_2)\cY(z_1q_3^{-1})$ induces another term corresponding to the Cartan part
\ba
\chi_2(z_1,z_2)\supset S(z_1/z_2)\frac{{\color{blue}Y(z_1)}}{Y(z_2q_1^{-1}q_2^{-1})},
\ea
and the remaining residues to complete the contour integral are collected from the insertion 
\begin{align}
\begin{tikzpicture}[baseline=(current  bounding  box.center)]
\draw (0,0.5)--(0,-0.5);
\draw (3,0.5)--(3,-0.5);
\draw (0,0)--(3,0);
\node at (1.5,0.3) {${\small \cY^{-1}(z_2)x^-(z_1)}$};
\end{tikzpicture}
\end{align}
The Ward identity for the above combination of operators $x^-(z_1)\cY^{-1}(z_2)$ covers exactly the remaining residues in the term (\ref{term-chi2}) in $\cY(z_1q_1^{-1}q_2^{-1})^{-1}$, and thus complete the contour integral for the whole (normalized) qq-character given by 
\ba
\bar{\chi}_{2}(z_1,z_2)=Y(z_1)Y(z_2)+S(z_2/z_1)\frac{Y(z_2)}{Y(z_1q_3)}+S(z_1/z_2)\frac{Y(z_1)}{Y(z_2q_3)}+\frac{1}{Y(z_1q_3)Y(z_2q_3)}.
\ea
An interesting observation may be made here that $S$-functions of the form $S(z_2/z_1)$ will be divergent in the limit $z_2\rightarrow z_1$, yet the above qq-character $\bar{\chi}_{2}(z_1,z_2)$ stays finite in the same limit. For example we have 
\ba
\lim_{z\rightarrow 1}\lt(S(z)+S(z^{-1})\rt)=\frac{1+q_1+q_2-6q_1q_2+q_1^2q_2+q_1q_2^2+q_1^2q_2^2}{(1-q_1q_2)^2}=:\mathfrak{c}(q_1,q_2).\label{def-cc}
\ea
The full expression of the superficially divergent part in $\bar{\chi}_{2}(z,z)$ can be evaluated to 
\ba
&&\lim_{z'\rightarrow z}\lt(S(z'/z)\frac{Y(z')}{Y(zq_3)}+S(z/z')\frac{Y(z)}{Y(z'q_3)}\rt)\nn\\
&&=\mathfrak{c}(q_1,q_2)\frac{Y(z)}{Y(zq_3)}-z\frac{(1-q_1)(1-q_2)}{1-q_1q_2}\partial_z \log\lt(Y(z)Y(zq_3)\rt)\frac{Y(z)}{Y(zq_3)},\label{limit-zz}
\ea
where the derivative term comes from the expansion of $Y$-functions around $z'=z$. Another similar limit is $z_2\rightarrow z_1q_1^{-1}q_2^{-1}$, where the problematic terms reduce to 
\ba
S(z/z')\frac{Y(z'q_3)}{Y(zq_3)}+S(zq_3^{-1}/z')\frac{Y(z)}{Y(z'q^2_3)},
\ea
where we used $S(zq_3)=S(z^{-1})$. The first term in the above appears to be divergent in the limit $z'\rightarrow z$, but in the contour integral we converted from to obtain the expression of the qq-character evaluated on the basis $\ket{v,\lambda}$, there is no such a divergence at all (unlike the case in (\ref{def-cc})). In contrast, in this limit, the contour integral we completed is over a constant $1$, and we do not need such a term in the qq-character\footnote{More precisely, we can see that $\oint_{C_1}{\rm d}x\frac{1}{z-x}S(z/z')\frac{Y(z'q_3)}{Y(xq_3)}+\oint_{C_2}{\rm d}x\frac{1}{z'-x}S(z/z')\frac{Y(xq_3)}{Y(zq_3)}=0$ by using $\oint_{C_1+C_2}{\rm d}x \frac{Y(z'q_3)}{Y(zq_3)}\rightarrow \oint_{C_1+C_2}{\rm d}x\ 1=0$ in the limit $z'\rightarrow z$.}. Therefore, the qq-charatcer $\bar{\chi}_{2}(z_1,z_2)$ in this special limit $z_2\rightarrow z_1q_1^{-1}q_2^{-1}$ is given by 
\ba
\bar{\chi}_{2}(z,zq_3)=Y(z)Y(zq_3)+S(q_3^{-1})\frac{Y(z)}{Y(zq^2_3)}+\frac{1}{Y(zq_3)Y(zq^2_3)}.\label{limit-zzq3}
\ea
The decoupling of the constant term can be interpreted as the subtraction of the singlet $\textbf{1} = \square$ from the tensor product $\textbf{2} \otimes \textbf{2} = \textbf{3} \oplus \textbf{1}$, which takes place only when $z_2/z_1=q_3$. A similar phenomenon is observed for $q$-characters (for example see \cite{Kuniba:2010ir}).

\subsection{Consistency check: spinor and adjoint representations of $D_4$}\label{a:qq-D4}

The web diagram for the $D_4$ quiver is given by\footnote{We note that the result will be the same to use the simplified web diagram shown in Figure \ref{fig_simpD} to derive the qq-character from the Ward identity approach.} 
\begin{align}
\begin{tikzpicture}[scale=2, baseline=(current  bounding  box.center)]
\draw (-2,0.5)--(-2,-0.5);
\draw (-1,0.5)--(-1,-0.5);
\draw (0,0.5)--(0,-0.5);
\draw (1,0.5)--(1,-0.5);
\draw (-1,0)--(0,0);
\draw (0,0.3)--(1,0.3);
\draw (0,-0.3)--(1,-0.3);
\draw (-1,0.2)--(-2,0.2);
\draw[ultra thick] (1-0.05,-0.3+0.05)--(1+0.05,-0.3-0.05);
\draw[ultra thick] (1+0.05,-0.3+0.05)--(1-0.05,-0.3-0.05);
\end{tikzpicture}
\end{align}
We label the nodes in the quiver diagram as follows for convenience. 
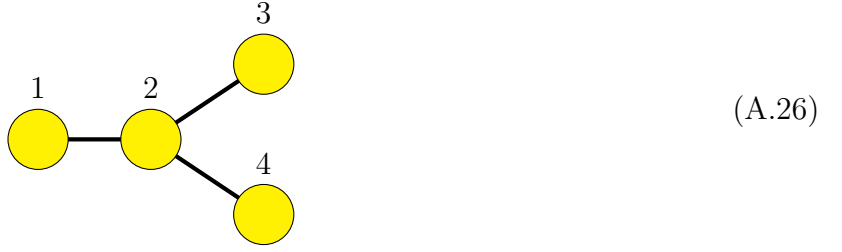
\begin{align}
\begin{tikzpicture}[baseline=(current  bounding  box.center)]
\draw [ultra thick] (0,0)--(1.5,1);
\draw [ultra thick] (0,0)--(1.5,-1);
\draw [ultra thick] (0,0)--(-1.5,0);
\draw [fill=yellow] (-1.5,0) circle (0.4);
\draw [fill=yellow] (0,0) circle (0.4);
\draw [fill=yellow] (1.5,1) circle (0.4);
\draw [fill=yellow] (1.5,-1) circle (0.4);
\node at (-1.5,0.4) [above] {$1$};
\node at (0,0.4) [above] {$2$};
\node at (1.5,1.4) [above] {$3$};
\node at (1.5,-0.6) [above] {$4$};
\end{tikzpicture}
\end{align}
The qq-character of the spinor representation starts from the highest-weight term $Y_3(z)$, which can be realized through the insertion 
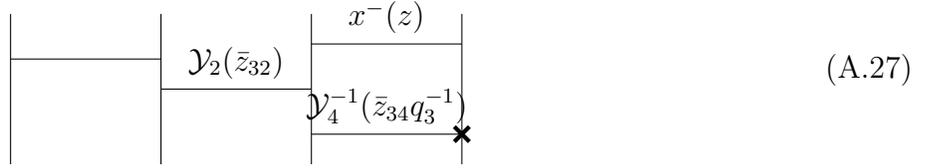
\begin{align}
\begin{tikzpicture}[scale=2, baseline=(current  bounding  box.center)]
\draw (-2,0.5)--(-2,-0.5);
\draw (-1,0.5)--(-1,-0.5);
\draw (0,0.5)--(0,-0.5);
\draw (1,0.5)--(1,-0.5);
\draw (-1,0)--(0,0);
\draw (0,0.3)--(1,0.3);
\draw (0,-0.3)--(1,-0.3);
\draw (-1,0.2)--(-2,0.2);
\draw[ultra thick] (1-0.05,-0.3+0.05)--(1+0.05,-0.3-0.05);
\draw[ultra thick] (1+0.05,-0.3+0.05)--(1-0.05,-0.3-0.05);
\node at (0.5,0.3) [above] {$x^-(z)$};
\node at (0.5,-0.3) [above] {$\cY^{-1}_4(\bar{z}_{34}q_3^{-1})$};
\node at (-0.5,0) [above] {$\cY_2(\bar{z}_{32})$};
\end{tikzpicture}
\end{align}
where we defined a new set of variables $\bar{z}_{ij}=z\mu_{ij}$ to shorten the notation involving bifundamental masses $\mu_{ij}$ in the calculation. The term we obtain from the other side of the equality in Ward identy of the above insertion reads 
\ba
\chi_{(0,0,1,0)}(z)\supset \frac{{\color{blue} Y_2(z\mu_{32})}}{Y_3(z q_1^{-1}q_2^{-1})}.
\ea
The reason why there is no $\cY_4$ in the above expression is that $\Phi$ and $\bar{\Phi}^\ast$ give exactly the same contribution in this calculation. 

In this way, we find the normalized qq-character of the spinor representation to be 
\ba
\bar{\chi}_{(0,0,1,0)}(z)=Y_3(z)+\frac{Y_2(z\mu_{32})}{Y_3(z q_1^{-1}q_2^{-1})}+\frac{Y_1(z\mu_{31})Y_4(z\mu_{34})}{Y_2(z\mu_{32}q_1^{-1}q_2^{-1})}+\frac{Y_4(z\mu_{34})}{Y_1(z\mu_{31}q_1^{-1}q_2^{-1})}\nn\\
+\frac{Y_1(z\mu_{31})}{Y_4(z\mu_{34}q_1^{-1}q_2^{-1})}+\frac{Y_2(z\mu_{32}q^{-1}_1q^{-1}_2)}{Y_1(z\mu_{31}q_1^{-1}q_2^{-1})Y_4(z\mu_{34}q_1^{-1}q_2^{-1})}\nn\\
+\frac{Y_3(zq_1^{-2}q_2^{-2})}{Y_2(z\mu_{32}q_1^{-2}q_2^{-2})}+\frac{1}{Y_3(zq_1^{-3}q_2^{-3})},
\ea
where we used the relation $\mu_{ij}=\mu_{ji}^{-1}q^{-1}_1q^{-1}_2$ for adjacent nodes $i$ and $j$ with the $i$-th node on the left to the $j$-th. It is equivalent to the qq-character of the vector representation under the permutation $1\leftrightarrow 3$, which is a manifestation of the triality symmetry in $D_4$. We remark that only the $\mathbb{Z}_2$ subgroup corresponding to exchanging the 3-rd and 4-th node in the Dynkin diagram of the triality can be manifestly seen from the web diagram, but the calculation to derive the qq-character is completely symmetric under the triality. 

Now we turn to the derivation of the qq-character of the adjoint representation of $D_4$. It initiates from the highest-weight term $Y_2(z)$, and the Ward identity approach tells us 
\ba
\chi_{(0,1,0,0)}(z)\supset T^+(z):=Y_2(z)+\frac{Y_1(z\mu_{21})Y_3(z\mu_{23})Y_4(z\mu_{24})}{Y_2(zq_1^{-1}q_2^{-1})}+\frac{Y_3(z\mu_{23})Y_4(z\mu_{24})}{Y_1(z\mu_{21}q_1^{-1}q_2^{-1})}\nn\\
+\frac{Y_1(z\mu_{21})Y_4(z\mu_{24})}{Y_3(z\mu_{23}q_1^{-1}q_2^{-1})}+\frac{Y_1(z\mu_{21})Y_3(z\mu_{23})}{Y_4(z\mu_{24}q_1^{-1}q_2^{-1})}+\frac{Y_2(zq_1^{-1}q_2^{-1})Y_4(z\mu_{24})}{Y_1(z\mu_{21}q_1^{-1}q_2^{-1})Y_3(z\mu_{23}q_1^{-1}q_2^{-1})}\nn\\
+\frac{Y_2(zq_1^{-1}q_2^{-1})Y_3(z\mu_{23})}{Y_1(z\mu_{21}q_1^{-1}q_2^{-1})Y_4(z\mu_{24}q_1^{-1}q_2^{-1})}+\frac{Y_2(zq_1^{-1}q_2^{-1})Y_1(z\mu_{21})}{Y_3(z\mu_{23}q_1^{-1}q_2^{-1})Y_4(z\mu_{24}q_1^{-1}q_2^{-1})}\nn\\
+\frac{{\color{blue} Y_1(z\mu_{21})Y_1(z\mu_{21}q_1^{-1}q_2^{-1})}}{Y_2(zq_1^{-2}q_2^{-2})}+\frac{{\color{blue} Y_3(z\mu_{23})Y_3(z\mu_{23}q_1^{-1}q_2^{-1})}}{Y_2(zq_1^{-2}q_2^{-2})}\nn\\
+\frac{{\color{blue} Y_4(z\mu_{24})Y_4(z\mu_{24}q_1^{-1}q_2^{-1})}}{Y_2(zq_1^{-2}q_2^{-2})}+\frac{{\color{blue} Y_2(zq_1^{-1}q_2^{-1})^2}}{Y_1(z\mu_{21}q_1^{-1}q_2^{-1})Y_3(z\mu_{23}q_1^{-1}q_2^{-1})Y_4(z\mu_{24}q_1^{-1}q_2^{-1})}.
\ea
We will obtain Cartan terms if we go further to consider the Ward identities for the blue terms, and  similar superficial divergences appear here as what we dealt with in (\ref{limit-zz}) and (\ref{limit-zzq3}). Substituting the results we obtained there, we have 
\ba
\chi_{(0,1,0,0)}(z)=T^+(z)+C^{D_4}_{(0,1,0,0)}(z)+T^-(z),
\ea
where 
\ba
&&C^{D_4}_{(0,1,0,0)}(z)=S(q_1q_2)\frac{Y_1(z\mu_{21})}{Y_1(z\mu_{21}q_1^{-2}q_2^{-2})}+S(q_1q_2)\frac{Y_3(z\mu_{23})}{Y_3(z\mu_{23}q_1^{-2}q_2^{-2})}+S(q_1q_2)\frac{Y_4(z\mu_{24})}{Y_4(z\mu_{24}q_1^{-2}q_2^{-2})}\nn\\
&&+\lt(\mathfrak{c}(q_1,q_2)-z\frac{(1-q_1)(1-q_2)}{1-q_1q_2}\partial_z \log\lt(\frac{Y_2(zq_1^{-1}q_2^{-1})Y(zq_1^{-2}q_2^{-2})}{Y_1(z\mu_{21}q_1^{-1}q_2^{-1})Y_3(z\mu_{23}q_1^{-1}q_2^{-1})Y_4(z\mu_{24}q_1^{-1}q_2^{-1})}\rt)\rt)\frac{Y_2(zq_1^{-1}q_2^{-1})}{Y_2(zq^{-2}_1q^{-2}_2)},\nn\\
\ea
and 
\ba
T^-(z)=\frac{Y_1(z\mu_{21}q_1^{-1}q_2^{-1})Y_3(z\mu_{23}q_1^{-1}q_2^{-1})Y_4(z\mu_{24}q_1^{-1}q_2^{-1})}{Y_2(zq_1^{-2}q_2^{-2})^2}+\frac{Y_2(zq_1^{-1}q_2^{-1})}{Y_1(z\mu_{21}q_1^{-1}q_2^{-1})Y_1(z\mu_{21}q_1^{-2}q_2^{-2})}\nn\\
+\frac{Y_2(zq_1^{-1}q_2^{-1})}{Y_3(z\mu_{23}q_1^{-1}q_2^{-1})Y_3(z\mu_{23}q_1^{-2}q_2^{-2})}+\frac{Y_2(zq_1^{-1}q_2^{-1})}{Y_4(z\mu_{24}q_1^{-1}q_2^{-1})Y_4(z\mu_{24}q_1^{-2}q_2^{-2})}\nn\\
+\frac{Y_3(z\mu_{23}q_1^{-1}q_2^{-1})Y_4(z\mu_{24}q_1^{-1}q_2^{-1})}{Y_2(zq_1^{-2}q_2^{-2})Y_1(z\mu_{21}q_1^{-2}q_2^{-2})}+\frac{Y_1(z\mu_{21}q_1^{-1}q_2^{-1})Y_4(z\mu_{24}q_1^{-1}q_2^{-1})}{Y_2(zq_1^{-2}q_2^{-2})Y_3(z\mu_{23}q_1^{-2}q_2^{-2})}\nn\\
+\frac{Y_1(z\mu_{21}q_1^{-1}q_2^{-1})Y_3(z\mu_{23}q_1^{-1}q_2^{-1})}{Y_2(zq_1^{-2}q_2^{-2})Y_4(z\mu_{24}q_1^{-2}q_2^{-2})}+\frac{Y_1(z\mu_{21}q_1^{-1}q_2^{-1})}{Y_3(z\mu_{23}q_1^{-2}q_2^{-2})Y_4(z\mu_{24}q_1^{-2}q_2^{-2})}\nn\\
+\frac{Y_1(z\mu_{21}q_1^{-1}q_2^{-1})}{Y_4(z\mu_{24}q_1^{-2}q_2^{-2})Y_3(z\mu_{23}q_1^{-2}q_2^{-2})}+\frac{Y_3(z\mu_{23}q_1^{-1}q_2^{-1})}{Y_4(z\mu_{24}q_1^{-2}q_2^{-2})Y_1(z\mu_{21}q_1^{-2}q_2^{-2})}\nn\\
+\frac{Y_2(zq_1^{-2}q_2^{-2})}{Y_1(z\mu_{21}q_1^{-2}q_2^{-2})Y_3(z\mu_{23}q_1^{-2}q_2^{-2})Y_4(z\mu_{24}q_1^{-2}q_2^{-2})}+\frac{1}{Y_2(zq_1^{-3}q_2^{-3})}.
\ea
It agrees with the known result presented in \cite{Kimura-r}, including the highly non-trivial Cartan part.

\subsection{Consistency check: fundamental representations of non-simply-laced quivers}\label{a:qq-BCFG}

Let us check the computation of qq-characters in BC$_2$ quiver from the Ward identity of DIM algebra to conclude this section. 

We start from the insertion of $x^-(z)$ in the 1-st node with $d_1=1$, 
\begin{align}
\begin{tikzpicture}[scale=2, baseline=(current  bounding  box.center)]
\draw[ultra thick](0,1)--(0,0.6);
\draw[ultra thick](0,0.4)--(0,-0.4);
\draw[ultra thick] (0,-0.6)--(0,-1);
\draw (0,0.5) circle (0.1);
\draw[ultra thick] (0.1,0.5)--(0.9,0.5);
\draw (1,0.5) circle (0.1);
\draw[ultra thick] (1,1)--(1,0.6);
\draw[ultra thick] (1,0.4)--(1,-0.5);
\draw (0,-0.5) circle (0.1);
\draw (-0.1,-0.5)--(-1,-0.5);
\draw (-1,0.5)--(-1,-1);
\node at (-0.5,-0.5) [above] {$x^-(z)$};
\end{tikzpicture}
\end{align}
This gives rise to the contribution 
\ba
\chi^{{\rm BC}_2}_{(1,0)}\supset Y^{(1)}(z)+\frac{{\cob Y^{(2)}}(z\mu_e)}{Y^{(1)}(zq_1^{-1}q_2^{-1})},
\ea
the term in blue is which we have not picked up its residue over its poles. To compensate this calculation, we insert 
\begin{align}
\begin{tikzpicture}[scale=2, baseline=(current  bounding  box.center)]
\draw[ultra thick](0,1)--(0,0.6);
\draw[ultra thick](0,0.4)--(0,-0.4);
\draw[ultra thick] (0,-0.6)--(0,-1);
\draw (0,0.5) circle (0.1);
\draw[ultra thick] (0.1,0.5)--(0.9,0.5);
\draw (1,0.5) circle (0.1);
\draw[ultra thick] (1,1)--(1,0.6);
\draw[ultra thick] (1,0.4)--(1,-0.5);
\draw (0,-0.5) circle (0.1);
\draw (-0.1,-0.5)--(-1,-0.5);
\draw (-1,0.5)--(-1,-1);
\node at (0.5,0.5) [above] {$x^-(z\mu_e)$};
\node at (-0.5,-0.5) [above] { $Y^{(1)}_{zq_1^{-2}q_2^{-1}}$};
\end{tikzpicture}
\end{align}
where the insertion of $Y^{(1)}(zq_1^{-2}q_2^{-1})$ is needed to convert $Y^{(d_2)}_{\cX_1}(zq_1^{-1}q_2^{-1})$ to $Y^{(d_1)}_{\cX_1}(zq_1^{-1}q_2^{-1})$. This leads to a new term needed to be compensated from another insertion, 
\ba
\chi^{{\rm BC}_2}_{(1,0)}\supset \frac{Y^{(2)}(z\mu_e)}{Y^{(1)}(zq_1^{-1}q_2^{-1})}+\frac{{\cob Y^{(1)}}(zq_1^{-2}q_2^{-1})}{Y^{(2)}(z\mu_e q_1^{-2}q_2^{-1})}.
\ea
The last insertion 
\begin{align}
\begin{tikzpicture}[scale=2, baseline=(current  bounding  box.center)]
\draw[ultra thick](0,1)--(0,0.6);
\draw[ultra thick](0,0.4)--(0,-0.4);
\draw[ultra thick] (0,-0.6)--(0,-1);
\draw (0,0.5) circle (0.1);
\draw[ultra thick] (0.1,0.5)--(0.9,0.5);
\draw (1,0.5) circle (0.1);
\draw[ultra thick] (1,1)--(1,0.6);
\draw[ultra thick] (1,0.4)--(1,-0.5);
\draw (0,-0.5) circle (0.1);
\draw (-0.1,-0.5)--(-1,-0.5);
\draw (-1,0.5)--(-1,-1);
\node at (-0.5,-0.5) [above] {$x^-(zq_1^{-2}q_2^{-1})$};
\node at (0.5,0.5) [above]{$Y^{(2)\ -1}_{z\mu_eq_1^{-2}q_2^{-1}}$};
\end{tikzpicture}
\end{align}
completes the calculation, and we arrive at 
\ba
\chi^{{\rm BC}_2}_{(1,0)}= Y^{(1)}(z)+\frac{Y^{(2)}(z\mu_e)}{Y^{(1)}(zq_1^{-1}q_2^{-1})}+\frac{Y^{(1)}(zq_1^{-2}q_2^{-1})}{Y^{(2)}(z\mu_e q_1^{-2}q_2^{-1})}+\frac{1}{Y^{(1)}(zq_1^{-3}q_2^{-2})}.\label{qq-BC2-10}
\ea

Similarly, starting from 
\begin{align}
\begin{tikzpicture}[scale=2, baseline=(current  bounding  box.center)]
\draw[ultra thick](0,1)--(0,0.6);
\draw[ultra thick](0,0.4)--(0,-0.4);
\draw[ultra thick] (0,-0.6)--(0,-1);
\draw (0,0.5) circle (0.1);
\draw[ultra thick] (0.1,0.5)--(0.9,0.5);
\draw (1,0.5) circle (0.1);
\draw[ultra thick] (1,1)--(1,0.6);
\draw[ultra thick] (1,0.4)--(1,-0.5);
\draw (0,-0.5) circle (0.1);
\draw (-0.1,-0.5)--(-1,-0.5);
\draw (-1,0.5)--(-1,-1);
\node at (0.5,0.5) [above] {$x^-(z)$};
\node at (-0.5,-0.5) [above] { $\substack{Y^{(1)}_{z\mu_e^{-1}q_1^{-1}q_2^{-1}}\\Y^{(1)}_{z\mu_e^{-1}q_1^{-2}q_2^{-1}}}$};
\end{tikzpicture}
\end{align}
we obtain the qq-character 
\ba
\chi^{{\rm BC}_2}_{(0,1)}=Y^{(2)}(z)+\frac{Y^{(1)}(z\mu_e^{-1}q_1^{-1}q_2^{-1})Y^{(1)}(z\mu_e^{-1}q_1^{-2}q_2^{-1})}{Y^{(2)}(zq_1^{-2}q_2^{-1})}+S(q_1)\frac{Y^{(1)}(z\mu_e^{-1}q_1^{-1}q_2^{-1})}{Y^{(1)}(z\mu_e^{-1}q_1^{-3}q_2^{-2})}\nn\\
+\frac{Y^{(2)}(zq_1^{-1}q_2^{-1})}{Y^{(1)}(z\mu_e^{-1}q_1^{-2}q_2^{-2})Y^{(1)}(z\mu_e^{-1}q_1^{-3}q_2^{-2})}+\frac{1}{Y^{(2)}(zq_1^{-3}q_2^{-2})},\label{qq-BC2-01}
\ea
where we used, 
\ba
\frac{S(q_1)}{S(q_1^{-2}q_2^{-1})}=1,
\ea
and $S(q_1^{-1})=S(q_2^{-1})=0$. 

Let us study the second example, G$_2$, whose web construction is given by 
\begin{align}
\begin{tikzpicture}[scale=2, baseline=(current  bounding  box.center)]
\draw[ultra thick](0,1)--(0,0.6);
\draw[ultra thick](0,0.4)--(0,-0.4);
\draw[ultra thick] (0,-0.6)--(0,-1);
\draw (0,0.5) circle (0.1);
\draw[ultra thick] (0.1,0.5)--(0.9,0.5);
\draw (1,0.5) circle (0.1);
\draw[ultra thick] (1,1)--(1,0.6);
\draw[ultra thick] (1,0.4)--(1,-0.5);
\draw (0,-0.5) circle (0.1);
\draw (-0.1,-0.5)--(-1,-0.5);
\draw (-1,0.5)--(-1,-1);
\node at (-0.1,0.5) [left] {$\Phi^{(d_2)}_{\cX_2}$};
\node at (0.5,0.5) [above] {$d_2=3$};
\node at (-0.5,-0.5) [below] {$d_1=1$};
\node at (0.1,-0.5) [right] {$\Phi^{\ast(d_2)}_{\cX_1}$};
\end{tikzpicture}
\end{align}
We examine the fundamental qq-characters of it. The only equation changes from the calculation for BC$_2$ is 
\ba
Y^{(d_2)}_{\cX_1}(z)=Y_{\lambda^{(1)}}(z)Y_{\lambda^{(1)}}(q_1^{-1}z)Y_{\lambda^{(1)}}(q_1^{-2}z).
\ea
The result is 
\ba
\chi^{{\rm G}_2}_{(1,0)}= Y^{(1)}(z)+\frac{Y^{(2)}(z\mu_e)}{Y^{(1)}(zq_1^{-1}q_2^{-1})}+\frac{Y^{(1)}(zq_1^{-2}q_2^{-1})Y^{(1)}(zq_1^{-3}q_2^{-1})}{Y^{(2)}(z\mu_eq_1^{-3}q_2^{-1})}+S(q_1)\frac{Y^{(1)}(zq_1^{-2}q_2^{-1})}{Y^{(1)}(zq_1^{-4}q_2^{-2})}\nn\\
+\frac{Y^{(2)}(z\mu_eq_1^{-2}q_2^{-1})}{Y^{(1)}(zq_1^{-3}q_2^{-2})Y^{(1)}(zq_1^{-4}q_2^{-2})}+\frac{Y^{(1)}(zq_1^{-5}q_2^{-2})}{Y^{(2)}(z\mu_eq_1^{-5}q_2^{-2})}+\frac{1}{Y^{(1)}(zq_1^{-6}q_2^{-3})},
\ea
and 
\ba
\chi^{{\rm G}_2}_{(0,1)}=Y^{(2)}(z)+\frac{Y^{(1)}(z\mu_e^{-1}q_1^{-1}q_2^{-1})Y^{(1)}(z\mu_e^{-1}q_1^{-2}q_2^{-1})Y^{(1)}(z\mu_e^{-1}q_1^{-3}q_2^{-1})}{Y^{(2)}(zq_1^{-3}q_2^{-1})}\nn\\
+S(q_1)S(q_1^2)\frac{Y^{(1)}(z\mu_e^{-1}q_1^{-1}q_2^{-1})Y^{(1)}(z\mu_e^{-1}q_1^{-2}q_2^{-1})}{Y^{(1)}(z\mu_e^{-1}q_1^{-4}q_2^{-2})}\nn\\
+S(q_1)S(q_1^2)\frac{Y^{(1)}(z\mu_e^{-1}q_1^{-1}q_2^{-1})Y^{(2)}(zq_1^{-2}q_2^{-1})}{Y^{(1)}(z\mu_e^{-1}q_1^{-3}q_2^{-2})Y^{(1)}(z\mu_e^{-1}q_1^{-4}q_2^{-2})}\nn\\
+\frac{Y^{(2)}(zq_1^{-1}q_2^{-1})Y^{(2)}(zq_1^{-2}q_2^{-1})}{Y^{(1)}(z\mu_e^{-1}q_1^{-2}q_2^{-2})Y^{(1)}(z\mu_e^{-1}q_1^{-3}q_2^{-2})Y^{(1)}(z\mu_e^{-1}q_1^{-4}q_2^{-2})}\nn\\
+S(q_1)S(q_1^2)\frac{Y^{(1)}(z\mu_e^{-1}q_1^{-1}q_2^{-1})Y^{(1)}(z\mu_e^{-1}q_1^{-5}q_2^{-2})}{Y^{(2)}(zq_1^{-5}q_2^{-2})}\nn\\
+S_3(q_1^{-1})\frac{Y^{(2)}(zq_1^{-2}q_2^{-1})}{Y^{(2)}(zq_1^{-4}q_2^{-2})}+S_3(q_1)\frac{Y^{(1)}(z\mu_e^{-1}q_1^{-5}q_2^{-2})Y^{(2)}(zq_1^{-1}q_2^{-1})}{Y^{(1)}(z\mu_e^{-1}q_1^{-2}q_2^{-2})Y^{(2)}(zq_1^{-5}q_2^{-2})}\nn\\
+S(q_1)S(q_1^2)S(q_1^{4}q_2^{1})\frac{Y^{(1)}(z\mu_e^{-1}q_1^{-1}q_2^{-1})}{Y^{(1)}(z\mu_e^{-1}q_1^{-6}q_2^{-3})}+\frac{Y^{(1)}(zq_1^{-3}q_2^{-2})Y^{(1)}(zq_1^{-4}q_2^{-2})Y^{(1)}(zq_1^{-5}q_2^{-2})}{Y^{(2)}(zq_1^{-4}q_2^{-2})Y^{(2)}(zq_1^{-5}q_2^{-2})}\nn\\
+S(q_1)S(q_1^2)\frac{Y^{(2)}(zq_1^{-1}q_2^{-1})}{Y^{(1)}(z\mu_e^{-1}q_1^{-2}q_2^{-2})Y^{(1)}(z\mu_e^{-1}q_1^{-6}q_2^{-3})}\nn\\
+S(q_1)S(q_1^2)\frac{Y^{(1)}(z\mu_e^{-1}q_1^{-3}q_2^{-2})Y^{(1)}(z\mu_e^{-1}q_1^{-4}q_2^{-2})}{Y^{(1)}(z\mu_e^{-1}q_1^{-6}q_2^{-3})Y^{(2)}(zq_1^{-4}q_2^{-2})}\nn\\
+S(q_1)S(q_1^2)\frac{Y^{(1)}(z\mu_e^{-1}q_1^{-3}q_2^{-2})}{Y^{(1)}(z\mu_e^{-1}q_1^{-5}q_2^{-3})Y^{(1)}(z\mu_e^{-1}q_1^{-6}q_2^{-3})}\nn\\
+\frac{Y^{(2)}(zq_1^{-3}q_2^{-2})}{Y^{(1)}(z\mu_e^{-1}q_1^{-4}q_2^{-3})Y^{(1)}(z\mu_e^{-1}q_1^{-5}q_2^{-3})Y^{(1)}(z\mu_e^{-1}q_1^{-6}q_2^{-3})}+\frac{1}{Y^{(2)}(zq_1^{-6}q_2^{-3})}.
\ea
The above results agree with the expressions for generators in $\cW^{G_2}$ obtained in \cite{Bouwknegt:1998da} (up to some apparent typos). 

We list several examples of the qq-characters for non-simply-laced quivers to conclude this section, where we always label the gauge nodes from left to right in the construction described in section \ref{s:BCFG}. 
\ba
\chi^{{\rm B}_3}_{(1,0,0)}= Y^{(1)}(z)+\frac{Y^{(2)}(z\mu_{12})}{Y^{(1)}(zq_1^{-1}q_2^{-1})}+\frac{Y^{(1)}(zq_1^{-2}q_2^{-1})Y^{(3)}(z\mu_{13})}{Y^{(2)}(z\mu_{12} q_1^{-2}q_2^{-1})}+\frac{Y^{(3)}(z\mu_{13})}{Y^{(1)}(zq_1^{-3}q_2^{-2})}\nn\\
+\frac{Y^{(1)}(zq_1^{-2}q_2^{-1})}{Y^{(3)}(z\mu_{13}q_1^{-2}q_2^{-1})}+\frac{Y^{(2)}(z\mu_{12}q_1^{-2}q_2^{-1})}{Y^{(1)}(zq_1^{-3}q_2^{-2})Y^{(3)}(z\mu_{13}q_1^{-2}q_2^{-1})}\nn\\
+\frac{Y^{(1)}(zq_1^{-4}q_2^{-2})}{Y^{(2)}(z\mu_{12}q_1^{-4}q_2^{-2})}+\frac{1}{Y^{(1)}(zq_1^{-5}q_2^{-3})}.\label{qq-B3-100}
\ea
\ba
\chi^{{\rm C}_3}_{(1,0,0)}=Y^{(1)}(z)+\frac{Y^{(2)}(z\mu_{12})}{Y^{(1)}(zq_1^{-1}q_2^{-1})}+\frac{Y^{(3)}(z\mu_{13})}{Y^{(2)}(z\mu_{12}q_1^{-1}q_2^{-1})}+\frac{Y^{(2)}(z\mu_{12}q_1^{-2}q_2^{-1})}{Y^{(3)}(z\mu_{13}q_1^{-2}q_2^{-1})}\nn\\
+\frac{Y^{(1)}(zq_1^{-3}q_2^{-2})}{Y^{(2)}(z\mu_{12}q_1^{-3}q_2^{-2})}+\frac{1}{Y^{(1)}(zq_1^{-4}q_2^{-4})}.
\ea

\section{Representations of DIM algebra}\label{a:rep-DIM}

The representations of the DIM algebra, as remarked in the beginning of this article, is labeled by two numbers, $p$ and $q$, which is induced from the following map of the central charges in DIM, 
\ba
(p,q):\ (\psi^+_0/\psi^-_0,\hat{\gamma})\mapsto (\gamma^{-2p},\gamma^q). 
\ea
In the correspondence with $(p,q)$ 5-branes, we set both of them to be integers, but generally we do not have to require so. To begin with, let us first write down the definition of the DIM algebra, 
\begin{eqnarray}
[\psi^\pm (z), \psi^\pm (w)] & = & 0,\label{pp-1}
\\
\psi^+ (z) \, \psi^- (w) & = & 
\frac{
g \lt( \gammah      z/w \rt) 
}{
g \lt( \gammah^{-1} z/w \rt) 
} 
\psi^- (w) \, \psi^+ (z)
\label{pp}
\\
\psi^\pm (z) \, x^+ (w) & = & g \lt( \gammah^{\pm \frac12} z/w \rt)    x^+ (w) \, \psi^\pm (z)
\label{px+}
\\
\psi^\pm (z) \, x^- (w) & = & g \lt( \gammah^{\mp \frac12} z/w \rt)^{-1} x^-(w) \, \psi^\pm (z)
\label{px-}
\\
x^\pm (z) \, x^\pm (w)  & = & g \lt( z/w \rt)^{\pm 1} x^\pm (w) \, x^\pm (z)
\label{xx1}
\\ 
\left[x^+(z), x^-(w)\right]  & = & 
\frac{(1-q_1) (1-q_2)}{  (1-q_3^{-1})}
\lt( \delta \lt( \gammah      w/z \rt)  \, \psi^+ \lt( \gammah^{  \frac12} w \rt) - 
    \delta \lt( \gammah^{-1} w/z \rt)  \, \psi^- \lt( \gammah^{- \frac12} w \rt)  
\rt), \nn\\
\label{xx2}
\end{eqnarray}
where the structure constant $g(z)$ is given by 
\ba
g(z)=\frac{(1-q_1z)(1-q_2z)(1-q_3z)}{(1-q_1^{-1}z)(1-q_2^{-1}z)(1-q_3^{-1}z)}=\frac{S(z)}{S(z^{-1})},\label{g-DI}
\ea
and in terms of mode expansion, $x^\pm(z)$ contains both positive and negative modes, while $\psi^\pm(z)$ only includes positive/negative modes, 
\ba
x^\pm(z)=\sum_{k\in\mathbb{Z}}x^\pm_kz^{-k},\quad \psi^\pm(z)=\sum_{k\in\mathbb{Z}_{\geq0}}\psi^\pm_{\pm k}z^{\mp k}.
\ea
The representation corresponding to $(p,q)=(n,1)$ is given by the following map to vertex operators, 
\ba
x^+(z)\mapsto u\gamma^nz^{-n}\eta(z),\quad x^-(z)\mapsto u^{-1}\gamma^{-n}z^n\xi(z),\quad \psi^\pm(z)\mapsto \gamma^{\mp n}\varphi^\pm(z),\quad \gammah\mapsto \gamma,
\ea
where 
\ba
\varphi^+(z)= \ :\eta(z\gamma^{1/2})\xi(z\gamma^{-1/2}):,\quad \varphi^-(z)= \ :\eta(z\gamma^{-1/2})\xi(z\gamma^{1/2}):,
\ea
and this class of representations give us the representation spaces to use in the unpreferred directions (even beyond the usual topological vertex formalism). 

$(p,q)=(1,0)$ representation is a special one we use for the preferred direction. The explicit expression reads 
\ba
&&x^+(z)\ket{v,\lambda}=b \sum_{x\in A(\lambda)}\delta(z/\chi_x)\Res_{z\rightarrow \chi_x}\frac{1}{z\cY_\lambda(z)}\ket{v,\lambda+x},\\
&&x^-(z)\ket{v,\lambda}=\gamma^{-1}b^{-1} \sum_{x\in R(\lambda)}\delta(z/\chi_x)\Res_{z\rightarrow \chi_x} z^{-1}\cY_\lambda(zq_3^{-1})\ket{v,\lambda-x},\\
&&\psi^\pm(z)\ket{v,\lambda}=\gamma^{-1}\Psi_\lambda(z)\ket{v,\lambda},
\ea
where $\psi^\pm(z)$ are respectively expansions with respect to $z^\mp$ of their matrix elements, and 
\ba
\Psi_\lambda(z)=\frac{\cY_\lambda(zq_3^{-1})}{\cY_\lambda(z)}.
\ea
This representation preserves the inner product $\langle v,\mu \ket{v,\lambda}=a_\lambda\delta_{\mu,\lambda}$ with respect to the dual representation acting on bra states as 
\ba
&&\bra{v,\lambda}\psi^\pm(z)=\gamma^{-1}\Psi_\lambda(z)\bra{v,\lambda},\\
&&\bra{v,\lambda}x^+(z)=-b\gamma^{-1}\sum_{x\in R(\lambda)}\bra{v,\lambda-x}\delta(z/\chi_x)\Res_{z\rightarrow \chi_x}\cY_\lambda(zq_3^{-1}),\\
&&\bra{v,\lambda}x^-(z)=-b^{-1}\sum_{x\in A(\lambda)} \bra{v,\lambda+x}\delta(z/\chi_x)\Res_{z\rightarrow \chi_x} \frac{1}{z\cY_\lambda(z)}.
\ea

\section{Contraction rules}

We list here all the constraction rules that are necessary in the computations for partition functions and AFS properties in this article: 

\ba
&&\contraction{}{\eta(z)}{}{\Phi_\lambda[u,v]}\eta(z)\Phi_\lambda[u,v]=\frac{1}{\cY_\lambda(z)}:\eta(z)\Phi_\lambda[u,v]:,\\
&&\contraction{}{\Phi_\lambda[u,v]}{}{\eta(z)}\Phi_\lambda[u,v]\eta(z)=-\frac{vq_3}{z}\frac{1}{\cY_\lambda(zq_3^{-1})}:\Phi_\lambda[u,v]\eta(z):,\\
&&\contraction{}{\xi(z)}{}{\Phi_\lambda[u,v]}\xi(z)\Phi_\lambda[u,v]=\cY_\lambda(\gamma^{-1}z):\xi(z)\Phi_\lambda[u,v]:,\\
&&\contraction{}{\Phi_\lambda[u,v]}{}{\xi(z)}\Phi_\lambda[u,v]\xi(z)=-\gamma^{-1}\frac{z}{v}\cY_\lambda(z\gamma^{-1}):\Phi_\lambda[u,v]\xi(z):,\\
&&\contraction{}{\varphi^-(z)}{}{\Phi_\lambda[u,v]}\varphi^-(z)\Phi_\lambda[u,v]= \ :\varphi^-(z)\Phi_\lambda[u,v]:,\\
&&\contraction{}{\Phi_\lambda[u,v]}{}{{\varphi^-}(z)}\Phi_\lambda[u,v]\varphi^-(z)=\gamma^2\frac{\cY_\lambda(z\gamma^{-1/2})}{\cY_\lambda(z\gamma^{-1/2}q_3^{-1})}:\Phi_\lambda[u,v]\varphi^-(z):,\\
&&\contraction{}{\varphi^+(z)}{}{\Phi_\lambda[u,v]}\varphi^+(z)\Phi_\lambda[u,v]=\frac{\cY_\lambda(z\gamma^{1/2} q_3^{-1})}{\cY_\lambda(z\gamma^{1/2})}:\varphi^+(z)\Phi_\lambda[u,v]:,\\
&&\contraction{}{\Phi_\lambda[u,v]}{}{{\varphi^+}(z)}\Phi_\lambda[u,v]\varphi^+(z)=\ :\Phi_\lambda[u,v]\varphi^+(z):,
\ea
\ba
&&\contraction{}{\eta(z)}{}{\Phi^\ast_\lambda[u,v]}\eta(z)\Phi^\ast_\lambda[u,v]=\cY_\lambda(z\gamma^{-1}):\eta(z)\Phi^\ast_\lambda[u,v]:,\\
&&\contraction{}{\Phi^\ast_\lambda[u,v]}{}{\eta(z)}\Phi^\ast_\lambda[u,v]\eta(z)=-\gamma^{-1}z/v\cY_\lambda(z\gamma^{-1}):\Phi^\ast_\lambda[u,v]\eta(z):,\\
&&\contraction{}{\xi(z)}{}{\Phi^\ast_\lambda[u,v]}\xi(z)\Phi^\ast_\lambda[u,v]=\frac{1}{\cY_\lambda(zq_3^{-1})}:\xi(z)\Phi^\ast_\lambda[u,v]:,\\
&&\contraction{}{\Phi^\ast_\lambda[u,v]}{}{\xi(z)}\Phi^\ast_\lambda[u,v]\xi(z)=-\frac{v}{z}\frac{1}{\cY_\lambda(z)}:\Phi^\ast_\lambda[u,v]\xi(z):,\\
&&\contraction{}{\varphi^-(z)}{}{\Phi^\ast_\lambda[u,v]}\varphi^-(z)\Phi^\ast_\lambda[u,v]= \ :\varphi^-(z)\Phi^\ast_\lambda[u,v]:,\\
&&\contraction{}{\Phi^\ast_\lambda[u,v]}{}{\varphi^-(z)}\Phi^\ast_\lambda[u,v]\varphi^-(z)=\gamma^{-2}\frac{\cY_\lambda(z\gamma^{-3/2})}{\cY_\lambda(z\gamma^{1/2})}:\Phi^\ast_\lambda[u,v]\varphi^-(z):,\\
&&\contraction{}{\varphi^+(z)}{}{\Phi^\ast_\lambda[u,v]}\varphi^+(z)\Phi^\ast_\lambda[u,v]=\frac{\cY_\lambda(z\gamma^{-1/2})}{\cY_\lambda(z\gamma^{-5/2})}:\varphi^+(z)\Phi^\ast_\lambda[u,v]:,\\
&&\contraction{}{\Phi^\ast_\lambda[u,v]}{}{\varphi^+(z)}\Phi^\ast_\lambda[u,v]\varphi^+(z)= \ :\Phi^\ast_\lambda[u,v]\varphi^+(z):.
\ea

\ba
&&\contraction{}{\eta(z)}{}{\bar{\Phi}^{\ast}_\lambda[u,v]}\eta(z)\bar{\Phi}^{\ast}_\lambda[u,v]=\frac{1}{\cY_\lambda(z\gamma)}:\eta(z)\bar{\Phi}^{\ast}_\lambda[u,v]:,\\
&&\contraction{}{\bar{\Phi}^{\ast}_\lambda[u,v]}{}{\eta(z)}\bar{\Phi}^{\ast}_\lambda[u,v]\eta(z)=-\frac{v}{z\gamma}\frac{1}{\cY_\lambda(z\gamma)}:\eta(z)\bar{\Phi}^{\ast}_\lambda[u,v]:,\\
&&\contraction{}{\xi(z)}{}{\bar{\Phi}^{\ast}_\lambda[u,v]}\xi(z)\bar{\Phi}^{\ast}_\lambda[u,v]=\cY_\lambda(z):\xi(z)\bar{\Phi}^{\ast}_\lambda[u,v]:,\\
&&\contraction{}{\bar{\Phi}^{\ast}_\lambda[u,v]}{}{\xi(z)}\bar{\Phi}^{\ast}_\lambda[u,v]\xi(z)=-\frac{zq_3}{v}\cY_\lambda(zq_3):\xi(z)\bar{\Phi}^{\ast}_\lambda[u,v]:,\\
&&\contraction{}{\varphi^-(z)}{}{\bar{\Phi}^\ast_\lambda[u,v]}\varphi^-(z)\bar{\Phi}^\ast_\lambda[u,v]= \ :\varphi^-(z)\bar{\Phi}^\ast_\lambda[u,v]:,\\
&&\contraction{}{\bar{\Phi}^\ast_\lambda[u,v]}{}{\varphi^-(z)}\bar{\Phi}^\ast_\lambda[u,v]\varphi^-(z)=\gamma^{2}\frac{\cY_\lambda(z\gamma^{5/2})}{\cY_\lambda(z\gamma^{1/2})}:\bar{\Phi}^\ast_\lambda[u,v]\varphi^-(z):,\\
&&\contraction{}{\varphi^+(z)}{}{\bar{\Phi}^\ast_\lambda[u,v]}\varphi^+(z)\bar{\Phi}^\ast_\lambda[u,v]=\frac{\cY_\lambda(z\gamma^{-1/2})}{\cY_\lambda(z\gamma^{3/2})}:\varphi^+(z)\bar{\Phi}^\ast_\lambda[u,v]:,\\
&&\contraction{}{\bar{\Phi}^\ast_\lambda[u,v]}{}{\varphi^+(z)}\bar{\Phi}^\ast_\lambda[u,v]\varphi^+(z)= \ :\bar{\Phi}^\ast_\lambda[u,v]\varphi^+(z):.
\ea

\ba
&&\contraction{}{\eta(z)}{}{\bar{\Phi}_\lambda[u,v]}\eta(z)\bar{\Phi}_\lambda[u,v]=\cY_\lambda(zq_3):\eta(z)\bar{\Phi}_\lambda[u,v]:,\\
&&\contraction{}{\bar{\Phi}_\lambda[u,v]}{}{\eta(z)}\bar{\Phi}_\lambda[u,v]\eta(z)=-\frac{z}{v}\cY_\lambda(z):\bar{\Phi}_\lambda[u,v]\eta(z):,\\
&&\contraction{}{\xi(z)}{}{\bar{\Phi}_\lambda[u,v]}\xi(z)\bar{\Phi}_\lambda[u,v]=\frac{1}{\cY_\lambda(z\gamma)}:\xi(z)\bar{\Phi}_\lambda[u,v]:,\\
&&\contraction{}{\bar{\Phi}_\lambda[u,v]}{}{\xi(z)}\bar{\Phi}_\lambda[u,v]\xi(z)=-\frac{v}{z\gamma}\frac{1}{\cY_\lambda(z\gamma)}:\xi(z)\bar{\Phi}_\lambda[u,v]:,
\ea
\ba
&&\contraction{}{\varphi^-(z)}{}{\bar{\Phi}_\lambda[u,v]}\varphi^-(z)\bar{\Phi}_\lambda[u,v]= \ :\varphi^-(z)\bar{\Phi}_\lambda[u,v]:,\\
&&\contraction{}{\bar{\Phi}_\lambda[u,v]}{}{\varphi^-(z)}\bar{\Phi}_\lambda[u,v]\varphi^-(z)=\gamma^{-2}\frac{\cY_\lambda(z\gamma^{-1/2})}{\cY_\lambda(z\gamma^{3/2})}:\bar{\Phi}_\lambda[u,v]\varphi^-(z):,\\
&&\contraction{}{\varphi^+(z)}{}{\bar{\Phi}_\lambda[u,v]}\varphi^+(z)\bar{\Phi}_\lambda[u,v]=\frac{\cY_\lambda(z\gamma^{5/2})}{\cY_\lambda(z\gamma^{1/2})}:\varphi^+(z)\bar{\Phi}_\lambda[u,v]:,\\
&&\contraction{}{\bar{\Phi}_\lambda[u,v]}{}{\varphi^+(z)}\bar{\Phi}_\lambda[u,v]\varphi^+(z)= \ :\bar{\Phi}_\lambda[u,v]\varphi^+(z):.
\ea

\ba
&&\contraction{}{\eta^{(d_i)}(z)}{}{\tilde\Phi^{\ast(d_i)}_{\cX_j}}\eta^{(d_i)}(z)\tilde\Phi^{\ast(d_i)}_{\cX_j}=Y^{(d_j)}_{\cX_j}(z\gamma_i/\gamma_j^{2}):\eta^{(d_i)}(z)\tilde\Phi^{\ast(d_i)}_{\cX_j}:,\\
&&\contraction{}{\tilde\Phi^{\ast(d_i)}_{\cX_j}}{}{\eta^{(d_i)}(z)}\tilde\Phi^{\ast(d_i)}_{\cX_j}\eta^{(d_i)}(z)=\lt(\prod_{k=0}^{d_i/d_j-1} -q_1^{-kd_j}\gamma_i^{-1}z/v_j\rt)Y^{(d_i)}_{\cX_j}(z\gamma_i^{-1}):\eta^{(d_i)}(z)\tilde\Phi^{\ast(d_i)}_{\cX_j}:,\\
&&\contraction{}{\xi^{(d_i)}(z)}{}{\tilde\Phi^{\ast(d_i)}_{\cX_j}}\xi^{(d_i)}(z)\tilde\Phi^{\ast(d_i)}_{\cX_j}=\frac{1}{Y^{(d_j)}_{\cX_j}(z\gamma_j^{-2})}:\xi^{(d_i)}(z)\tilde\Phi^{\ast(d_i)}_{\cX_j}:,\\
&&\contraction{}{\tilde\Phi^{\ast(d_i)}_{\cX_j}}{}{\xi^{(d_i)}(z)}\tilde\Phi^{\ast(d_i)}_{\cX_j}\xi^{(d_i)}(z)=\lt(\prod_{k=0}^{d_i/d_j-1} -q_1^{-kd_j}z/v_j\rt)^{-1}\frac{1}{Y^{(d_i)}_{\cX_j}(z)}:\xi^{(d_i)}(z)\tilde\Phi^{\ast(d_i)}_{\cX_j}:.
\ea

\section{Converting formulae for trivalent vertex}\label{a:triv-vert}

When trivalent vertices are involved in the web construction, we need to know how to convert an element $x^\pm(z)$ acting from the left-hand side to them to the action from the right-hand side. 

The first class of operators we want to consider are $\ket{\Omega^{(n)},\vec{\alpha}}\rangle:=\sum_\lambda a_\lambda(v) \ket{v,\lambda} \otimes \ket{v\alpha_1,\lambda}\otimes \ket{v\alpha_2,\lambda} \otimes \dots \ket{v\alpha_n,\lambda}$, which can be constructed from trivalent vertices plus one reflection operator. When $n=1$, $\ket{\Omega^{(n)},\vec{\alpha}}\rangle$ coincides with the reflection operator $\ket{\Omega,\alpha}\rangle$. The converting formula it can be derived as 
\ba
(x^+(z)\otimes 1)\sum_\lambda a_\lambda(v)\ket{v,\lambda}\otimes \ket{v\alpha,\lambda}=\sum_\lambda\sum_{x\in A(\lambda)}\delta(z/\chi_x)a_\lambda(v)\Res_{z\rightarrow \chi_x}\frac{1}{z\cY^{(1)}_\lambda(z)}\ket{v,\lambda+x}\otimes\ket{v\alpha,\lambda}\nn\\
=\sum_\lambda\sum_{x\in R(\lambda)}\delta(z/\chi_x)\frac{a_{\lambda-x}(v)}{a_{\lambda}(v)}a_\lambda(v)\Res_{z\rightarrow \chi_x}\frac{1}{z\cY^{(1)}_{\lambda-x}(z)}\ket{v,\lambda}\otimes\ket{v\alpha,\lambda-x}\nn\\
=\sum_\lambda\sum_{x\in R(\lambda)}\delta(z/\chi_x)\gamma^{-1}a_\lambda(v)\Res_{z\rightarrow \chi_x}z^{-1}\cY^{(1)}_{\lambda}(zq_3^{-1})\ket{v,\lambda}\otimes\ket{v\alpha,\lambda-x}\nn\\
=\sum_\lambda\sum_{x\in R(\lambda)}\delta(z\alpha/\chi_x\alpha)\gamma^{-1}a_\lambda(v)\Res_{z\rightarrow \chi_x\alpha}z^{-1}\cY^{(2)}_{\lambda}(z q_3^{-1})\ket{v,\lambda}\otimes\ket{v\alpha,\lambda-x}\nn\\
=(1\otimes x^-(z\alpha))\sum_\lambda a_\lambda(v)\ket{v,\lambda}\otimes \ket{v\alpha,\lambda},\nn
\ea
where we used 
\ba
\Res_{z\rightarrow \chi_x}\frac{\cY_{\lambda}(z)}{\cY_{\lambda-x}(z)}=\chi_x\frac{(1-q_1)(1-q_2)}{(1-q_3^{-1})}.
\ea
That is to say, 
\ba
(x^+(z)\otimes 1)\ket{\Omega,\alpha}\rangle=-(1\otimes x^-(z\alpha))\ket{\Omega,\alpha}\rangle.
\ea

Note that whenever a trivalent vertex or its generalization appears in the web construction, all its legs are  contracted with some vertex operators labeled respectively by a Young diagram. Let us consider the case that all the legs of $\ket{\Omega^{(n)},\Phi^{(n_i)},\vec{\alpha}}\rangle$ are attached to the vertices $\Phi$, and we define 
\ba
\ket{\Omega^{(n)},\Phi^{(n_i)},\vec{\alpha}}\rangle:=\sum_\lambda a_\lambda(v) \ket{v,\lambda} \otimes \ket{v\alpha_1,\lambda}\otimes \bigotimes_{i=2}^n\Phi^{(n_i)}_{\lambda}[u_i,v\alpha_i].
\ea
The converting formula for it is given by 
\ba
(x^+(z)\otimes 1)\ket{\Omega^{(n)},\Phi^{(n_i)},\vec{\alpha}}\rangle=-\lt(\prod_{i=2}^n-(u_iv\alpha_i)^{-1}(z\alpha_i/\gamma)^{n_i+1}\rt)(1\otimes x^-(z\alpha_1))\ket{\Omega^{(n)},:\Phi^{(n_i)}\eta(z\alpha_i)^{-1}:,\vec{\alpha}}\rangle.\nn\\
\ea

Another class of operators we need as the generalization of trivalent vertex take the form $I^{(n)}(\vec{\alpha}):=\sum_\lambda a_\lambda(v) \ket{v,\lambda} \otimes \bra{v\alpha_1,\lambda}\otimes \bra{v\alpha_2,\lambda}\ \otimes \dots \bra{v\alpha_n,\lambda}$. For $n=1$, this operator is merely a twisted identity operator, i.e. $I^{(1)}(1)={\bf 1}$, and we have in general 
\ba
x^+(z)I^{(1)}(\alpha)=I^{(1)}(\alpha)x^+(z\alpha).
\ea
To generalize this identity, we similarly define 
\ba
I^{(n)}(\Phi^{\ast(n_i)},\vec{\alpha}):=\sum_\lambda a_\lambda(v) \ket{v,\lambda} \otimes \bra{v\alpha_1,\lambda}\otimes\bigotimes_{i=2}^n \Phi^{\ast(n_i)}[u_i,v\alpha_i],
\ea
and its converting formula reads 
\ba
(x^+(z)\otimes 1)I^{(n)}(\Phi^{\ast(n_i)},\vec{\alpha})=\lt(\prod_{i=2}^n u_i\gamma^{1+n_i}(z\alpha_i)^{-n_i}\rt) I^{(n)}(:\Phi^{\ast(n_i)}\xi(z\alpha_i)^{-1}:,\vec{\alpha})(1\otimes x^+(z\alpha_1)).
\ea

\section{AFS property with respect to $x^-$}\label{AFS-xm}

As we only listed the AFS properties associated to $x^+(z)$ acting from the unpreferred direction, we list here those associated to $x^-(z)$. 

For $\bar{\Phi}^\ast$, we have 
\begin{align}
\begin{tikzpicture}[baseline=(current  bounding  box.center)]
\draw (-1,1)--(-1,-1);
\draw (-2,0)--(-1,0);
\draw[ultra thick] (-1-0.1,0.1)--(-1+0.1,-0.1);
\draw[ultra thick] (-1+0.1,0.1)--(-1-0.1,-0.1);
\node at (-1,1) [above] {$x^-(z)$};
\draw (0.4,0)--(0.6,0);
\draw (3.5,1)--(3.5,-1);
\draw (2.5,0)--(3.5,0);
\draw[ultra thick] (3.5-0.1,0.1)--(3.5+0.1,-0.1);
\draw[ultra thick] (3.5+0.1,0.1)--(3.5-0.1,-0.1);
\node at (3.5,-1) [below] {$q^{-1}_3x^-(z)$};
\node at (2.5,0) [left] {$\psi^+(zq_3)$};
\draw (4.9,0.05)--(5.1,0.05);
\draw (4.9,-0.05)--(5.1,-0.05);
\draw (9,1)--(9,-1);
\draw (8,0)--(9,0);
\draw[ultra thick] (9-0.1,0.1)--(9+0.1,-0.1);
\draw[ultra thick] (9+0.1,0.1)--(9-0.1,-0.1);
\node at (8,0) [left] {$-x^+(zq_3)$};
\node at (9,1) [above] {$\gammah^{-p}:\xi(z)\bar{\xi}(zq_3):$}; 
\end{tikzpicture}
\end{align}

For $\bar{\Phi}$, we have 
\begin{align}
\begin{tikzpicture}[baseline=(current  bounding  box.center)]
\draw (0.5,1)--(0.5,-1);
\draw (0.5,0)--(1.5,0);
\draw[ultra thick] (0.5-0.1,0.1)--(0.5+0.1,-0.1);
\draw[ultra thick] (0.5-0.1,-0.1)--(0.5+0.1,0.1);
\node at (0.5,1) [above] {$q_3^{-1}x^-(z)$};
\draw (2.9,0)--(3.1,0);
\draw (4,1)--(4,-1);
\draw (4,0)--(5,0);
\draw[ultra thick] (4-0.1,0.1)--(4+0.1,-0.1);
\draw[ultra thick] (4-0.1,-0.1)--(4+0.1,0.1);
\node at (4,-1) [below] {$x^-(z)$};
\draw (5.9,0.05)--(6.1,0.05);
\draw (5.9,-0.05)--(6.1,-0.05);
\draw (7,1)--(7,-1);
\draw (7,0)--(8,0);
\draw[ultra thick] (7-0.1,0.1)--(7+0.1,-0.1);
\draw[ultra thick] (7-0.1,-0.1)--(7+0.1,0.1);
\node at (8,0) [right] {$\gamma^{-1}x^+(z\gamma)$};
\node at (7,1) [above] {$\xi(z)\bar{\eta}^{-1}(z\gamma)$};
\end{tikzpicture}
\end{align}

At last, for the half-blood vertex, we write down 
\begin{align}
\begin{tikzpicture}[scale=2,yscale=-1, baseline=(current  bounding  box.center)]
\draw[ultra thick](0,0)--(0,-0.4);
\draw[ultra thick] (0,-0.6)--(0,-1);
\draw (0,-0.5) circle (0.1);
\draw (-0.1,-0.5)--(-0.5,-0.5);
\node at (0,-1) [above] {$:\prod_{k=0}^{d_i/d_j-1}x^-(zq_1^{-kd_j}\gamma_j^2/\gamma_i^2):$};
\node at (-0.5,-0.5) [left] {$\psi^-(z)^{d_i/d_j}$};
\draw (0.9,-0.5)--(1.1,-0.5);
\draw (1,-0.6)--(1,-0.4);
\draw[ultra thick](4.5,0)--(4.5,-0.4);
\draw[ultra thick] (4.5,-0.6)--(4.5,-1);
\draw (4.5,-0.5) circle (0.1);
\draw (4.4,-0.5)--(4,-0.5);
\node at (4,-0.5) [left] {$\sum_{l=0}^{d_i/d_j-1}x^-(zq_1^{-ld_j}\gamma_i)$};
\node at (4.5,-1) [above] {$:\tilde{\sigma}^{+(d_i)}_j(zq_1^{-ld_j}\gamma_i^{1/2})\prod_{\substack{k=0\\k\neq l}}^{d_i/d_j-1}x^-(zq_1^{-kd_j}\gamma_j^2/\gamma_i^2):$};
\draw (5.4,-0.475)--(5.6,-0.475);
\draw (5.4,-0.525)--(5.6,-0.525);
\draw[ultra thick](7,0)--(7,-0.4);
\draw[ultra thick] (7,-0.6)--(7,-1);
\draw (7,-0.5) circle (0.1);
\draw (6.9,-0.5)--(6.5,-0.5);
\node at (7,0) [below] {$x^-(z)$};
\end{tikzpicture}
\end{align}
where 
\ba
\tilde{\sigma}^{+(d_i)}_j= \ :\xi^{(d_i)}(z)\tilde{\xi}^{(d_i)}_j(z)^{-1}:.
\ea

\bibliography{BCFG}

\end{document}